\documentclass[chap]{thesis}
\usepackage{graphicx}  
\usepackage{amssymb}   
\usepackage{color}

\usepackage[hang]{caption}      
\begin{document}

%
    
%
\thesistitle{\bf First Principles Study of Intein Reaction Mechanisms}        
\author{Philip Shemella}        
\degree{Doctor of Philosophy}        
\department{Physics}        
\signaturelines{6}     
\thadviser{Saroj K. Nayak}
\memberone{Georges Belfort}        
\membertwo{Marlene Belfort}        
\memberthree{Angel E. Garc\'{i}a}
\memberfour{Shekhar Garde}
\memberfive{Gy\"{o}rgy Korniss}
\submitdate{May 2008\\(For Graduation May 2008)}        
\copyrightyear{2008}   

%
\titlepage     
\copyrightpage         
\tableofcontents        
\listoftables          
\listoffigures         

 
\specialhead{ACKNOWLEDGMENT}
\small
To my doctoral committee:  Professors M. Belfort, G. Belfort, Garde, Garc\'{i}a, Korniss, and Nayak. Thank you for taking the time to comment and criticize and to make this work as coherent as possible.

To those who continue to provide scientific support. For the many helpful dialogues I would like to thank the groups of Marlene Belfort, Georges Belfort, Shekhar Garde, Patrick Van Roey, Chunyu Wang, Angel Garc\'{i}a; and in particular Gil Amitai and Brian Pereira for helpful discussions as I gradually become acclimatized to biology.

To my research group: Yiming, Kala, Xihong, Li, Yu and Brock; thank you for the suggestions and support you've provided.  I am very grateful for the direction, guidance, and friendship that Professor Nayak has offered and I am very fortunate to have him as an adviser.

To those who have provided computational support: Thank you to those at NCSA, SCOREC, and CCNI and in particular Mike Kupferschmid, Harriet Borton, and Adam Todorski, who have shared their expertise.

To the staff of the Physics Department:  Nicole, Jun, Susan, Mary Ellen, and Joan - thank you for the constant support.

To my grad school friends: Dave, Matt, Mike, Ben, Amanda, Paul, Tom, Churamani, Scott, Bal\'{a}zs, Anthony, Johnny, and to friends from other eras: Matt, Josh, Louis, Ellis. Thank you all for the wonderful and necessary distractions you've provided.

To Caterina, we met the first day of grad school and as it does, time went too fast.  Thank you for being my friend.  I wish you the best things in life.

To my family. My parents Ted and Debby, thank you for constantly encouraging independence and offering passive support, whether I realized it or not.  To my brothers Nathan, David, and Stephen; thanks for being such great guys most of the time.
\\

\begin{center}
\small
\textit{I know but one freedom and that is the freedom of the mind.}\\
-- Antoine de Saint Exup\'{e}ry
\end{center}

 
\specialhead{ABSTRACT}
Protein splicing is an autocatalytic reaction where two flanking sequences (exteins) are excised and ligated. The enzymatic protein sequence that lies between the exteins, known as the intein, is extremely efficient at protein splicing and has been utilized for biotechnological applications. The characterization of intein reaction mechanisms is important for understanding how and why certain mutations may be used to control the splicing and cleavage reactions, as well as tuning the reaction rate without affecting the mechanism.  In combination with crystallographic structures as well as both site-directed and random mutagenesis, we have studied the reaction mechanisms for intein splicing as well as for cleavage at the C-terminus using first principle quantum mechanical simulations.
	Previous experimental studies have shown that mutation at a critical N-terminal residue of the intein resulted in splicing inhibition. Despite this inhibition of the overall splicing reaction, peptide bond cleavage isolated at the C-terminal may still occur independently. With an aspartate to glycine mutation, the ``cleavage mutant'' was found to react more rapidly in a low pH environment.  We have characterized the pH dependent C-terminal cleavage reaction and studied the effect of mutation on the energy barrier, and have provided for the first time an atomic level understanding of this important process.
	Next, we have extended our computational study to address the overall intein splicing mechanism. The splicing reaction is a highly synchronized chemical process where the effect of mutation can accelerate, decelerate, or partially or completely inhibit steps along the reaction.  We have focused our study on the energetic effect of mutation on the reaction profile and corresponding protein structure. From this, we have made a prediction for the splicing mechanism that utilizes the highly conserved amino acids and explicitly describes the behavior of protons. An explanation for the experimental inhibition of splicing with the aspartate to glycine mutation is also presented.
	In summary, with a series of quantum mechanical calculations ranging from gas phase, to an implicit solvent scheme, to combined quantum/classical simulations, we have provided insight into some of the key steps of intein reactions. These studies may be exploited for many applications involving inteins including molecular switches and sensors as well as controlled drug delivery.

\chapter{INTRODUCTION}

\section{Computer simulations in chemistry}
Nestled between experiment and pure theory, computational chemistry has become an integral tool for researchers working in physics, chemistry, and biology, as well as nanotechnology and biotechnology.  The role of simulation is to verify and confirm or to predict and suggest experimental studies.  Computer simulations allow the researcher to access states both visible and invisible to experiment, and make predictions based on this knowledge.  A chemical reaction may be quantified by the amount of reactants, the amount of products, and the time elapsed. To explain a mechanism and molecular structure and energies on the atomic level, computational methods are important. 

The field of computational chemistry spans length and time scales, and the desired level of accuracy is important to know prior to running simulations.  To simulate protein folding, which requires an extremely long simulation trajectory, amino acids may be ``coarse-grained,'' where the atomic description of each side chain is aggregated into a composite value. To achieve long trajectories this approximation as well as others are essential.  However, to calculate the pK$_a$ of a side chain or the chemical shifts \textit{via} nuclear magnetic resonance (NMR), not only will an atomic level description be necessary, but also a method that can calculate observable properties from first principles is required.

\section{Computational quantum mechanics}
The energies associated with bond breakage and formation are an essential property for an enzymatic processes. For example, a change in energy barrier of $\sim$1.4 kcal/mol corresponds to an order of magnitude change in the reaction rate. States observed at equilibrium may be predicted based on relative energies between structures.  To computationally access the energy of the system, and to do so not only for  equilibrium structures but also for transition states, first principles electronic structure calculations are required.  Using an all-electron method, the electron orbitals are considered variable and flexible, and they depend on neighboring atoms and environment.  This is important because the chemistry at transition states may vary greatly from equilibrium structures: instead of four bonds, carbon atoms may have three or five bonds during a chemical reaction.  Transition states are where quantum mechanical principles dominate.  By solving the Schr\"{o}dinger equation for all electrons, and relaxing their orbital positions and therefore allowing the electron density to vary, an accurate description of the system can be obtained that is useful for understanding fundamental chemistry both near and far from equilibrium.

\section{Inteins}
\subsection{Overview}
\begin{figure}
\centering   
\includegraphics[scale=0.7]{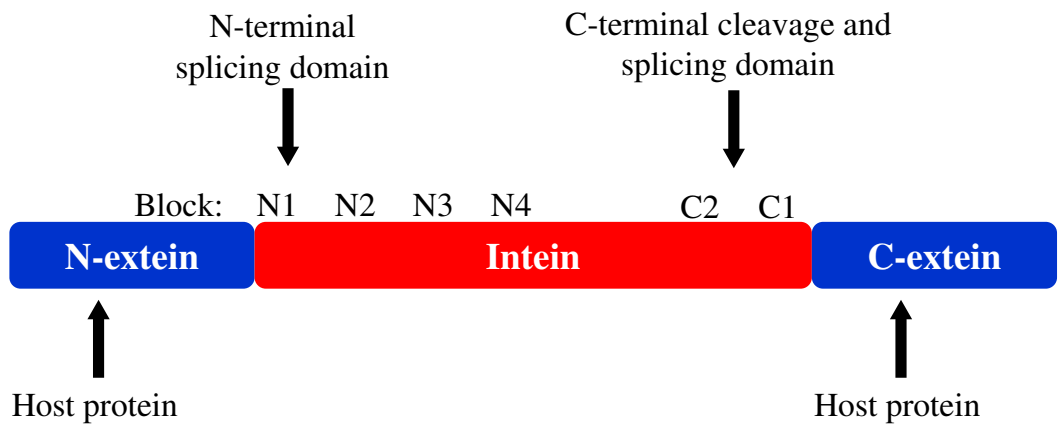}
\caption{\label{fig:motif} Schematic intein and N- and C-exteins.  Splicing motifs contain highly conserved amino acids, such as N1-Cys1, N3-His10, C2-Asp5, and C1-His7, C1-Asn8, C1-Cys+1}
\end{figure}

Protein splicing involves the autocatalytic release of a peptide segment, termed an intein, with the joining of two flanking protein sequences (exteins) \cite{belforts, perler1994pse}.  Inteins are autocatalytic proteins that exist in all three domains of life. Experiments have identified key reaction steps in protein splicing whereas sequence comparisons have revealed the conserved amino acids required for this reaction. Figure \ref{fig:motif} shows a schematic for conserved intein residues and their corresponding block (C or N) designation.  Experimental mutational studies have been carried out to further control the protein splicing reaction \cite{hiraga2005mas,shingledecker2000rcr}. For example, by mutating the first residue at the N-terminus (N1 block) of the intein from Cys to Ala (N1-Cys1Ala), the first step of the splicing reaction, namely the N-terminal N--S shift\footnote{Atoms are annotated with one letter, \textit{i.e.} H = hydrogen.  Amino acids are annotated with three letters, \textit{i.e.} His = histidine.}, is inhibited, thus isolating the C-terminal cleavage reaction \cite{paulus2000psa}.  Mutation schemes that control the reaction rate and/or the specific products could be exploited in many biotechnological applications such as bioseparations \cite{miao2005ssa,banki2005sbu}, drug development \cite{paulus2003itp}, and molecular sensors \cite{mootz2002pst, muralidharan2006ple}.  

\subsection{Computational methodology}
First principles density functional theory was used to study the electronic structure of protein systems.  Various methods were utilized for accurate description of the intein model system:  the electrostatic environment was either neglected (gas phase), approximated (implicit solvent), or treated explicitly (QM/MM).  Various levels of theory were used:  classical molecular mechanics, semi-empirical quantum mechanics, and first principles quantum mechanics.

\subsection{Results: Splicing}
We have studied the protein splicing mechanism for inteins.  The role of mutations for the \textit{Mtu} recA intein is considered, especially the C2-Asp5Gly mutation that inhibits splicing and creates the C-terminal cleavage mutant (CM).  A mechanism is proposed that is consistent with crystal \cite{vanroey2007cam, mizutani2002psr, poland2000sii, klabunde1998csg, duan1997csp, ichiyanagi2000csa, sun2005csi} or NMR structures \cite{johnson2007nsk}  and mutagenesis results \cite{wood1999gsy, southworth2000aps, chen2000psa, mills2004psp,brian}, and includes an atomic-level description of the steps of intein splicing.  From this mechanism we can gain a detailed understanding of a reaction that is useful for biotechnology applications.

\begin{figure}
\centering   
\includegraphics[scale=0.6]{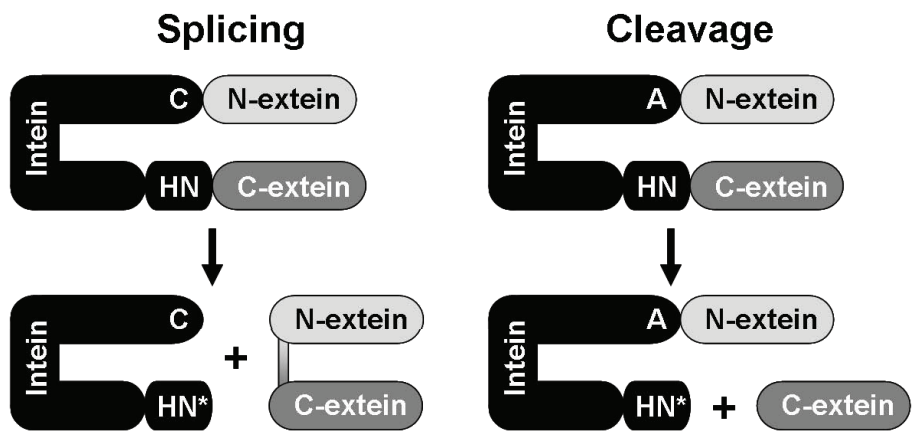}
\caption{\label{fig:conserved1} Intein reactions: splicing and cleavage (after N1-Cys1Ala mutation). C, A, H, N, and N* represent Cys, Ala, His, Asn, and succinimide, respectively}
\end{figure}


Experiments have identified key reaction steps in protein splicing as well as revealed the conserved amino acids required for this reaction. The intein splicing reaction in terms of conserved amino acids is shown in Figure \ref{fig:conserved1}.  Intein splicing starts with an N-S shift and thioester formation at the N-terminal N1-Cys1 (see Appendix A for all natural amino acid structures).   Next is transesterification, where the C-terminal Cys+1 attacks the carbon atom of the N-terminal thioester.  Following that, C1-Asn8 cyclization occurs and finally the S-N shift leads to separate products of fused N- and C-exteins and the released intein sequence.  

\subsection{Results: C-terminal cleavage}
The cleavage of the peptide bond between the intein and the C-extein was investigated.  This is one step that was isolated by inhibiting other steps in the overall splicing reaction, and was found to occur more rapidly in a low pH environment.  Our first principles calculations provide an atomic-level description of C-terminal cleavage that uses a hydronium ion to include a pH--induced reaction.  Various model systems are used and results from these calculations are compared with experiment with good agreement.

\begin{figure}
\centering    
\includegraphics[scale=0.5]{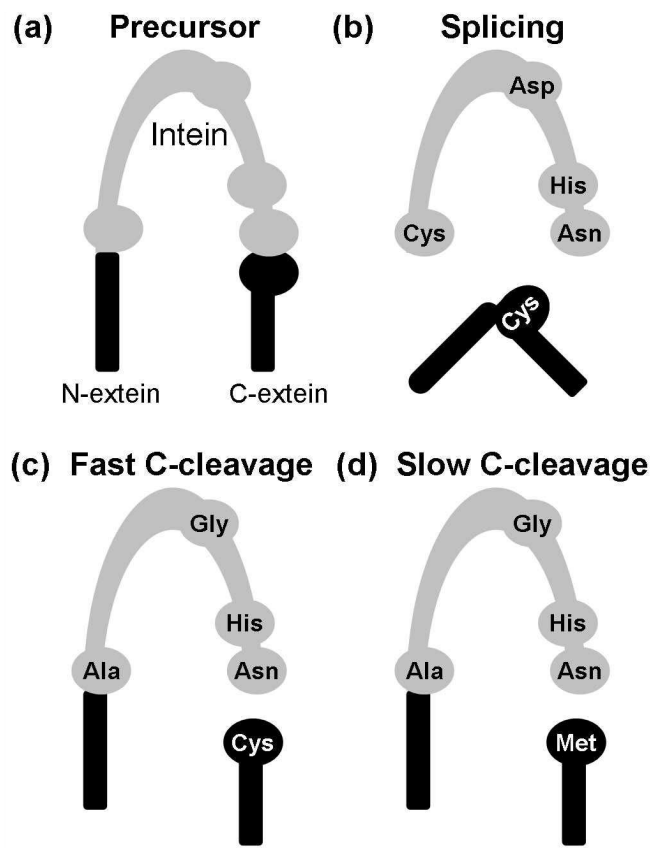}
\caption{\label{fig:drawing} Different reaction rates and reaction products are observed for various mutations to the intein or extein sequence. Intein and extein precursor (top left) and two possible products: splicing product (top right), and fast (bottom left) and slow (bottom right) C-terminal cleavage.}
\end{figure}

Experimental mutagenesis studies have been carried out to further control the protein splicing reaction (see Figure \ref{fig:motif} and Appendix \ref{tab:sequence} for intein sequences); for example, by mutating the first residue at the N-terminus of the intein from N1-Cys1Ala, the first step of the splicing reaction, namely the N-terminal N-S shift, is inhibited, thus isolating the C-terminal cleavage reaction (see Figure \ref{fig:drawing}) \cite{paulus2000psa}.  

\subsubsection{Environmental effects of single amino acid mutation}

In a related context, experimental results have shown for the \textit{Mycobacterium tuberculosis} (\textit{Mtu}) recA mini-intein that the C2-Asp5Gly mutation\footnote{C2-Asp5 indicates aspartate at the fifth position of the C2-block.  C2-Asp5Gly indicates that aspartate is mutated to glycine. This mutation was previously referred to as D422G.} creates the highly active C-terminal cleavage mutant (CM) that was experimentally reported to be more active in a low pH environment \cite{wood1999gsy,wood2000oss,belfort2005gsa,woodthesis}.

Interestingly, once splicing is inhibited, the C-terminal Cys+1 residue (which is the first amino acid of the C-terminal extein or C-extein) is found to be functionally unnecessary. Wood \textit{et al.} have found that this amino acid regulates the reaction rate but does not alter the mechanism \cite{wood2000oss}.  Furthermore, since the CM is found to be exceedingly reactive in a low pH environment, they have utilized Met, which is the native N-terminus of the C-extein sequence, to slow down the reaction by an order of magnitude. In this experiment, three proteins of various sizes were analyzed with the Cys(+1)Met C-extein mutation: Thymidylate Synthase (31.5 kDa), Hfq Protein (18 kDa), and rh aFGF (14 kDa). For these proteins the ratios of reaction rates between the Cys(+1)Met mutants are found to be 12.0, 5.0, and 7.86, respectively \cite{wood2000oss}.  Figure \ref{fig:drawing} shows a schematic of the intein precursor and products based on these results \cite{muralidharan2006ple,wu2002imp}, although the exact mechanisms that govern the splicing and cleavage reactions are not understood at the atomic level.  

In order to obtain an atomic-level understanding on the reaction mechanisms as well as on the effect of mutation on the reaction barrier, we have carried out detailed quantum mechanical simulations on intein C-terminal cleavage reactions.  We describe pH dependent C-terminal cleavage calculations for the \textit{Mtu} recA intein; performed with semi-empirical, QM gas phase, QM implicit solvent, and combined QM/MM calculations \cite{shemella, shemella2}. Harnessing the C-terminal cleavage reaction may allow for an intein-based delivery device, where the reaction is triggered by a certain stimulus.

\section{Research outline}
The goal of our research is to characterize intein splicing and cleavage reaction mechanisms for use in biotechnology applications, and to compliment the experimental efforts of collaborators at RPI and the Wadsworth Center.  Understanding the splicing and cleavage mechanisms on the atomic level will provide input for the engineering of smaller and faster intein reactions, as well as a controllable reaction.  In addition, the possibility of a synthetic intein is appealing for biotechnological application due to the ability to control parameters such as size, reaction, and function \cite{lew1998psv}.

Our computational results indicate that certain mutations either inhibit or enhance specific reaction steps of the overall splicing reaction, a conclusion that is consistent with experiment. With quantum mechanical simulations, intermediate states may be isolated and studied in the context of altering the molecular triggers and inhibitors that impact protein splicing with inteins.  The ability to study precursor, intermediate, and post-reaction product states is extremely useful and carried out with first principles methods.


\chapter{COMPUTATIONAL METHODS}

\section{Introduction}
\begin{figure}[h]
\centering    
\includegraphics[scale=0.5]{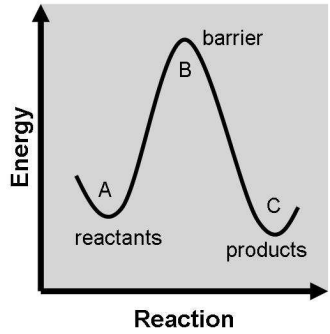}
\caption{\label{fig:barrier} Illustrative energy barrier for an enzymatic process (one dimensional reaction coordinate).  The difference in energy between the reactants ($E_{A}$) and the barrier ($E_{B}$) is known as the energy of activation, $\Delta E^{a}$, and governs the rate of reaction. The difference in energy between the reactants and the products ($E_{C}$) may be positive (endothermic) or negative (exothermic), which is useful for predicting which state (reactants or products) is more likely at equilibrium.}
\end{figure}

The field of computational chemistry is extremely broad and includes many methods that encompass a wide range of length and time scales.  For example, to simulate large-scale protein structural rearrangement, each atom can be quantized as a partial point charge centered on the atom.  By calculating forces from the potential energy of the `sea' of point charges, and then by integrating Newton's equations of motion, atomic trajectories over long time scales are achievable. One issue with parameterization of the nucleus and the electrons is that once the parameters are set, typical calculations do not include the possibility of ``on the fly'' re-parameterization. To polarize the partial charge or to break or form new covalent bonds adds an additional level of complexity. In enzyme catalysis, the breaking and forming of bonds is critical, and during these reactions the atoms pass from energetically stable states through the transition states; entering a new stable state.  The chemistry involved in the intermediate states is atypical ({\textit i.e.} C atoms may have more or less than four bonds), hence the elevated energy of the system.  Figure \ref{fig:barrier} shows a sample reaction energy profile for enzyme catalysis.  

\section{Many-body quantum mechanics}
In order to understand both the geometric and electronic structure at the intermediate states, which is essential for accurate prediction of energy barriers and reaction rates, we have used computational quantum mechanics.  With these first principles calculations, the all-electron wave function is optimized in a self-consistent manner in order to predict chemical properties and structures.  Approximate solutions to the many-electron system will be discussed first by starting with the single electron Hamiltonian.

\subsection{One-electron Hamiltonian}
For non-interacting electrons, we can write the total system Hamiltonian as the sum of one-electron Hamiltonians: 
\begin{equation}
H = \sum_{i=1}^{N}h_{i}
\end{equation}
where $N$ is the total number of electrons.  The classical one-electron Hamiltonian is given by
\begin{equation}
h_{i} = -\bigtriangledown^{2}_{i} - \sum^{M}_{k=1}\frac{Z_{k}}{r_{ik}}
\label{one1}
\end{equation}
which is composed of the electron kinetic energy and the nuclear-electron attraction ($M$ is the total number of nuclei).  Eigenfunctions $\psi_{i}$ must satisfy 
\begin{equation}
h_i \psi_i=\epsilon_i \psi_i
\label{one2}
\end{equation}
which is the one-electron Schr\"{o}dinger equation.  The many-electron eigenfunction is then the product of the one-electron eigenfunctions

\begin{equation}
\Psi = \prod_{i=1}^{N} \psi_{i}
\end{equation}
Because each one-electron Hamiltonian ($h_{i}$) acts only on its corresponding eigenfunction ($\psi_{i}$), the overall system Hamiltonian may be written as
\begin{equation}
H\Psi = \sum^{N}_{i=1}h_{i}\prod_{i=1}^{N} \psi_{i} = (\sum_{i=1}^{N}\epsilon_{i})\Psi
\end{equation}

\subsection{The Born-Oppenheimer approximation}
Protons and electrons experience an electrostatic force on the same order of magnitude due to an equal magnitude of charge.  Because of this, the change in their momenta must also be the same.  The mass of a nucleon is 1836 times that of an electron, and from this, the nuclei (which consists of many nucleons) will have a much smaller velocity compared an the electron.  On the time scale of nuclear motion, we can consider that the electrons will relax to the ground state electronic structure.  The total wave function for nuclei and electrons, $\tilde\Psi$ can be separated into the electronic wave function, $\Psi$, and the nuclear wave function $\Phi$:
\begin{equation}
\tilde \Psi (\{\textbf{r}_i\},\{\textbf{r}_\alpha\})= \Psi (\{\textbf{r}_i\},\{\textbf{r}_\alpha\}) \Phi (\{\textbf{r}_\alpha\})
\end{equation}
The Born-Oppenheimer approximation treats the nuclei as classical and stationary particles, and the electrons are said to on the Born-Oppenheimer surface.  Nuclear kinetic energy is neglected when determining the electronic structure. 

This approximation does not work for light nuclei at low temperatures, where a quantum mechanical description of the nuclei is required and usually considered by path integral Monte Carlo (PIMC) or molecular dynamics (PIMD).  For the purposes of our biological systems at standard pressure and biological temperatures, the Born-Oppenheimer approximation is sufficient for accurate electronic structure calculations.

\section{Density functional theory}
First principles density functional theory (DFT) \cite{hohenberg1964ieg,kohn1965sce} has been used in our study of intein splicing and cleavage mechanisms.  In this all-electron approximation to the Schr\"{o}dinger equation, the electron density can be expressed in terms of the wave function:
\begin{equation}
\rho(\textbf{r}) = \mid\psi\mid^{2},
\end{equation}
and from the density the number of electrons can be determined:
\begin{equation}
N = \int \rho(\textbf{r})d\textbf{r}.
\end{equation}
DFT allows for the ground state properties of a chemical system to be determined without considering the many-electron wave function.  Instead, N electrons are grouped into one composite density, which is determined in a static potential of nucleus-electron and electron-electron interactions.

\subsection{Classical beginnings}
So far, we have written the one-electron Hamiltonian for non-interacting electrons in a lattice of nuclei (Equation \ref{one1}).  To add explicit electron-electron interactions, the energy is separated into potential and kinetic terms; starting from classical energy equations and eventually adding quantum terms.  The classical potential energy between nuclei and electrons ($V_{ne}$) and between electrons and electrons ($V_{ee}$) are given by the functionals of the density $\rho(\textbf{r})$:

\begin{equation}
V_{ne}[\rho(\textbf{r})] = \sum_{k}^{nuclei}\int \frac{Z}{\mid\textbf{r}-\textbf{r}_{k}\mid}  \rho(\textbf{r})d\textbf{r} 
\end{equation}

\begin{equation}
V_{ee}[\rho(\textbf{r})] = \frac{1}{2} \int \int \frac{\rho(\textbf{r}_1) \rho(\textbf{r}_2)}{\mid\textbf{r}_1-\textbf{r}_2\mid}d\textbf{r}_1 d\textbf{r}_2. 
\end{equation}
The classical kinetic energy of non-interacting electrons is used: 
\begin{equation}
T_{ni} = -\frac{1}{2}\bigtriangledown^{2}_{i}.
\label{kinetic}
\end{equation}

\subsection{Quantum corrections to classical terms}
The actual kinetic energy is difficult to quantify due to electron correlation effects: specifically, we have not explicitly calculated the correlation interaction due to wave functions for excited electronic states.  But, there does exist some electron correction to this classical kinetic energy which we call the correlation energy, $E_c$.  Also, there are the exchange forces due to spin, which have not been included.  These interactions are called the exchange energy $E_x$ and can be written exactly as
\begin{equation}
   E_{x} = \int d\textbf{r$_{1}$} d\textbf{r$_{2}$} \psi^\star_a(\textbf{r$_{1}$}) \psi_b(\textbf{r$_{1}$}) r_{12}^{-1} \psi^\star_b(\textbf{r$_{2}$}) \psi_a(\textbf{r$_{2}$}).
   \label{exchange}
\end{equation} 
The electron exchange arises from the Pauli exclusion principle for fermions.  The total wave function for a fermionic system, which is the tensor product of the spatial wave function and the spin wave function, must be anti-symmetric.  If two electrons are in the same spatial state (\textit{i.e.} 1s orbital), then their spin wave function must be antiparallel.  From this, there is a force, the exchange force, that keeps electrons with opposite spin in the same spatial orbital.  Likewise, electrons with parallel spin experience a force that keeps them from entering the same spatial orbital.  

The exact exchange shown in Equation \ref{exchange} is rarely calculated explicitly with current DFT methods due to computational expense of the integration, so the these two quantum mechanical corrections are combined to be the exchange-correlation energy $E_{xc}$, which is cleverly defined as the difference between the actual total energy and the classical total energy.  Feynman called the exchange-correlation energy the ``stupidity energy'' due to our inability to solve it exactly for non-arbitrary systems \cite{feynman1998sm}.  
\begin{equation}
E_{xc}[\rho(\textbf{r})] = E_{total}[\rho(\textbf{r})] - T_{ni}[\rho(\textbf{r})] - V_{ne}[\rho(\textbf{r})] - V_{ee}[\rho(\textbf{r})]
\end{equation}
A great deal of theoretical effort has been and continues to be spent on creating exchange-correlation functionals that are useful for a broad range of chemical systems.

\subsection{Hohenberg-Kohn theorems}
As a result of the Born-Oppenheimer approximation, the Coulomb potential of the nuclei can be considered a static external potential:
\begin{equation}
V_{ext}(\textbf{r}) = -\sum_{k}^{nuclei}\int \frac{Z}{\mid\textbf{r}-\textbf{r}_{k}\mid}  d\textbf{r}.
\end{equation}
The Hamiltonian for electrons can then be written as
\begin{equation}
\hat F = -\frac{1}{2}\sum_{i}\bigtriangledown^{2}_{i}+\frac{1}{2}\sum_{i}\sum_{j\neq i}\frac{1}{\mid\textbf{r}_i-\textbf{r}_j\mid}
\label{eff}
\end{equation}
so that tht total Hamiltonian is 
\begin{equation}
\hat H = \hat F + \hat V_{ext}.
\end{equation}

\subsubsection{Density determines unique potential}
Returning to the external potential, it can be shown that it is uniquely determined by the ground state electronic density.  This proof, which is by \textit{reductio ad absurdum}, was first shown by Hohenberg and Kohn \cite{hohenberg1964ieg}.   The external potential $V_{ext}(\textbf{r})$ is related to a ground state wave function $\left| \Psi_0 \right>$ and density $\rho(\textbf{r})$.  Now consider a second external potential $V_{ext}'(\textbf{r})$ which corresponds to a different ground state wave function $\left| \Psi_0' \right>$ but the same density $\rho(\textbf{r})$.  The ground state energies of the two systems are 
\begin{equation}
E_0 = \left< \Psi_0 \right| \hat H \left| \Psi_0 \right> 
\end{equation}
\begin{equation}
E_0' = \left< \Psi_0' \right| \hat H' \left| \Psi_0' \right>.
\end{equation}

We then use $\left| \Psi_0' \right>$ as a trial wave function for the Hamiltonian $\hat H$.  Because $\left| \Psi_0' \right>$ is not an eigenfunction for $\hat H$, we know that 
\begin{eqnarray}
E_0  <  \left< \Psi_0' \right| \hat H \left| \Psi_0' \right>  \label{ineq1} \\
E_0'  <  \left< \Psi_0 \right| \hat H' \left| \Psi_0 \right>.
\label{ineq2}
\end{eqnarray}
By adding both $H'$ and $-H'$ to the right side of the equation, we can write:
\begin{eqnarray}
\left< \Psi_0' \right| \hat H \left| \Psi_0' \right> & = & \left< \Psi_0' \right| \hat H' \left| \Psi_0' \right> + \left< \Psi_0' \right| (\hat H - \hat H') \left| \Psi_0' \right> \\
 &   = &  E_0' + \int d\textbf{r} \rho(\textbf{r})[V_{ext}(\textbf{r})-V_{ext}'(\textbf{r})].
  \label{hk1}
\end{eqnarray}
Also, we can reverse the conditions and take $\left| \Psi_0 \right>$ as a trial wave function for the Hamiltonian $\hat H'$:
\begin{eqnarray}
 \left< \Psi_0 \right| \hat H' \left| \Psi_0 \right> & = & \left< \Psi_0 \right| \hat H \left| \Psi_0 \right> + \left< \Psi_0 \right| (\hat H' - \hat H) \left| \Psi_0 \right> \\
 &   = &  E_0 + \int d\textbf{r} \rho(\textbf{r})[V_{ext}(\textbf{r})-V_{ext}'(\textbf{r})].
  \label{hk2}
\end{eqnarray}
Re-writing the inequalities from Equations \ref{ineq1} and \ref{ineq2} in terms of the forms shown in Equations \ref{hk1} and \ref{hk2}:
\begin{eqnarray}
 E_0' & <  &  E_0 + \int d\textbf{r} \rho(\textbf{r})[V_{ext}(\textbf{r})-V_{ext}'(\textbf{r})] \\
 \label{hk3}
 E_0 & <  &  E_0' + \int d\textbf{r} \rho(\textbf{r})[V_{ext}'(\textbf{r})-V_{ext}(\textbf{r})]
 \label{hk4}
\end{eqnarray}
Then, by taking the summation of Equations \ref{hk3} and \ref{hk4}, we find that 
\begin{equation}
E_0 + E_0' < E_0 + E_0',
\end{equation}
which is clearly a contradiction. Therefore, the external potential $V_{ext}(\textbf{r})$ is uniquely determined by the ground state density $\rho(\textbf{r})$. In addition, the number of electrons can be determined by integrating the density over all space.
\begin{equation}
N = \int d\textbf{r} \rho(\textbf{r}).
\end{equation}
The energy of the system can be written as
\begin{equation}
E_V[\rho(\textbf{r})] = F[\rho(\textbf{r})] + \int d\textbf{r} V(\textbf{r}) \rho(\textbf{r}),
\end{equation}
where $F$ is the electronic Hamiltonian from Equation \ref{eff}.

\subsection{Kohn-Sham equations}
The variational problem for $N$ electrons using the Hohenberg-Kohn density functional method can be written as,
\begin{equation}
\delta \Big[ F[\rho(\textbf{r})] + \int d\textbf{r} V_{ext}(\textbf{r})\rho(\textbf{r}) - \mu \Big(\int d\textbf{r} \rho(\textbf{r}) - N\Big)\Big]=0,
\label{diff1}
\end{equation}
where $\mu$ is a Lagrange multiplier.  First shown by Kohn and Sham, $F[\rho(\textbf{r})]$ can be separated:
\begin{equation}
F[\rho(\textbf{r})] = T_{ni}[\rho( \textbf{r})] + \frac{1}{2} \int d\textbf{r}d\textbf{r'}\frac{\rho(\textbf{r})\rho(\textbf{r'})}{\mid \textbf{r}-\textbf{r'}\mid} + E_{xc}[\rho(\textbf{r})],
\label{diff2}
\end{equation}
where $T_{ni}$ is the classical kinetic energy from Equation \ref{kinetic}.  From Equations \ref{diff1} and \ref{diff2}, we  rearrange so that
\begin{equation}
\frac{\delta T_{ni}[\rho( \textbf{r})]}{\delta \rho( \textbf{r})} + V_{KS}(\textbf{r}) = \mu,
\label{mu}
\end{equation}
where the Kohn-Sham potential is given by
\begin{equation}
V_{KS}(\textbf{r}) = \int d\textbf{r}\frac{\rho(\textbf{r})}{\mid \textbf{r}-\textbf{r'}\mid} + V_{xc}(\textbf{r}) + V_{ext}(\textbf{r}).
\end{equation}
The exchange-correlation potential is related to the exchange-correlation functional by,
\begin{equation}
V_{xc}(\textbf{r}) = \frac{\delta E_{xc}[\rho(\textbf{r})]}{\delta \rho(\textbf{r})}.
\end{equation}
It is important to note that Equation \ref{mu} would also be valid for a system of non-interacting particles experiencing an external potential $V_{KS}(\textbf{r})$.

In order to find the ground state density, $\rho_0(\textbf{r})$, which corresponds to the energy minimum of the electronic system, we use the simple one-electron Schr\"{o}dinger equation based on Equations \ref{one1} and \ref{one2}.  This can be rewritten as
\begin{equation}
\Big[-\frac{1}{2}\bigtriangledown^{2}_{i}+V_{KS}(\textbf{r}) \Big] \psi_i(\textbf{r}) = \epsilon_i \psi_i(\textbf{r}).
\label{schrod}
\end{equation}
The ground state density can be found from the wave functions by this relation:
\begin{equation}
\rho(\textbf{r}) = 2\sum_{i=1}^{N/2}\mid\psi_i(\textbf{r})\mid^2,
\label{density}
\end{equation}
where the factor of two is due to spin degeneracy from the assumption that the orbitals are singly-occupied.  

\subsubsection{Variational principle}
We have shown in the above section that the external potential and ground state wave function are uniquely determined by the ground state density, and now we will show how to determine the ground state density.  The variational principle dictates that every trial wave function except the unique ground state wave function will give an energy higher than the ground state energy:
\begin{equation}
\left< H \right> = \frac{\left< \psi \right| H \left| \psi \right>}{\left< \psi \mid \psi \right>} \geq E_0
\end{equation}
where $E_0$ is the smallest eigenvalue of $H$.
Expressed in terms of the electron density, $\rho(\textbf{r})$, the calculated energy, $E_V[\rho(\textbf{r})]$, is larger than the ground state energy, $E_0$, which is a minimum:
\begin{equation}
E_V[\rho(\textbf{r})] \geq E_0.
\end{equation}
This variational principle on the energy can be proven by using $\rho(\textbf{r})$ which determines $V_{ext}(\textbf{r})$ and ground state $\left| \Psi \right>$.  Using this state as a trial state for the external potential $V(\textbf{r})$, we can write the total energy of the test system,
\begin{eqnarray}
\left< \Psi \right| \hat H \left| \Psi \right> & = &  \left< \Psi \right| \hat F \left| \Psi \right> + \left< \Psi \right| \hat V \left| \Psi \right> \\
& = & F[\rho(\textbf{r})] + \int d\textbf{r} V(\textbf{r}) \rho(\textbf{r}) \\
& = & E_V [\rho(\textbf{r})] \geq E_0.
\end{eqnarray}
By minimizing the functional $E_V[\rho(\textbf{r})]$, the energy will eventually approach but never meet the ground state energy, $E_0$.

\subsection{Self-consistent procedure}
Because the Kohn-Sham potential $V_{KS}(\textbf{r})$ depends on the density $\rho(\textbf{r})$, the equations must be solved self-consistently.  First, a guess is made for density. The Schr\"{o}dinger equation is solved (Equation \ref{schrod}) and a new set of orbitals $\{\psi_i(\textbf{r})\}$ is determined.  From these orbitals, a new density is found based on Equation \ref{density}.  This new density then becomes the next guess for the Schr\"{o}dinger equation, which gives new orbitals and another new density.  This process is continued until the output and input densities are equivalent, indicating the ground state energy.

\subsection{DFT implementation of exchange and correlation}
So far, the exact nature of the exchange-correlation functional $E_{xc}$ has not been discussed other than that it is the ``quantum correction'' to the classical kinetic and potential energies, and that it should include energetic corrections for fermionic spin exchange.  We write $E_{xc}$ as a function of $\epsilon_{xc}$, the `energy density'
\begin{equation}
E_{xc}[\rho(\textbf{r})] \equiv \int d\textbf{r} \epsilon_{xc}[\rho(\textbf{r})] \rho(\textbf{r})
\end{equation}

\subsubsection{Local density approximation (LDA)}
If $\epsilon_{xc}$ is computed based solely on the local position $\textbf{r}$, then the method is called the Local Density Approximation (LDA).  Found by Thomas-Fermi theory and Quantum Monte Carlo methods for a homogeneous electron gas \cite{ceperley1980gse}, the exchange-correlation energy for LDA is written:
\begin{equation}
\epsilon_{xc}^{LDA}[\rho(\textbf{r})] = \epsilon_{x}[\rho(\textbf{r})] + \epsilon_{c}[\rho(\textbf{r})] = -\frac{0.458}{r_s} - \frac{0.44}{r_s+7.8},
\label{lda}
\end{equation}
where
\begin{equation}
r_s = \left(\frac{3}{4\pi \rho(\textbf{r})}\right)^{1/3}.
\end{equation}

\subsubsection{Generalized gradient approximation (GGA)}
Because this electron-gas model does not regularly predict accurate chemical bonding properties, the next term in the expansion of the density is typically used.  This is called the Generalized Gradient Approximation (GGA), in which
\begin{equation}
\epsilon^{GGA}_{xc}[\rho(\textbf{r})] = \epsilon^{LDA}_{xc}[\rho(\textbf{r})] + \Delta\epsilon_{xc}\left[\frac{\mid \bigtriangledown \rho(\textbf{r})\mid}{\rho^{4/3}(\textbf{r})}\right].
\end{equation}
GGA has shown remarkable flexibility for various chemical systems \cite{martin2004esb}, especially covalent bond energies and distances, although known shortcomings are a poor description of van der Waals bonding \cite{wu2001tea} and incorrect unoccupied orbital energies leading to underestimation of band gaps for semi-conductors \cite{vanschilfgaarde2006qsc}.

\subsubsection{Hybrid DFT methods}
Primarily, we have used B3LYP: the Becke three-parameter hybrid functional \cite{becke1993}. The exchange (x) and correlation (c) energy is written as $E^{B3LYP}_{xc}$, where 
\begin{equation}
E^{B3LYP}_{xc} = (1-a)E^{LDA}_{x} + aE^{HF}_{x} + b \Delta E^{Becke}_{x} + E^{LDA}_{c} + c \Delta E^{LYP}_{c},
\end{equation} 
and the coefficients were optimized to match extensive molecular data sets (\textit{a}=0.20, \textit{b}=0.72, and \textit{c}=0.81) \cite{becke1993}.
 The first term in the hybrid method is $E^{LDA}_{x}$, which is the LDA exchange term from Equation \ref{lda}.  $E^{HF}_{x}$ is Hartree-Fock exchange integral, which is exact and is given in Equation \ref{exchange}. Becke's B88 exchange term \cite{becke1988dfe} is based on empirical results, and is written as,
\begin{equation}
E_{x}^{Becke}[\rho(\textbf{r})] = - \beta \int d\textbf{r} \rho(\textbf{r})^{4/3} \frac{\alpha^2}{(1+6\beta \sinh^{-1} \alpha)} 
\end{equation}
where $$\alpha = \frac{\mid \bigtriangledown \rho(\textbf{r})\mid}{\rho(\textbf{r})^{4/3}}.$$ Found by matching molecular data sets, $\beta$ was found to be $0.0042$ Hartree. Correlation functionals are from the LDA \cite{perdew1992aas} and from Lee, Yang, and Parr (LYP) \cite{lee1988dcs,miehlich1989roc}, the latter based on an empirically determined model of the correlation energy of electrons in a helium atom.

Implemented with Gaussian code \cite{gauss03}, this hybrid gradient-corrected method is considered one of the most accurate exchange-correlation functionals and has been used with great success in other biological systems \cite{zhang2005dmc,zhang2006rmg}. Calculations with post-Hartree-Fock M\o ller-Plesset perturbation theory (MP2) \cite{moeller1934nat,headgordon1988mee,frisch1990dmg,frisch1990sda} were conducted to test the accuracy of the B3LYP method for this system, and the energy barrier calculations were consistent. 

\subsection{Basis sets}
Any numerical basis set may be used, although for computational efficiency only a several are implemented.  A primitive Gaussian-type orbital (GTO) in atom-centered Cartesian coordinates is
\begin{equation}
\phi(x,y,z;\alpha,i,j,k) = \left(\frac{2\alpha}{\pi}\right)^{3/4} \left[\frac{(8\alpha)^{i+j+k}i!j!k!}{(2i)!(2j)!(2k)!}\right]^{1/2} x^i y^j z^k e^{-\alpha(x^2+y^2+z^2)}
\end{equation}
where $\alpha$ controls the width of the GTO, and $i$, $j$, and $k$ are non-negative integers that define the angular momentum of the orbital.  For example, for an s-type GTO (orbital angular momentum $l=0$), the integers are $i=j=k=0$, and $\phi$ is spherically symmetric.  For a p-type GTO (orbital angular momentum $l=1$), there are three possibilities for $i$, $j$, and $k$ that lead to the Cartesian prefactor of either $x,y,z$.  For d-type GTOs (orbital angular momentum $l=2$), there are six possibilities for the index values ($i,j,k$).  Specifically, the Cartesian prefactors are $x^2,y^2,z^2,xy,xz,yz$.  All of these Cartesian prefactors are multiplied by the Gaussian term in order to obtain proper atomic orbital shape.

GTOs are easily integrated by computational schemes, but they do not accurately enough resemble atomic orbitals (AOs). Primitive GTOs are smooth and differential at the nucleus ($r=0$).  But for a hydrogenic system, the actual atomic orbital (AO) will have a non-continuous derivate at the $r=0$, which would be better described $r^{-1}$, a less integrable function (called Slater-type orbital, or STO).  In order to use a basis function with the proper radial shape, and one that when squared is easily integrable \textit{via} numerical methods, we have used contracted Gaussian functions, which are the linear combination of primitive GTOs designed to resemble an STO.  A general contracted GTO may be written as
\begin{equation}
\varphi(x,y,z;\{\alpha\},i,j,k) = \sum_{a=1}^M c_a \phi(x,y,z;\alpha_a,i,j,k)
\end{equation}
where $M$ is the number of GTOs used in the linear combination.  The coefficients $c_a$ are chosen for normalization as well as to optimize the basis function shape, which are then called a linear combination of atomic orbitals (LCAO).  

We have used the double-$\zeta$ basis set, 6-31G(d,p), for geometry optimizations during initial reaction path sampling \cite{hehre2003scm}, where the `6' represents six GTOs for core electrons and the `31' represents split GTOs for valence electrons: specifically three and one.  Split-valence basis sets allow for a more accurate description of chemical bonding due to increased flexibility to fit valence electrons into molecular orbitals, and are the norm when using a Gaussian-type basis set. The `(d,p)' indicates that we are using polarization functions that allow for a shift in the wave function away from the atomic center.  We have also used the triple-$\zeta$ basis set, 6-311++g(d,p), for calculations of the local minima and transition states found with the first basis set \cite{mclean2006cgb}.  Diffuse functions for long range interactions are represented with a `+', and are especially important for anions.  Basis sets of similar size are typically used for systems with similar number of electrons, and our test calculations as well as the work of others have shown these basis sets to be sufficient for similar atom types \cite{zhang2005dmc, zhang2006rmg}.

\section{Calculation methodology}
In addition to highly efficient computational codes available and tuned for parallel processing, we have written various useful computer codes for data analysis and manipulation, especially for trajectories of large QM/MM systems that contain thousands of atoms which would be extremely cumbersome otherwise.  The electronic structure calculations are performed with robust computer code widely used in the field, and each method is described below.

\subsection{Gas phase}
Calculations \textit{in vacuo} are the simplest way to understand the energies associated with a reaction mechanism, since they treat the quantum mechanical active site as an isolated molecule or group.  Gas phase calculations capture the amount of energy necessary to transverse the energy barrier, which is typically then lowered by electrostatic interactions with the remaining protein and/or solvent.  By studying mechanisms in the gas phase, we can separate the energy related to a particular bond breaking/forming from other factors such as structural rearrangement and polarization effects.  One drawback of using a small model system with gas phase calculations is that the protein backbone is either rigid with the starting configuration, or completely relaxed and prone to unlikely rearrangement in the protein context.  Despite these limitation, reaction profiles calculated with gas phase calculations are an important part in describing the energies of a reaction.

\subsection{Semi-empirical PM3}
Semi-empirical PM3 \cite{stewart1989ops} calculations were performed in order to obtain a two-dimensional potential energy surface in the $xy$ plane for two constrained reaction coordinates.  Geometry optimizations with PM3 are computationally efficient and sufficiently accurate for the present purposes.   Deficiencies of PM3 with respect to chemical structures, such as the ``flattening'' of small and medium sized rings \cite{ferguson1992vzap},  nitrogen atoms with a lone pair having pyramidal geometries \cite{csonka1993acr},  and inaccurate hydrogen bonding distances are well known.  Nevertheless, PM3 is useful for efficiently scanning many geometries and locating general locations of transition states, as is done here.

\subsection{Implicit solvent}
One method for approximating the environmental electrostatic effect is to use an implicit solvent.  In this scheme, the active site is polarized by the dielectric medium which is itself polarizable.  The Polarizable Continuum Model (PCM) \cite{miertus1981eis}  was used to simulate solvent effects in the detailed calculations. The numerical Integral Equation Formalism  \cite{cances1997ese} (IEFPCM) was used because it allows for interlocking atomic spheres to represent the extent of the system in solution, which is important for protons that are in between atoms during a chemical reaction and at or around the energy barrier.   

Non-dimensional dielectric constants are defined by $\epsilon_{r} = \epsilon_{s}/\epsilon_0$, where $\epsilon_0$ is the vacuum permittivity and $\epsilon_s$ is the static dielectric constant for the dielectric. For the gas phase, $\epsilon_{r}=1$.  For water, $\epsilon_{r}=78.39$. Geometry optimizations were performed in implicit solvent and results are compared with gas phase calculations.

\subsection{Multiscale modeling}
\subsubsection{Combined quantum and classical mechanics}
The Quantum Mechanics/Molecular Mechanics (QM/MM) layering method is used, and involves treating the protein active site and critical solvent molecules with first principles methods while treating the remaining full-protein system with classical force fields \cite{maseras1995ini, vreven2003goq}. 
Similar multiscale methods have been used with good success \cite{thompson1995esb, schoneboom2002eos, gao2002qmm, cui2002tap, torrent2002epe}.
The classical system is periodic and is truncated to include the protein (intein and exteins) as well as all interior waters and certain exterior water molecules that are within a range of 7.0 \AA{} to the protein surface.  All atoms are relaxed, and each calculation includes at least 6000 atoms, roughly 2350 of which belong to the protein.  The full-protein plus solvent system, termed the real system, is treated only with the classical MM method. Within the real system, the active site model system is partitioned and is treated independently by QM and MM methods.   Dangling bonds that are introduced by partitioning the model system are passivated with hydrogen atoms. With normal QM/MM energy calculations and geometry optimizations, protein and solution outside the model system is typically only included as a mechanical perturbation.  For this reason, it is critical that the model system should include protein segments and solution molecules that are interacting electrostatically.  The combined Hamiltonian may be written:  
\begin{equation}
E^{QM:MM}_{ONIOM} = E^{QM}_{Model} - E^{MM}_{Model} + E^{MM}_{Real}.
\end{equation}

\subsubsection{Charge embedding}
In addition to the mechanical perturbation on the QM Hamiltonian, the electrostatic contribution from the partial charges of the MM region can be included as a perturbation on the QM Hamiltonian.  For this scheme the partial charges are those used in the MM calculation and are scaled by the default manner where atoms bonded to the inner-most four layers and atoms outside that threshold are not included \cite{vreven2003goq}.  Typically, we report $E^{QM}_{Model}$, which represents the QM active site energy.  The other energy terms, including the combined $E^{QM:MM}_{ONIOM}$ involves classical parameters that have no relevance to the energies of bond forming and breaking at transition states.  

\subsection{Geometry minimization}
Due to the complexity of biomolecular reactions, a rigorous multidimensional search over local conformational space is essentially required although not computationally feasible for large systems \cite{dellago2006tps}.  Due to the time expense for each calculation, we have used the constant minimization procedure.  For intermediate states along the reaction path, one coordinate is constrained while the remaining system is relaxed. The constrained internal coordinate, called the Asn cyclization distance, was the atomic distance between the Asn side chain N atom and the carbonyl C of Asn on the scissile peptide bond. In calculations with a hydronium ion (H$_3$O$^+$), the three O-H bond distances were often constrained to 0.98 \AA{} to avoid spontaneous proton donation observed otherwise.

\subsection{Free energy calculation}
Thermal and entropic contributions calculated with a harmonic approximation for the optimized geometries at the B3LYP/6-31G(d,p) level were combined with the electronic energy to obtain free-energy profiles in the gas phase and in the implicit solvent. Zero point energies were found to differ by between 0.04 and 1.33 kcal/mol, which are within the expected error for the calculation. The approximate entropic components of the free energy include contributions from translational, electronic, rotational, and vibrational degrees of freedom \cite{irikura1998ctp,ochterski2000tg} and were obtained from frequency calculations at room temperature. Thermal corrections do not include imaginary frequencies of vibrational modes for transition states.


\chapter{C-TERMINAL CLEAVAGE}
\section{Introduction: C-terminal cleavage}
\subsection{Experimental motivation}
C-terminal cleavage involves the isolated excision of the C-extein from the intein substrate. In the experimental study by Wood \textit{et al.} on the \textit{Mycobacterium tuberculosis} (\textit{Mtu}) recA cleavage mutant mini-intein ($\Delta$I-CM), a decrease in solution pH from 7.5 to 6.0 was found to lead to a significant increase in the rate of C-terminal cleavage \cite{wood1999gsy,wood2000oss}. The higher C-terminal cleavage activity at lower pH for this intein as well as for the \textit{Ssp} DnaB intein are, however, inconsistent with the currently available details of the mechanism proposed for intein cleavage within the context of splicing.  For example, Ding \textit{et al.} have suggested that for the reaction of the \textit{Ssp} dnaB mini-intein, C2-His12 (F-block) acts as a base, deprotonating the nitrogen of the C-terminal C1-Asn8 side chain \textit{via} a vicinal water molecule. Experiments on short peptides in solution also show an increased tendency of Asn to be ionized and to cyclize (leading to succinimide formation) over the pH range of 7.4-13.8 \cite{geiger1987dia,capasso1996kam,catanzano1997tae, peters2006adp}.  For the side chain of His to act as a base and accept protons, one of its two imidazole side chain nitrogen atoms must be deprotonated on average. At lower pH, especially below the pK$_a$, His is more likely to be present in the doubly protonated and positively charged state, diminishing its ability to accept protons.  Thus, the mechanistic details underlying the increased C-terminal cleavage activity observed in experiments of Wood \textit{et al.} \cite{wood1999gsy,wood2000oss}  are expected to be different from those proposed by Ding \textit{et al.} \cite{ding2003csm}. 


\subsection{Reaction mechanism}
We have studied the cleavage mechanism of the \textit{Mtu} recA intein using information from available intein crystal structures and mutagenesis experiments. Part of the intein structure close to the C-terminal reaction site was considered in both gas phase as well as implicit solvent model calculations.  

\begin{figure}
\centering
\includegraphics[scale=0.9]{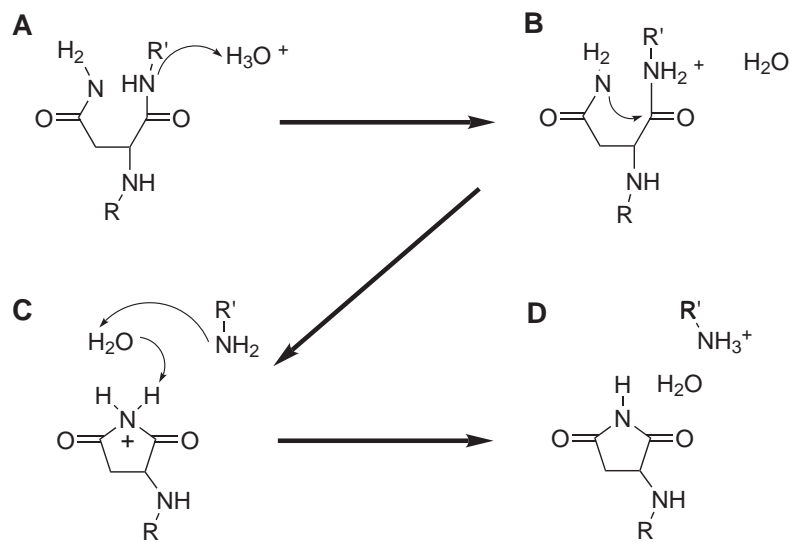}
\caption{\label{fig:mechanism} The proposed N-protonation reaction scheme for Asn cyclization to succinimide; \textbf{R} and \textbf{R'} represent the intein and C-extein, respectively.}
\end{figure}

Based on a combination of semi-empirical and first principles computational analyses, and with gas phase, implicit solvent, and QM/MM calculations, we proposed new details of the mechanism of Asn cyclization catalyzed by the cleavage mutant of the \textit{Mtu} recA mini-intein that account for an increase in activity of C-terminal cleavage at low pH \cite{shemella}. Figure \ref{fig:mechanism} shows important reaction steps.  This mechanism involves the protonation of the nitrogen of the scissile peptide bond by a vicinal hydronium ion ($H_{3}O^{+}$). This leads to stretching of that peptide bond due to the loss of $\pi$-bond resonance and consequent increase in carbon electrophilicity \cite{milnerwhite1997pcn}.  Asn cyclization and subsequent succinimide formation occur,  resulting in peptide bond cleavage.  Our results are consistent with the experimental observations of Wood \textit{et al.} that indicate a simple proton-catalyzed reaction \cite{wood1999gsy, wood2000oss}. The frozen internal coordinate, called the Asn cyclization distance, is the atomic distance between the Asn side chain N atom and the carbonyl C of Asn on the scissile peptide bond (shown with an arrow in Figure \ref{fig:mechanism}B).

Given current computational resources, detailed quantum mechanical calculations are limited in system size. In the present context, where the mechanism of enzymatic catalysis is of interest, one may choose on the order of $10^2$ atoms and perform only a few calculations, or choose on the order of $10^1$ atoms that is relevant part of the system, and explore a variety of possible
reaction pathways. Inclusion of a larger number of atoms not only adds to the computational expense, but could allow conformational rearrangements that are inconsistent with the protein structural context being studied. As a compromise,
and based on available crystal structure knowledge, our initial system includes 25 atoms: the C-terminal C1-Asn8 side chain, the backbone atoms of the penultimate C1-His7 and of the C-terminal Cys+1 (the dangling bonds are passivated with hydrogen atoms), and one explicit water molecule. Additional calculations with larger systems containing 50+ atoms, and a full protein quantum mechanics/molecular mechanics (QM/MM) treatment indicated energetic results will be described in Chapter 4.

\section{Semi-empirical analysis}
\subsection{Description of model system}
\begin{figure}
\centering
\includegraphics[scale=1.0]{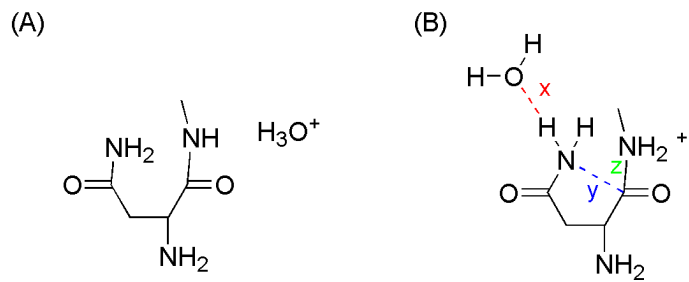}
\caption{\label{fig:pm3xyz} PM3 reaction coordinate.  The system before N-protonation is shown (A), and the reaction coordinate for the energy profile shown in Figure \ref{fig:pm3} is highlighted (B).}
\end{figure}

The PM3 Hamiltonian was used for scanning the potential energy surface in two dimensions. Given that the doubly protonated state of the amide nitrogen is a likely starting point of the C-terminal cleavage reaction at low pH, we performed geometry optimizations using the PM3 method for 428 independent points in the two dimensional space based on reaction coordinates $x$ and $y$ (shown in Figure \ref{fig:pm3xyz}).  Coordinate $x$ represents the distance between the oxygen atom of the water molecule and the hydrogen atom of the Asn side chain, referred to as the Asn ionization distance. Coordinate $y$, the Asn cyclization distance, is the separation between the Asn side chain nitrogen and the peptide carbonyl carbon. 

\begin{figure}
\centering
\includegraphics[scale=0.4]{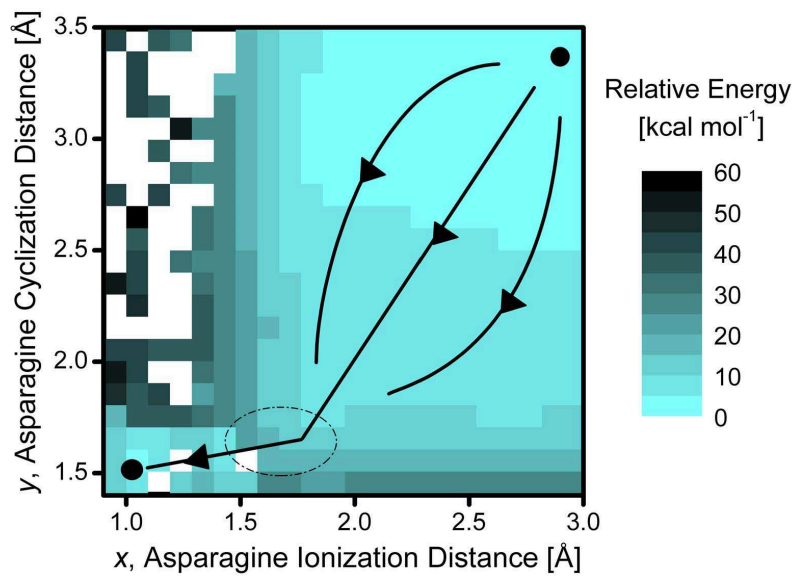}
\caption{\label{fig:pm3} Semi-empirical (PM3) energy surface shown for the reaction coordinates shown in Figure \ref{fig:pm3xyz}.  Filled circles indicate reactant and product state, and the arrows indicate proposed reaction path.  The dotted ellipse indicates the general location of the energy barrier.}
\end{figure}

\subsection{Reaction coordinate space}
The space of reaction coordinates for the semi-empirical calculation is shown in Figure \ref{fig:pm3xyz} and corresponding energies for the two-dimensional energy scan are shown in Figure \ref{fig:pm3}.  The Asn ionization distance, $x$, ranges from 0.9 to 3.0 \AA{} ($x$=2.0 \AA{} corresponds to a typical hydrogen bond, whereas $x$=1.0 \AA{} indicates the deprotonated Asn side chain and a re-formed hydronium ion).  The Asn cyclization distance, $y$, ranges from 1.4 to 3.5 \AA{} ($y$=3.5 \AA{} is the relaxed distance in the initial state, whereas $y$=1.5 \AA{} indicates fully cyclized Asn). The initial state ($x$=3.0, $y$=3.5 \AA{}) located on the top-right of the plot is chosen to be zero for the relative reaction energy.  The final products state is located in the bottom left corner of the graph ($x$=1.0, $y$=1.5 \AA{}) and corresponds to cyclized Asn (succinimide), a re-formed hydronium ion, and a cleaved peptide bond.  

\subsection{Reaction path}
The path marked by arrows on Figure \ref{fig:pm3} indicates the likely path followed by the C-terminal cleavage reaction.  Along that path, $y$ is reduced significantly first, and is then followed by a reduction in the value of $x$. The reduction in $y$ may happen along combinations of paths shown in Figure \ref{fig:pm3} because that region of the energy landscape is relatively featureless.  In any case, cyclization of Asn appears to be almost complete before the ionization of the side chain nitrogen takes place. The barrier region is located near $x=1.6$, $y=1.6$ \AA{} indicated by the ellipse in Figure \ref{fig:pm3} and has energy of about 25 kcal/mol higher than the reference state.  Figure \ref{fig:pm3} shows that alternate paths, \textit{i.e.}, Asn ionizes before its cyclization, and are highly unlikely as they sample regions of considerably high energies.

We note that PM3 calculations for geometry optimizations do not converge for certain choices of $x$ and $y$ (open squares in Figure \ref{fig:pm3}). Most of these points are neighbored by points of higher energies, and therefore, should not affect the general conclusions drawn above.

\section{First principles reaction study}
\subsection{Gas phase molecular system}
\begin{figure}
\centering
\includegraphics[scale=1.0]{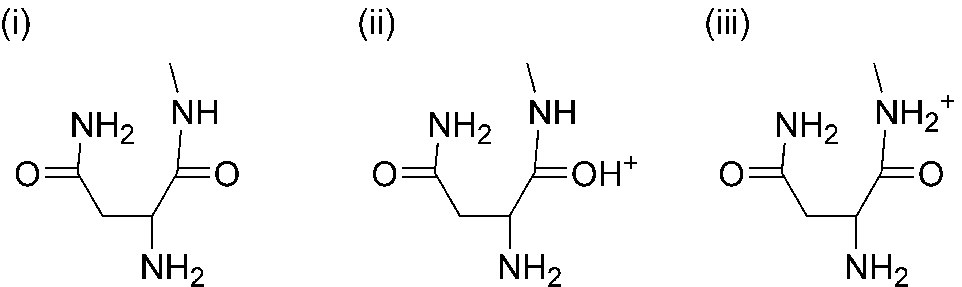}
\includegraphics[scale=0.04]{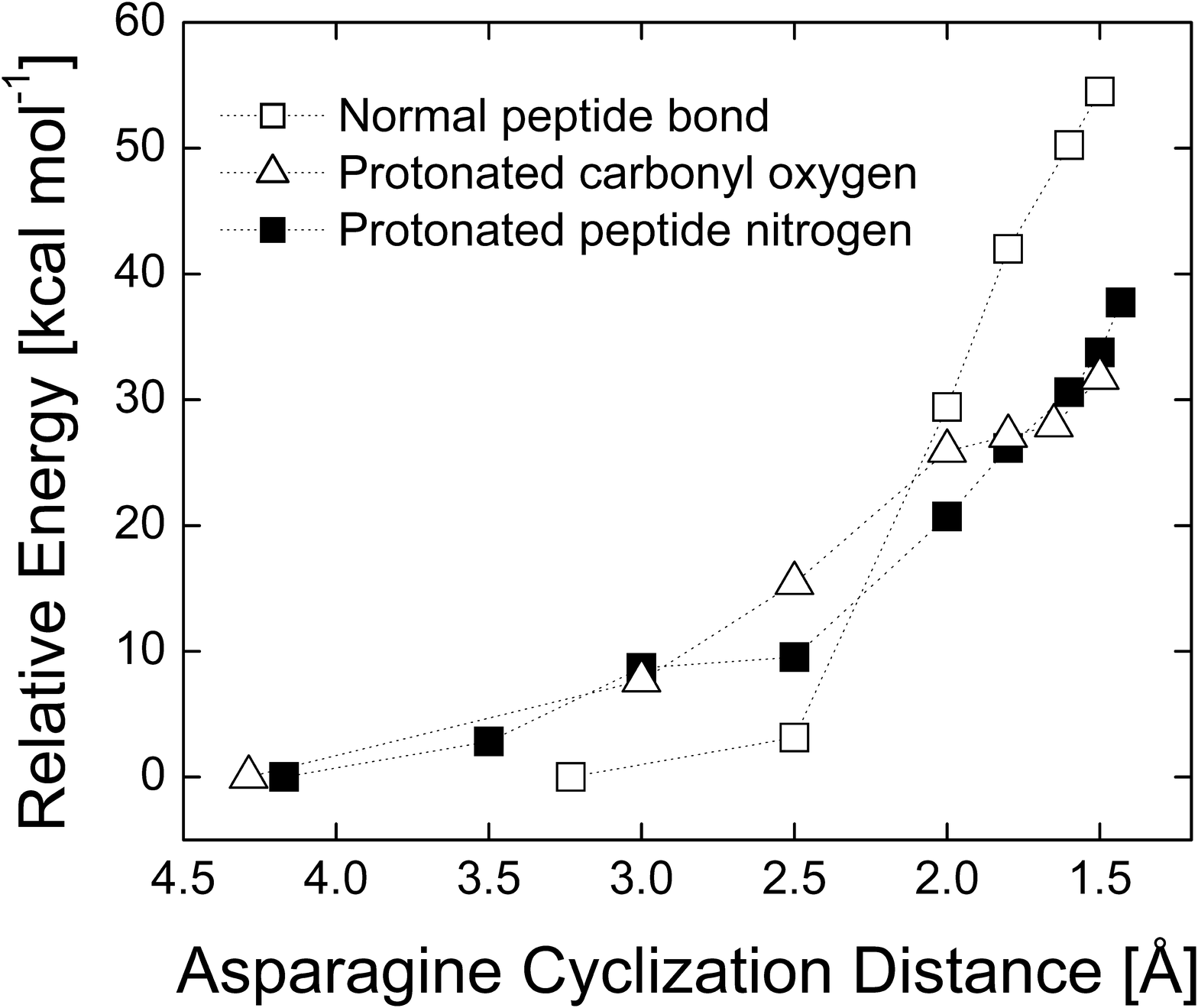}
\includegraphics[scale=0.04]{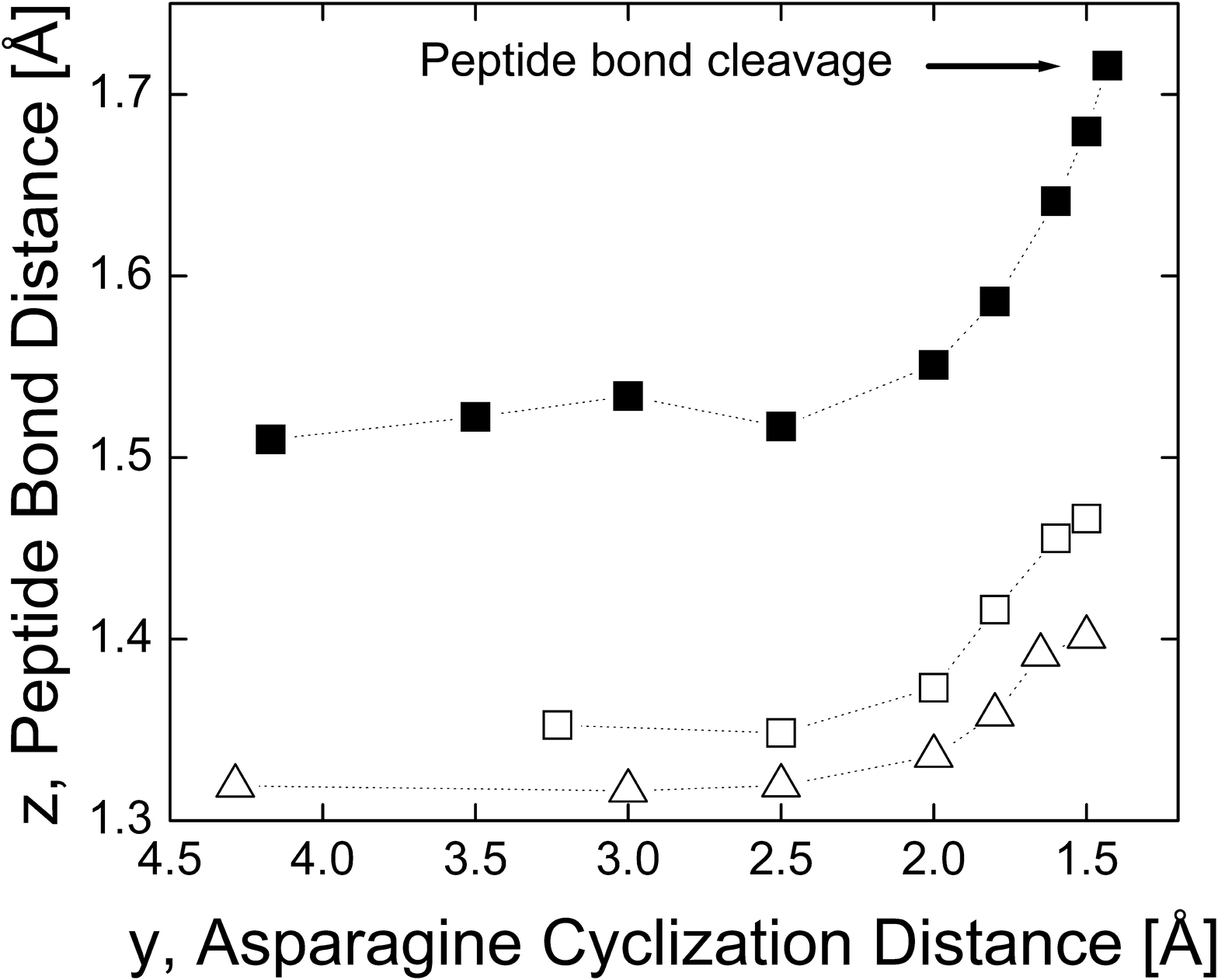}
\caption{\label{fig:cyclization} Schematic of the three scenarios considered for Asn
cyclization: i), normal peptide, ii), O-protonation, and iii), N-protonation.
Asn cyclization energies (top graph) and peptide bond stretching (bottom graph) versus the Asn cyclization distance, $y$, for the neutral peptide bond system ($\square$), for the system where the carbonyl oxygen is protonated ($\triangle$), and
where the peptide nitrogen atom is protonated ($\blacksquare$). Gas phase structures are optimized at the B3LYP/6-31G(d,p) level.}
\end{figure}

As will be apparent in the discussion below, protonation of the backbone nitrogen of the scissile peptide bond is a necessary first step in the reaction at low pH. To this end, Figure \ref{fig:cyclization} considers three scenarios corresponding to i), a neutral peptide case in which neither of the atoms are protonated, ii), O-protonation of the carbonyl oxygen, and iii), the N-protonation of the peptide nitrogen \cite{milnerwhite1997pcn} Specifically, we study consequences of these three scenarios on the system energy and the equilibrium length of the scissile peptide bond. In a neutral peptide (scenario i above), the relaxed peptide bond length is 1.35 \AA{}. As Asn cyclization proceeds (\textit{i.e.}, as $y$ is reduced), the system energy increases significantly (Figure \ref{fig:cyclization}, bottom graph). Correspondingly, there is only a slight increase in the peptide bond length, indicating that it remains essentially intact. 

When the carbonyl oxygen atom is protonated (scenario ii), the relaxed peptide bond length in fact decreases to 1.32 \AA{}, as expected from the increased $\pi$-conjugation (or the double bond character of the bond) between C and N. Asn cyclization energy (Figure \ref{fig:cyclization}, top plot) in this case is lower than that for the neutral peptide case; however, the peptide bond is significantly more stable and remains essentially intact as Asn cyclization proceeds. 

In contrast, when the peptide nitrogen atom is protonated (scenario iii), the relaxed peptide bond length increases to 1.51 \AA{}, indicating the weakening of that bond. As Asn cyclization proceeds, that distance increases further and leads to breaking of that bond, resulting from the fact that a doubly protonated nitrogen makes a good leaving group (Figure \ref{fig:cyclization}, bottom plot). The cyclization energy (Figure \ref{fig:cyclization}, top plot) in this case is lower than that for the neutral peptide case and similar to that for oxygen protonation, which does not lead to peptide bond cleavage (see above). Collectively, these preliminary calculations indicate that protonation of the amide nitrogen is an important first step for C-terminal cleavage in low pH environments.

The protonation of the nitrogen of the scissile peptide bond proposed above makes that nitrogen  atom transiently doubly protonated. In the broader context of enzymatic catalysis, this proposal is not new. Indeed, in the hydrolysis of a peptide bond by serine proteases, the nitrogen of the scissile peptide bond accepts a proton from the His of the catalytic triad \cite{voet1995b}. In their study of enzymatic reaction catalyzed by the HIV-1 protease, Trylska \textit{et al.} found that protonation of the amide nitrogen was essential for peptide bond cleavage \cite{trylska2004rhb}. Similarly, the protonation of the amide nitrogen was found to be the essential step in the hydrolysis of a formamide molecule, which was used as a computational model for peptide bond hydrolysis \cite{krug1992tsn}. 

\begin{figure}
\centering
\vspace{2.0in}
\includegraphics[scale=1.0]{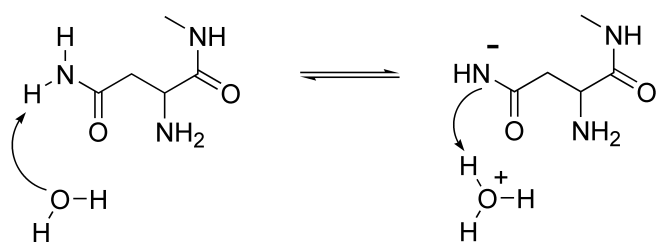}
\includegraphics[scale=0.05]{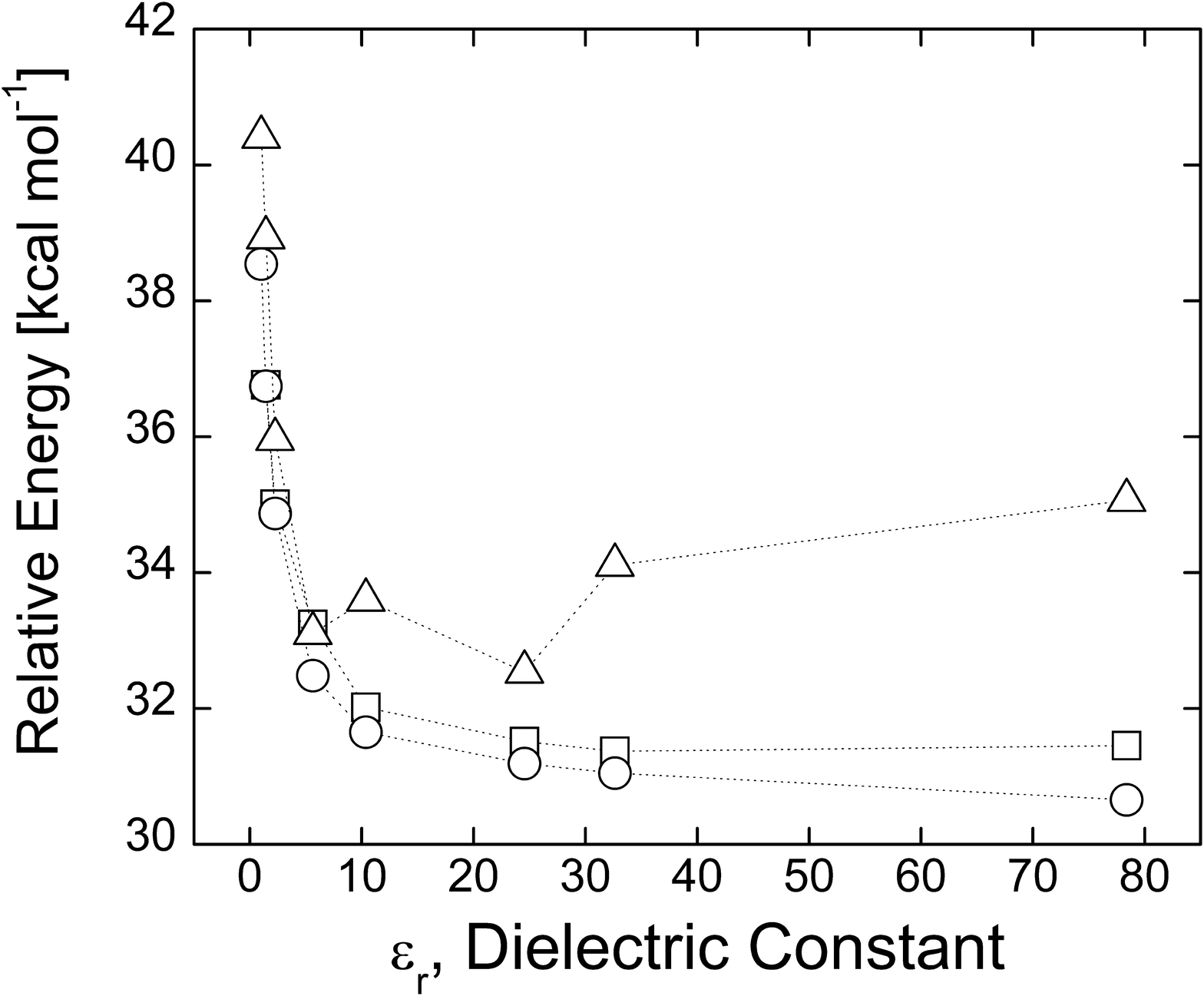}
\caption{\label{fig:ionization} Energy of ionization of Asn side-chain nitrogen (top panel) calculated in solvents of different dielectric constants: vacuum ($\epsilon_{r}=$1.000), argon ($\epsilon_{r}=$1.43), benzene ($\epsilon_{r}=$2.247), chlorobenzene ($\epsilon_{r}=$5.621), dichloroethane ($\epsilon_{r}=$10.36), ethanol ($\epsilon_{r}=$24.55), methanol ($\epsilon_{r}=$32.63), and water ($\epsilon_{r}=$78.39). Each implicit solvent has unique parameters such as radius and density. Note that the peptide nitrogen has only one proton in this case. Gas phase coordinates, energy in solvent ({\small{$\bigcirc$}}); solvent optimized coordinates, energy ($\Box$); and Gibbs free energy ($\bigtriangleup$).}
\end{figure}

The observation made above that the Asn cyclization proceeds before the ionization of its side chain is supported independently by high-level quantum calculations at the B3LYP/6-31G(d,p) level. Specifically, we followed the ionization of the Asn by gradually transferring the proton from the side-chain nitrogen to the vicinal water molecule (Figure \ref{fig:ionization}, top panel). These high-level calculations also included effects of the dielectric constant of the local environment, which was assumed to be equal to 1 in the (gas phase) PM3 calculations.

\subsection{Energetic results}
The graph in the bottom panel of Figure \ref{fig:ionization} shows that both the electronic energy (E) and Gibbs free energy (G) for the ionization of the Asn side chain are rather high, equal to $\sim$31 kcal/mol and $\sim$35 kcal/mol, respectively, even in the highly polar medium such as water. The intein active site is expected to have a dielectric constant lower than that of water, and therefore, the relative energy of ionization will be even higher. We note that the Asn cyclization distance, $y$, is relaxed in these calculations and does not reduce significantly. Thus, Asn cyclization will require additional energy. In contrast, as shown later, the side-chain ionization is almost spontaneous once the Asn side chain has undergone cyclization
and formed a succinimide, consistent with the experimental observations of enhanced cleavage at low pH by Wood \textit{et al}.
\cite{wood1999gsy,wood2000oss}.

\begin{figure}
\centering
\includegraphics[scale=1.0]{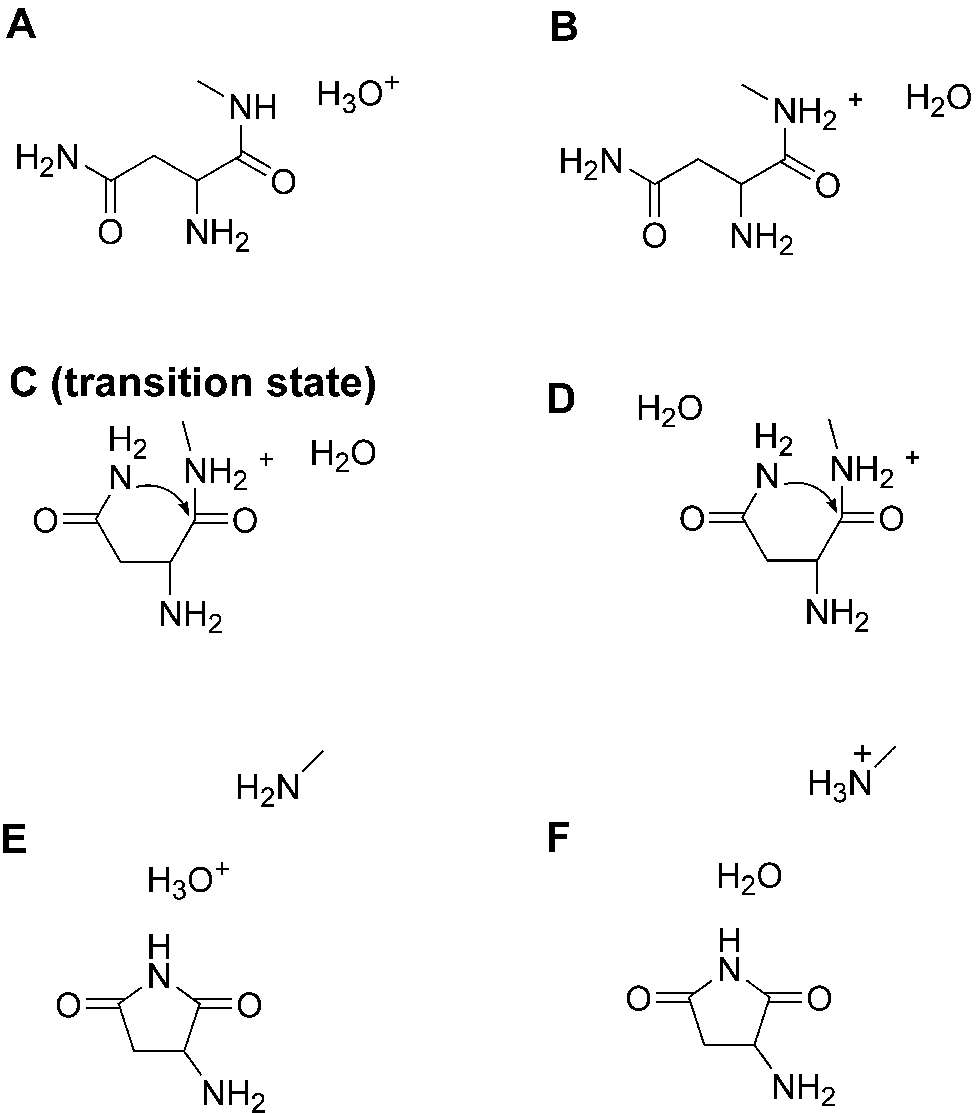}
\caption{\label{fig:reaction6} Important states in the proposed hydronium ion catalyzed Asn cyclization and peptide bond cleavage mechanism (see text for details).}
\end{figure}

\subsection{Mechanistic details}
Collectively, the above calculations allow us to propose a somewhat detailed C-terminal cleavage reaction mechanism at low pH, in which six states shown in Figure \ref{fig:reaction6} are particularly important. Figure \ref{fig:reaction6}A shows the hydronium ion in the context of the relevant part of our intein system. The second state involves the donation of a proton by the hydronium ion to the peptide nitrogen, resulting in water and N-protonated state (Figure \ref{fig:reaction6}B). Asn cyclization is shown in Figure \ref{fig:reaction6}, C and D, where the Asn side chain still has two protons. The explicit water molecule is adjacent to the peptide nitrogen in one case (Figure \ref{fig:reaction6} C) and moves to accept a proton from the forming succinimide in another (Figure \ref{fig:reaction6}D). The formed succinimide with the proton passed back to the hydronium
ion is shown in Figure \ref{fig:reaction6}E, whereas the final product is shown in Figure \ref{fig:reaction6}F. Water is re-formed and the extein segment leaves an a terminal -NH$_3^+$ group.

\begin{figure}
\centering
\vspace{2.0in}
\includegraphics[scale=0.4]{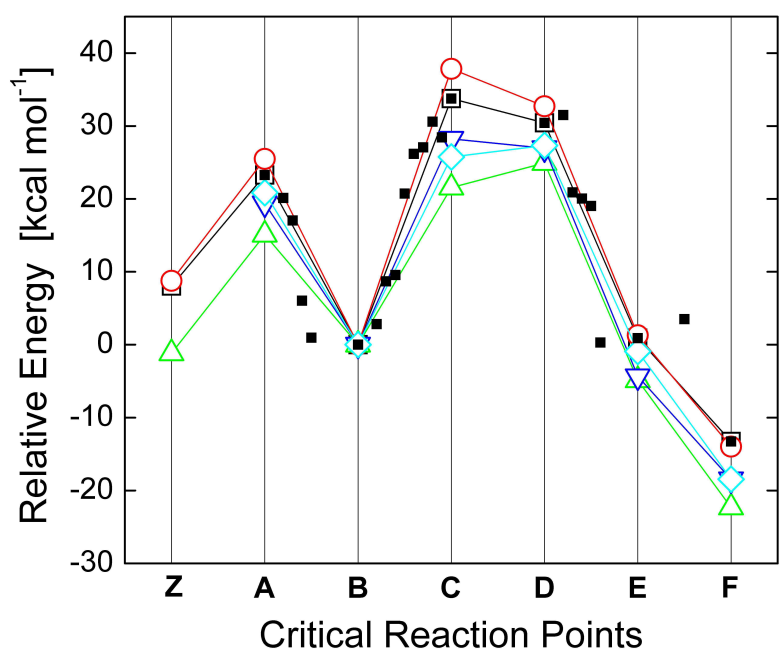}
\caption{\label{fig:energies6} Electronic energies and approximate Gibbs free energies for states A through F in Figure \ref{fig:reaction6} (energies are relative to state B). State Z is shown as a reference and corresponds to a positively charged His and a water molecule (Z). Figure key: vacuum optimized coordinates energy ($\Box$) and Gibbs free energy (\textcolor{red}{{\small{$\bigcirc$}}}); water optimized coordinates, energy (\textcolor{green}{$\bigtriangleup$}); and Gibbs free energy (\textcolor{blue}{$\bigtriangledown$}). Single point energies calculated with B3LYP/6-311++g(2d,p) using the coordinates optimized in implicit water without the diffuse functions (\textcolor{cyan}{{\large$\Diamond$}}). Energies of intermediate structures between the states in Figure \ref{fig:reaction6} optimized in the gas phase ({\small{$\blacksquare$}}) are also shown.}
\end{figure}

The vicinal water molecule plays an important role in this mechanism and is used both as an acid (state A/B) and a base (state D/E). Indeed, succinimide with NH$_2$ is highly acidic due to the resonance effect of amide bonds on either side of that nitrogen. As a result, the nitrogen readily gives a proton to a nearby water molecule (state F).

The energies corresponding to the various states (from A to F) presented in Figure \ref{fig:reaction6} are shown in Figure \ref{fig:energies6}. State Z in Figure \ref{fig:energies6} is shown as a reference. The transition from Z to A corresponds to a positively charged His and a water molecule (Z) forming a neutral His and a hydronium ion (state A). The energy barrier from B to C/D corresponds to the Asn cyclization, where the second proton is still attached to the Asn side chain. Indeed the normal mode corresponding to the single imaginary frequency in state C was in the direction of bond formation, as expected. For this system, the energetics suggest that the nitrogen will give its second proton to water to reform the hydronium ion (Figure \ref{fig:reaction6}E). As discussed above, in the absence of protonation of the peptide nitrogen (\textit{e.g.}, in the case of a neutral peptide), the Asn cyclization has a higher barrier.

Figure \ref{fig:reaction6} also highlights the effects of taking into account the dielectric constant of the environment of the active site on intein. As expected, the energy barrier ($\sim$33 kcal/mol) in the gas phase is reduced to $\sim$25 kcal/mol in implicit solvent having a high dielectric constant. When tested with MP2, the energy barrier was found to be 29.4 kcal/mol at the MP2/aug-cc-pVDZ level. 
Since the crystal structures of active inteins do not include exteins, the conformation of the exteins and N- and C-terminal active sites is unknown. Hence, the intein plus extein system used in the QM/MM calculation will require additional verification to be comparable to precursor inteins used in experiment. The actual dielectric constant of the protein interior is between that of bulk water and vacuum, and hence, our calculations suggest that the energy barrier for C-terminal cleavage lies between 25 and 33 kcal/mol. The experimental value is $\sim$21 kcal/mol at pH 6.0 \cite{wood1999gsy,wood2000oss}.

We note that these initial calculations have several limitations. In the actual intein system, the overall protein structure (including both the intein and exteins) that surrounds the active site provides a significantly greater structural as well as
chemical context for the reaction to occur. Also, there will likely be more than one water molecule in the vicinity of the
active site that could mitigate the C-terminal cleavage reaction. Our system, in contrast, is significantly smaller due to
computational limitations. In addition, our calculations are by necessity static in nature, and ignore the conformational
and water exchange dynamics that are important in enzymatic catalysis \cite{eppler2006wda}. These types of concerns are shared by
most (if not all) quantum calculations of enzymatic reactions. Nevertheless, these initial calculations provide a plausible mechanism for C-terminal cleavage that are tested in the next chapter by using larger computational systems and application of better multi-scale methods.

It should be noted that the possibility of protonation by His is not excluded by the mechanism proposed here. For example, positively charged His could donate a proton to the peptide nitrogen via a water molecule; the reaction can then follow steps similar to the ones outlined here.


\chapter{C-TERMINAL CLEAVAGE - EFFECT OF +1 EXTEIN RESIDUE}
The previous chapter described preliminary results in determining the low pH reaction mechanism for intein C-terminal cleavage.  Here, the C-terminal cleavage reaction mechanism is further discussed in the context of a larger computational system (at least 45 atoms) as well as within the framework of the coupled Hamiltonian approach of quantum mechanics/molecular mechanics (QM/MM).  

\section{Non-essential mutation}
Once splicing was inhibited, the downstream Cys residue (which was the first amino acid of the C-terminal extein or C-extein) was found to be functionally unnecessary for the C-terminal cleavage mechanism. Interestingly, Wood \textit{et al.} observed that this amino acid regulated the reaction rate but did not alter the mechanism \cite{wood2000oss}.  Furthermore, since the CM was found to be exceedingly reactive at low pH values, Wood {\it et al.} \cite{wood2000oss} utilized Met, which was the native N-terminus of the protein that formed the C-extein sequence, to decrease the reaction rate by an order of magnitude. In this experiment, three proteins of various sizes were contrasted with only the Cys/Met C-extein mutation: Thymidylate synthase (31.5 kDa), Hfq Protein (18 kDa), and rh aFGF (14 kDa). For these proteins, the Cys to Met mutation resulted in a decrease of the reaction rate by a factor of 12.0, 5.0, and 7.8, respectively \cite{wood2000oss,woodthesis}.  Figure \ref{fig:drawing} shows a schematic of the intein precursor and products based on these results \cite{muralidharan2006ple,wu2002imp}, although the exact mechanisms that govern the splicing and cleavage reactions are not understood at the atomic level.  In particular, the effect of the single amino acid mutation at C+1, flanking the conserved C1: His7-Asn8 dipeptide at the intein terminus, on the reaction rate is not understood.

\begin{figure}
\centering    
\includegraphics[scale=0.5]{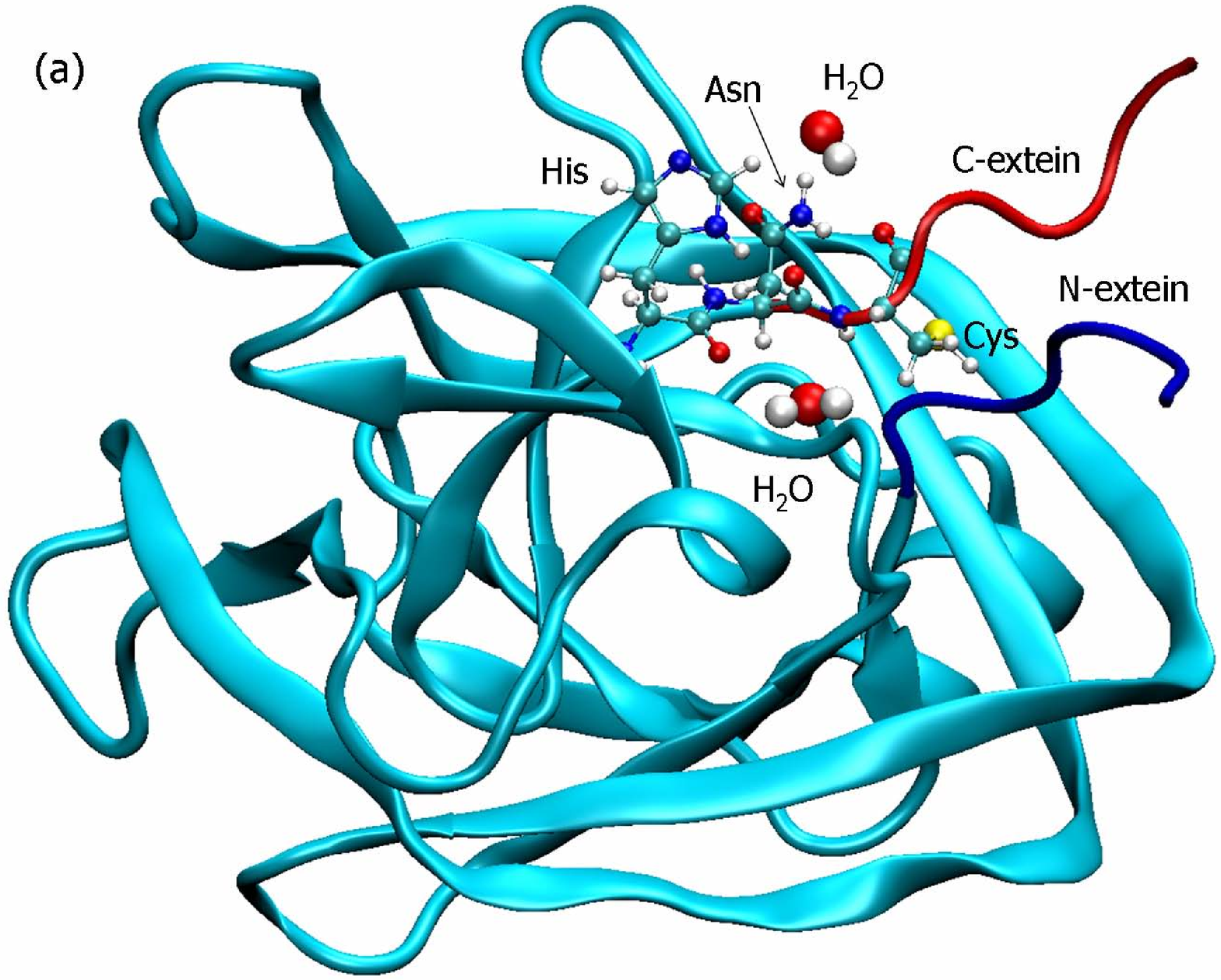}
\includegraphics[scale=0.5]{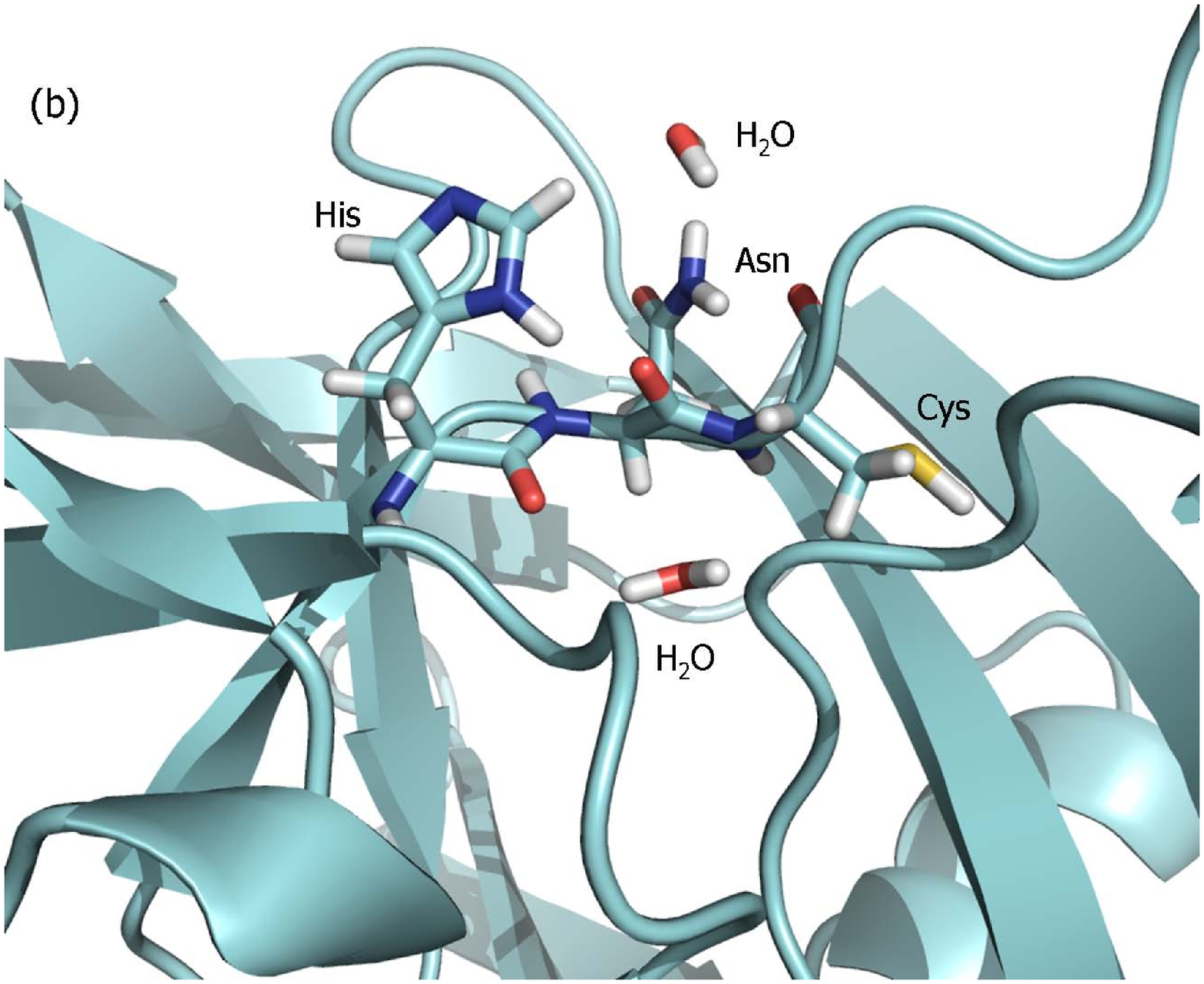}
\caption{\label{fig:protein} The intein cleavage mutant (CM) crystal structure (PDB code 2IN8) with computationally added exteins (a). The C-terminal catalytic site (His--Asn--Cys + two water molecules) is highlighted (b).}
\end{figure}

In order to obtain an atomic-level understanding of the effect of mutation on the reaction barrier, detailed quantum mechanical calculations on the intein C-terminal cleavage reaction have been carried out \cite{shemella}. Simulations were based on both full quantum mechanical molecular analysis as well as a hybrid quantum mechanics and molecular mechanics (QM/MM) approach where the entire protein and solvent are treated classically with parameterized force fields in a molecular mechanics (MM) calculation as shown in Figure \ref{fig:protein}(a).  The 53 atom C-terminal catalytic site (C1-block: His--Asn--Cys, or His--Asn--Xxx, where Xxx is an alternate amino acid) was treated with quantum mechanics (QM) and is shown in Figure \ref{fig:protein}(b). 

The computational energy barrier was smaller for the C-terminal sequence His--Asn--Cys than for that of the His--Asn--Met mutant, consistent with experimental observations \cite{wood2000oss,woodthesis}.  The difference in energy barrier between Cys/Met residues was due to the difference in electron affinity of the amino acids. In addition to Cys and Met, several other amino acids at the first C-extein position (C+1) were studied here.  The energy barrier for C-terminal cleavage, calculated with a larger model system, is confirmed to match experiment.

\subsection{Classical protein system}
Starting with the intein crystal structure for the \textit{Mtu} recA intein, ($\Delta\Delta$Ihh-CM, PDB code 2IN8) \cite{vanroey2007cam}, a product protein without exteins, N- and C-terminal exteins were computationally added and then equilibrated with classical molecular dynamics (MD) simulations.  The N-extein sequence consisted of Ace-Val-Val-Lys-Asn-Lys and the C-extein sequence consisted of Cys-Ser-Pro-Pro-Phe-Nme, both based on the native extein sequences \cite{davis1991nsr}.  Ace and Nme were capping residues for the N and C-terminal exteins, respectively.  AMBER force field parameters \cite{cornell1995sgf} were implemented with GROMACS code \cite{vanderspoel2005gff}. MD simulations were carried out for 4 ns (0.5 ns equilibration, 3.5 ns production run) with temperature T = 298 K, pressure = 1 bar, and number of water molecules = 9548 for Cys and 9549 for Met systems.

\begin{figure}
\centering         
\includegraphics[scale=1.2]{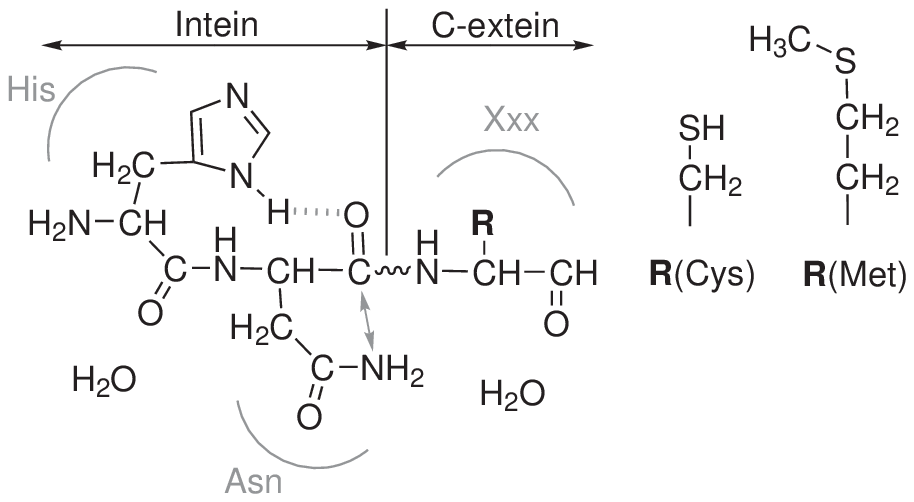}
\caption{\label{fig:tripeptide} The C1-block His--Asn--Xxx active site is shown. The highly conserved H-bond is shown with a dotted line, the cyclization coordinate of Asn is shown with an arrow, and the scissile peptide bond is shown with a wavy line.  Side chains for Cys and Met are shown, although Ala, Val, Thr and Ser were also considered.}
\end{figure}

\subsection{Tripeptide subsystem}
\subsubsection{Description of model system}
The tripeptide active site system (His--Asn--Cys) is highlighted in the view of the full intein crystal structure in Figure \ref{fig:protein}(b).  Gas phase calculations were used to study the effect of site-directed mutagenesis (see Figure \ref{fig:tripeptide}).  Intein crystal structures usually include a hydrogen bond between the N$^{\delta}$--H of the (penultimate) His side chain and the carbonyl O of Asn, the final amino acid of the intein \cite{mizutani2002psr, poland2000sii, klabunde1998csg, duan1997csp, ichiyanagi2000csa}.  Although the penultimate intein His residue has been previously assumed to be the proton donor for C-terminal cleavage reaction in the context of splicing \cite{ding2003csm}, further inspection revealed that this was not the case for pH dependent C-terminal cleavage.  For a simple proton-catalyzed reaction, there is an inverse linear rate dependence on the pH, which was observed experimentally for the C-terminal cleavage reaction \cite{wood2000oss}.  Since the ability of His to act as an acid is based on its local pK$_a$ value, the expected pH-rate curve should be non-linear, specifically sigmoidal in shape, which is in contrast to the linearity observed experimentally. 

The proposed N-protonation mechanism begins with the protonation of the peptide N by a hydronium ion (H$_{3}$O$^{+}$, see Figure \ref{fig:mechanism}).  This in turn causes the scissile peptide bond to elongate, and hence reduces the energy necessary for peptide bond cleavage after Asn cyclization.  After Asn cyclization and aminosuccinimide formation, the extra proton passes to the cleaved C-extein N-terminus (-NH$_{2}$), which is excised and leaves with a positive charge (-NH$_{3}^{+}$, see Figure \ref{fig:mechanism}D).  Although O-protonation was more energetically favorable for a generic or average peptide that was fully solvent exposed, in the case of the intein C-terminal active site, the carbonyl O was strongly hydrogen bonded to the N$^{\delta}$--H of His and was also pointed inward, toward the core of the protein and away from the main body of solvent.  The Asn cyclization reaction after O-protonation instead of N-protonation has been shown to require more energy and does not lead to cleavage of the peptide bond \cite{shemella}.  

Prior to the QM/MM full protein study, the His--Asn--Cys tripeptide system (Figure \ref{fig:tripeptide}) was studied with an isolated gas phase reaction \footnote{Gas phase energy barriers are typically higher than barriers that include electrostatic contributions such as implicit solvent calculations.}.  Certain constraints were included to ensure that the backbone structure reflects that of the protein crystal structure:  both terminal backbone atoms were geometrically fixed in the crystal structure configuration, both dihedral angles are constrained to values from the crystal structure and throughout the classical molecular dynamic trajectories, and the hydrogen bond between N$^{\delta}$--H of His and the carbonyl O of Asn was constrained at a distance of $1.8$ \AA{}.  Without these constraints, the subsystem would likely rearrange into a structure that does not represent the intein C-terminal structure but does minimize the gas phase energy.  By contrasting the effects of mutations, electronic structure properties at critical points were studied, including those at the purely quantum mechanical transition state.

\subsubsection{Energetic results}
For the N-protonation mechanism calculated with the tripeptide system, the computational energy barrier for the His--Asn--Cys system in the gas phase was $27.95$ kcal/mol, in good agreement with the experimental results of {\small$\sim$}$21$ kcal/mol \cite{wood2000oss}.  For a system roughly 30 atoms smaller, the previous gas phase energy barrier was {\small$\sim$}$33$ kcal/mol \cite{shemella}.  This difference indicates that even the most basic approximation of the tertiary structure is important for accurate prediction of certain reaction energy barriers, as we will see with the QM/MM reaction.  Additionally, we have tested and confirmed that the hydrogen bond between N$^{\delta}$--H of His and the carbonyl O of Asn (dashed line in Figure \ref{fig:tripeptide}) caused O to not accept a proton from H$_{3}$O$^{+}$.  This hydrogen bond is usually found at the C-terminus of inteins and is important for reducing the possibility of proton transfer to the carbonyl O.  In fact, the normally highly exothermic reaction for H$_{3}$O$^{+}$ to donate a proton to the carbonyl O atom is endothermic for cases where O is hydrogen bonded with another group \cite{shemella1}.

\begin{table}
\caption{Tripeptide energy barriers ($\Delta$E) for various C-extein mutations (His--Asn--Xxx), percent change ($\% \Delta$E) from His--Asn--Cys energy barrier, and expected change in reaction rate $k_{rel}$ compared to His--Asn--Cys. Structures were geometrically optimized  with the B3LYP/6-311++G(d,p) level of theory. The percent change in the energy barrier, $\% \Delta E \equiv \frac{\Delta E_{Xxx}-\Delta E_{Cys}}{\Delta E_{Xxx}}*100\%$. Reaction rates \textit{k} are relative to the His--Asn--Cys wildtype at $T=310.15$ K (37 $\,^{\circ}\mathrm{C}$.  The Arrhenius equation was used to compare the relative reaction rates between two mutants: $k=k_{1}/k_{2}= e^{- (\Delta E_{1}- \Delta E_{2})/RT}$, where $k_{i}$ and $\Delta E_{i}$ were the reaction rate and energy barrier for the $i^{th}$ mutant, respectively; $R$ was the gas constant and $T$ was the temperature in Kelvin.}
\label{tab:tribarrier} 
\begin{center}
\begin{tabular}{cccc} 
Mutant (Xxx) & $\Delta$E [kcal/mol]	&  $\% \Delta$E  & $k_{rel}$  \\
\hline 
\textbf{Cys}	& \textbf{27.95}	&  0.00	 &  \textbf{1}  \\ 
Thr & 27.56 & -1.39 & 1.88 \\
Ser	& 27.75 & -0.71	& 1.38	\\   
Ala	& 28.64	& 2.46	& 0.32	\\   
Val	& 28.97	& 3.64	& 0.19 \\   
\textbf{Met}	& \textbf{29.58}	& \textbf{5.83}	& \textbf{0.07}	\\   
\end{tabular}
\end{center}
\end{table}

Table \ref{tab:tribarrier} summarizes the calculated energy barriers and relative rate constants for the gas phase tripeptide system with several His--Asn--Xxx mutations. By including additional atoms, the gas phase energy barrier with Xxx = Cys ($27.95$ kcal/mol) was less than the previously calculated barrier for a smaller system ($33$ kcal/mol \cite{shemella}) due to polarity and geometrical effects.  The larger system used here was expected to more closely match the experiment of $21$ kcal/mol, which is does, because of the additional mechanical and electronic influences of nearby protein and solvent groups. 

The energy barrier of the His--Asn--Met system was 1.63 kcal/mol higher than the His--Asn--Cys system, which corresponds to a 5.83\% increase in the energy barrier.
When Cys was mutated to Met, the relative C-terminal reaction rate was predicted to be $0.07$ as fast, or decreased by more than an order of magnitude ($14.0$), which is consistent with experimental results \cite{wood2000oss,woodthesis}.  Interestingly, this model predicts that Thr and Ser instead of Cys will be slightly more effective at pH-dependent C-terminal cleavage, a prediction that is consistent with the +1 position being occupied by Cys, Thr, or Ser in nature, and will be tested in experiment.  In the context of splicing, experiments have shown that Cys, Ser, and Thr are the only amino acids with the ability to complete the transesterification step of splicing \cite{paulus2000psa}, which is consistent because they also are the most efficient at C-terminal cleavage according to the calculations presented here.

\begin{figure}
\centering
     \includegraphics[scale=1.0]{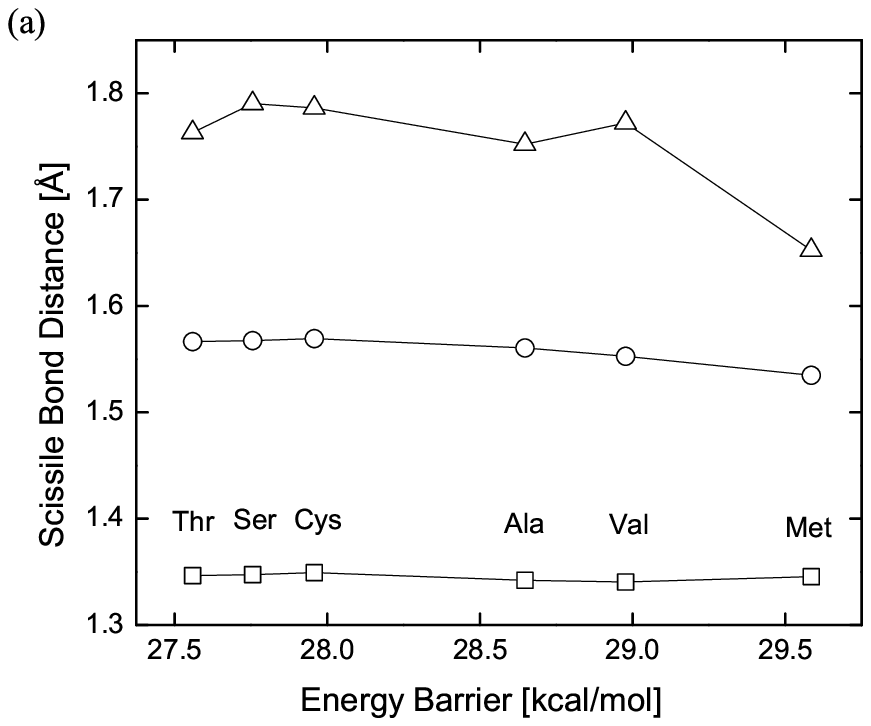}
     \includegraphics[scale=1.0]{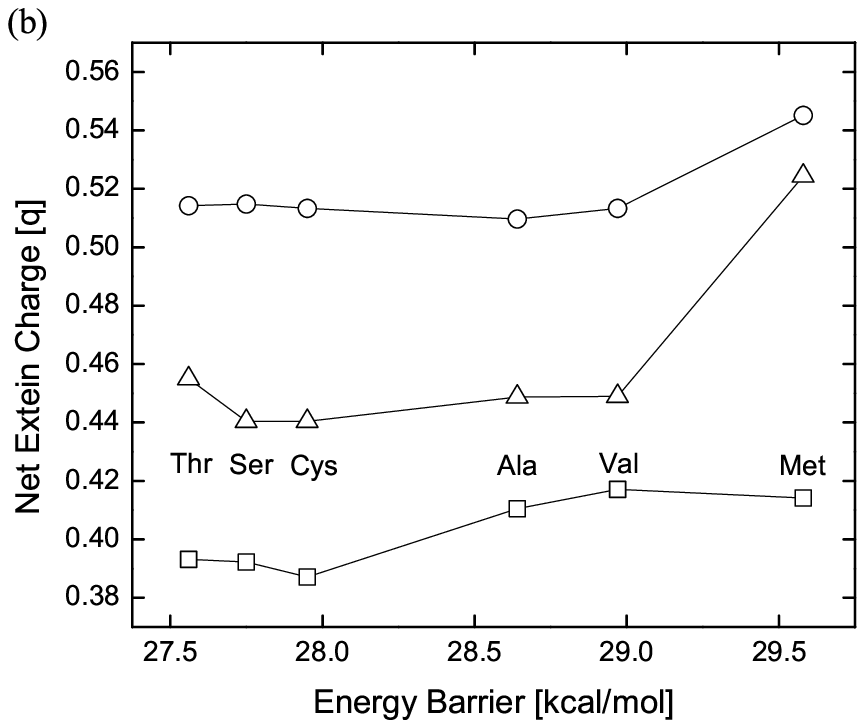}
\caption{\label{fig:bond-stretch} Relaxed scissile peptide bond distance (a) and NPA charges summed for atoms on the C-extein (b) for the tripeptide gas phase system, His--Asn--Xxx (Xxx = Thr, Ser, Cys, Ala, Val, Met). Both the scissile bond distance and the net charge for the C-extein amino acid (Xxx) are plotted as a function of the specific mutant's energy barrier and are shown for the normal amide, ($\square$);  the N-protonated amide, ({\small{$\bigcirc$}}); and the Asn cyclization transition state ($\bigtriangleup$).}
\end{figure}

\subsubsection{Charge analysis}
Natural Populations Analysis (NPA) \cite{reed1988iin} was used to study the electron population and the partial atomic charges. Figure \ref{fig:bond-stretch}A illustrates the effect of amino acid mutation on the scissile peptide bond distance and Figure \ref{fig:bond-stretch}B shows the sum of the NPA charges for the mutated C-extein residue, starting with the -NH at the scissile junction and including the side chain. The scissile bond distance and charge results are shown as a function of each mutant's energy barrier, and include the normal amide, the N-protonated amide, and the transition state corresponding to the pH dependent C-terminal cleavage reaction.  For the neutral amide, the C--N scissile peptide bond distance was $1.3492$ \AA{}  for Cys, which decreased to $1.3455$ \AA{} for Met.  Although this change was extremely small, it does confirm that the amino acid side chain played a small but perceptible role in the properties of a normal peptide bond (which is well known from proton exchange experiments \cite{bai1993pse}).  For the N-protonation step and then the Asn cyclization transition state, the correlation between short scissile bond distance and high energy barrier was more apparent: a shorter peptide bond implied more $\pi$-bond resonance between C and N, less $\pi$-bond resonance between C and O, and more energy was required to break the C--N bond.  An elongated peptide bond implied less $\pi$ bonding between C and N and less energy necessary for peptide bond cleavage \cite{milnerwhite1997pcn}.

A correlation between the energy barrier and the net charge can be seen (Figure \ref{fig:bond-stretch}B), especially for the Cys/Met mutation, signifying that the residues that were able to accept more electrons exhibit a reduced energy barrier whereas the residues that were less likely or unable to accept electrons displayed an increased energy barrier.

\subsection{Single amino acid molecules} 
\subsubsection{Electron affinity and ionization potential analysis} 
To further elucidate the effect of the mutation of the first C-extein amino acid side chain on the energy barrier, the isolated Cys and Met amino acids were studied.  The electron affinities (EA) and ionization potentials (IP) for each were calculated with the B3LYP/6-311++G(d,p) level of theory.  The EA for Cys, (the amount of energy gained or lost when the system goes from neutral to negatively charged), was $6.79$ kcal/mol.  For Met, the EA was $8.27$ kcal/mol, signifying that the side chain of the gas phase Cys residue was more electronegative than for Met.  The reason that Cys was more stable with charge than Met was due to the bonding for each S atom.  Although each side chain contained an S atom, for Cys the S atom was bonded to one methyl group and one H atom.  For Met, both bonds of the S atom were to methyl groups, hence different electron occupation properties.  In changing from neutral to negatively charged, the partial charge of S for Cys changed from $-0.01051$ to $-0.11874$ units of charge, corresponding to the addition of $0.10823$ electrons.  For Met, the charge went from  $0.16894$ to $0.12532$ units of charge, corresponding to the gain of only $0.04362$ electrons.  The S of Cys was able to accommodate more than twice the amount of delocalized electron population as compared to Met, indicating more energetic stability in the negatively charged system.  The difference in ionization potential (IP) for the same isolated Cys and Met amino acids was calculated. The removal of one electron from Cys required $203.05$ kcal/mol while that for Met was $191.14$ kcal/mol.  Combining the fact that Met was more stable when an electron was removed, and the fact that Cys was more stable when an electron was added, we conclude that the ``electron pulling'' and ``electron pushing'' properties of the first C-extein amino acid side chain must have an effect on the actual properties of the scissile peptide bond.

\subsubsection{Energetic analysis of molecular orbitals near the Fermi energy}
\begin{figure}
\centering
\includegraphics[scale=1.0]{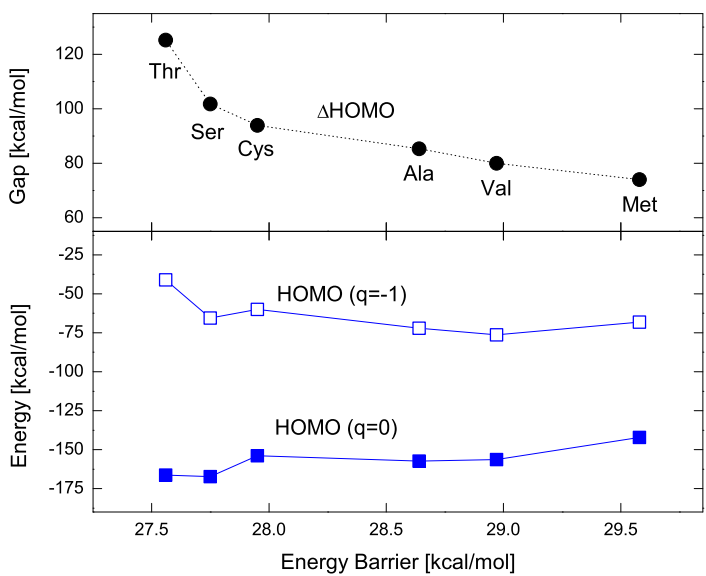}
\caption{\label{fig:HOMO-LUMO} Energies for the highest occupied molecular orbital (HOMO) for the neutral system ($\blacksquare$) and the negatively charged system ($\square$) for the isolated amino acid molecules (Thr, Ser, Cys, Ala, Val, Met), shown in order of their energy barrier found independently for the tripeptide reaction calculation.  The difference between these energies is the energy gap ({\Large $\bullet$}) and is clearly dependent to the energy barrier for the given mutant.} 
\end{figure}

For the isolated amino acids (Thr, Ser, Cys, Ala, Val, and Met), the highest occupied molecular orbital (HOMO) for the neutrally charged system as well as the negatively charged system was compared. The difference in energy between the HOMO of the electron doped (negatively charged) and the neutral system is termed the energy gap, and is shown in Figure \ref{fig:HOMO-LUMO}.
From this analysis of the negatively charged amino acids (geometrically optimized with neutral charge), the isolated amino acids are ranked in order of the energy barrier found when they are the mutant for the tripeptide system, and there was a clear trend in the energy gap between the neutral and negatively charged molecules.  The energy gap was closely related to the electron affinity of the molecule:  as the energy barrier increased for a particular mutant, the gap decreased.  
This single amino acid analysis is of particular interest because from the electronic structure properties of an isolated molecule representing an amino acid side chain, calculated properties such as the electron affinity, the ionization potential, and the molecular orbital energy levels may explain and perhaps predict the relative reaction rate for an unknown mutant at the first C-extein position.

\begin{figure}
\centering
\includegraphics[scale=0.5]{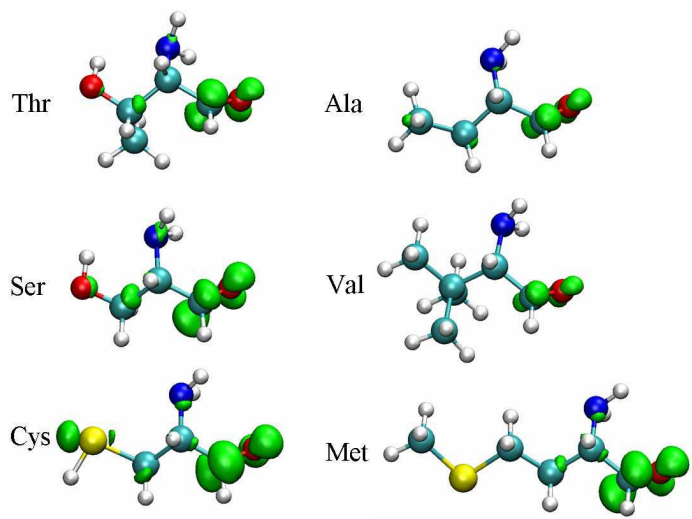}
\caption[The electron affinity (EA) density for single amino acid molecules (Thr, Ser, Cys, Ala, Val, and Met). The electron density surface describes the delocalization of the electron affinity when an electron is added to the system, thus going from neutral to negatively charged ($\Delta \rho$). For downstream amino acids that were efficient at C-terminal cleavage (Thr, Ser, Cys), the EA density extended to the side chain. For amino acids that were less efficient (Ala, Val, Met), the EA density remained on the peptide-like part of the molecule, and away from the side chain.  Atom colors are as follows:  carbon is cyan, nitrogen is blue, oxygen is red, sulfur is yellow, and hydrogen is white; the electron density surface is green.]{\label{fig:electron-affinity} The electron affinity (EA) density for single amino acid molecules (Thr, Ser, Cys, Ala, Val, and Met). The electron density surface describes the delocalization of the electron affinity when an electron is added to the system, thus going from neutral to negatively charged ($\Delta \rho$). For downstream amino acids that were efficient at C-terminal cleavage (Thr, Ser, Cys), the EA density extended to the side chain. For amino acids that were less efficient (Ala, Val, Met), the EA density remained on the peptide-like part of the molecule, and away from the side chain.  Atom colors are as follows:  carbon is cyan, nitrogen is blue, oxygen is white, sulfur is yellow, and hydrogen is white; the electron density surface is green \cite{vmd}.}
\end{figure}

The localization of the EA densities found for molecules characterized in Figure \ref{fig:HOMO-LUMO} is plotted as a volumetric surface in Figure \ref{fig:electron-affinity}, which shows the difference in electron density between the neutral (optimized geometry) and negatively charged (single point geometry) single amino acid residues (Thr, Ser, Cys, Ala, Val, and Met). The presence of electrons on the molecular side chain was observed for amino acids that are more efficient when downstream of the scissile peptide bond in intein C-terminal cleavage.

\subsubsection{Tripeptide analysis}
Returning to the tripeptide system shown in Figure \ref{fig:tripeptide}, Table \ref{tab:population} shows electron population analysis for orbitals with $l=1$ angular momentum ($2s$ orbital), as well as total occupation for $l=0,1$ ($2s$ and $2p$ orbitals).  From the analysis of target atoms belonging to the scissile peptide bond, the expected differences in electron population between Cys/Met mutants were observed.  Specifically, the N atom for Met was generally more occupied with electrons than Cys, which gave it a greater negative charge.

\begin{table}
\caption{Atomic orbital populations for the $2s$ and net $2p$ orbitals as well as the total electronic occupation for the peptide N atom in the gas phase tripeptide calculation.  N is generally less occupied by electrons for Cys as compared to Met, which is consistent with single amino acid electron affinity results. The sum of electron occupation for the $2p_{x}$, $2p_{y}$, and $2p_{z}$ orbitals is written as $2p$. The NPA charge is calculated by subtracting the total electron occupation from the atomic number; a larger electron occupation signifies a more negative charge.}
\label{tab:population}        
\begin{center}
\begin{tabular}{ccccc}
\bf{Orbital} & \bf{Mutant} &  &  \bf{Occupation} & \\ 
      & & Neutral ground state & N-Protonated & Transition state \\
\hline
[$2s$]					&Cys & 1.250 & 1.359	 &	1.386  \\   
						&Met &1.259  & 1.360	 &	1.376	\\   

[$2p$] 					&Cys & 4.341 &	4.285 & 	4.277 \\
						&Met & 4.357 & 4.329 &      4.299 \\

Total 					& Cys & 7.616	& 7.660	& 7.684 \\
					    & Met & 7.641	& 7.710	& 7.699 \\
\end{tabular}
\end{center}
\end{table}

For both mutants, the N atom showed a considerable increase of $2s$ electrons, which corresponded to C and other atoms  returning $\sigma$ electrons to N when the C-N bond was elongated after N-protonation.  A similar situation with $\sigma$ electron back-transfer to N was found for peptide bond rotation, where at the transition state of $90^\circ$ the N atom lost $\pi$ electrons although there was an increase in $\sigma$ electrons to N \cite{milnerwhite1997pcn}; this phenomenon explains why N actually became more negative as similarly seen in the present study. The $2p$ orbitals for N showed distinct differences for the Cys/Met mutations -- even for the neutral ground state which was a normal amide system, a distinction that signified the side chains of adjacent amino acids were important in dictating the exact properties of the peptide bond.  

For the normal amide, the charge of the peptide N for Cys was $-0.616$ and for Met the charge was $-0.641$. For the N-protonation case, the charge of N for the Cys case was $-0.660$, where for Met the charge was $-0.710$.  For the transition state, the charge on N for Cys was $-0.684$, and for Met was $-0.699$. For all three cases the charge of N for Met was more negative than for Cys, which was consistent with the electron affinity calculation described previously.  The side chain plays a subtle yet important role in the electrostatic environment during the cleavage reaction. By having less charge on N, the -NH$_{2}$ group is more energetically favored to leave. From this electron population analysis, differences in the electronic structure of the scissile peptide bond for Cys and Met were observed, which explained why the energy barrier for Cys and Met mutants would be distinct despite an identical mechanism.

\section{Reaction analysis with QM/MM calculations}
\begin{figure}
\centering
\includegraphics[scale=1.0]{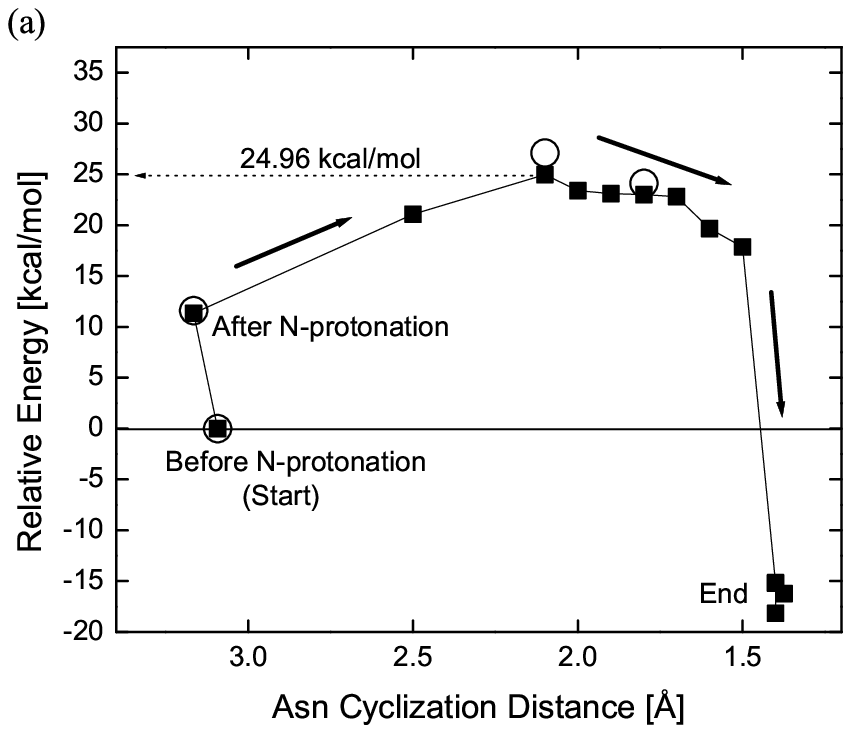}
\includegraphics[scale=1.0]{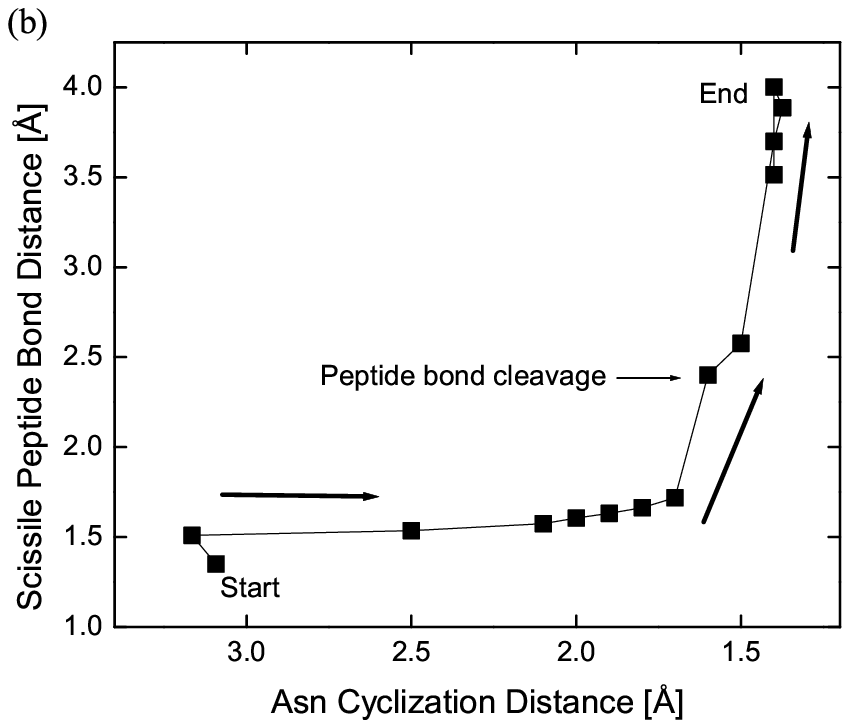}
 \caption{\label{fig:cys-qmmm} Combined QM/MM reaction energy profile (a) and distance of the scissile peptide bond during breakage (b) for His--Asn--Cys plus two water QM system.  QM/MM geometry optimization, ($\blacksquare$).  QM/MM + charge embedding single point energies, ({\small{$\bigcirc$}}).}
\end{figure}

The full protein QM/MM reaction profile was initially calculated with the QM active site region of His--Asn--Cys, and two water molecules (2346 protein atoms, 4161 water atoms, and total 53 QM atoms) \cite{pereira2006qpc}.  Figure \ref{fig:cys-qmmm} shows the QM/MM energy barrier with and without electrostatic embedding.  The energy barrier was $24.96$ kcal/mol for the QM/MM calculation with geometry optimization, in excellent agreement with the $21$ kcal/mol measured experimentally \cite{wood2000oss}.

\subsection{Effect of mutation on energy barriers}
The energy barrier difference for the Cys/Met mutation is of interest in the context of a QM/MM calculation, but because the Met side chain was too spatially extended to simply replace the smaller Cys side chain, additional classical MD simulations were performed (starting from the initial intein plus extein structure) but with Met at the C-extein +1 residue.  Once the full protein system was equilibrated, the QM active site was partitioned to be His--Asn--Met plus the two water molecules in the same location as before (59 total QM atoms).  The Asn cyclization reaction coordinate was scanned after N-protonation by H$_{3}$O$^{+}$.  To compare the effect of the Met/Cys mutation directly, the smaller Cys was substituted for Met, and the geometry was again relaxed.  By doing this, the change in reaction energies may be compared directly because the original protein structures were common for both Met and Cys residues.

These structures were in near total overlap, with the exception of the side chain of the (+1) amino acid, either -CH$_{2}$-SH for Cys, or -(CH$_{2}$)$_{2}$-S-CH$_{3}$ for Met.  Using the B3LYP/6-31G(d,p) level of theory, independent reaction profiles for the Met/Cys mutation were calculated.   For Met the barrier was $27.07$ kcal/mol and for Cys was $26.17$ kcal/mol.  The His--Asn--Met QM active site (as part of the QM/MM system) had an energy barrier of $0.90$ kcal/mol higher than His--Asn--Cys, which corresponded to ratio between reaction rates of $k=k_{Cys}/k_{Met}=0.22$, in good agreement with experimental results and consistent with the tripeptide system conclusions \cite{wood2000oss,woodthesis}. 

\subsection{Effect of mutation on electron occupation}
In addition to energy barriers, the Mulliken charge \cite{mulliken1955epa} was calculated for critical atoms\footnote{Natural Population Analysis (NPA) is not implemented with QM/MM at this time.}.  For the N atom of the scissile bond and for the ground state, the partial charge was $-0.538$ for Cys and for Met was $-0.545$.  For the N-protonation state the partial charge of N was $-0.609$ for Cys and was $-0.615$ for Met.  At the transition state, the charge for Cys was $-0.584$ and for Met was $-0.598$.  In all cases the partial charge of the N atom for the Met mutant was more negative, which was consistent with the tripeptide results, and is explained by using the electron affinity and ionization potential for the isolated Cys and Met amino acids.  When the net Mulliken charge was summed for the C-extein residue (Cys or Met) in the QM/MM context for the normal amide ground state, for Met the net charge was $0.225$, and for Cys the net charge was $0.209$.  

Within the QM/MM system, the charge for the backbone and side chain of the first C-extein residue was added. The net charge of Cys was more negative than Met, which is in agreement with the model QM calculations described in the preceding paragraphs.

By combining model system QM calculations and full-protein QM/MM simulations, the non-mechanistic regulation of reaction rate regulation for single amino acid mutations near to the active site was confirmed, explained, and predicted.  Similar methods are also useful for testing an unknown mechanism based on the correlated experimental results of kinetic data (from non-essential amino acid site-directed mutagenesis).

\section{Conclusions: C-terminal cleavage} 

The C-terminal cleavage reaction and the previously proposed N-protonation mechanism were tested by increasing the QM system size by 30 atoms to at least 53 atoms.  In addition, full-protein QM/MM analysis was performed.  The pH dependent C-terminal cleavage reaction undergoes simple proton-catalysis by a hydronium ion that protonates the peptide N atom.  The peptide bond, now resonance destabilized, is elongated and the peptide C atom is open for attack by the Asn side chain.  During Asn cyclization, the peptide bond cleaves while an aminosuccinimide ring is formed.  The final step involves the donation of the extra proton on the aminosuccinimide to the -NH$_{2}$ leaving group \textit{via} water, thus making the leaving group positively charged.  Our QM/MM results included the effects from the protein interior, both mechanical and electrostatic.

The ``non-mechanistic'' role of the first amino acid of the C-extein was confirmed.  This amino acid, although not necessary for C-terminal cleavage, did have an effect on the reaction rate by about an order of magnitude, as measured by Wood \textit{et al.} \cite{wood1999gsy,wood2000oss,woodthesis}.  In this study, the precise energy barrier for C-terminal cleavage (and hence reaction rate) was shown to be dependent on the side chain of the amino acid downstream from the scissile bond.  Explained by the electron occupation and partial atomic charges for each residue at the C+1 position, considerable differences that led to a distinction in energy barriers were calculated and found to be in agreement with experimentally observed reaction rates.


\chapter{SPLICING}

\section{Splicing introduction}
\begin{figure}
\centering
\label{fig:conserved}
\includegraphics[scale=0.7]{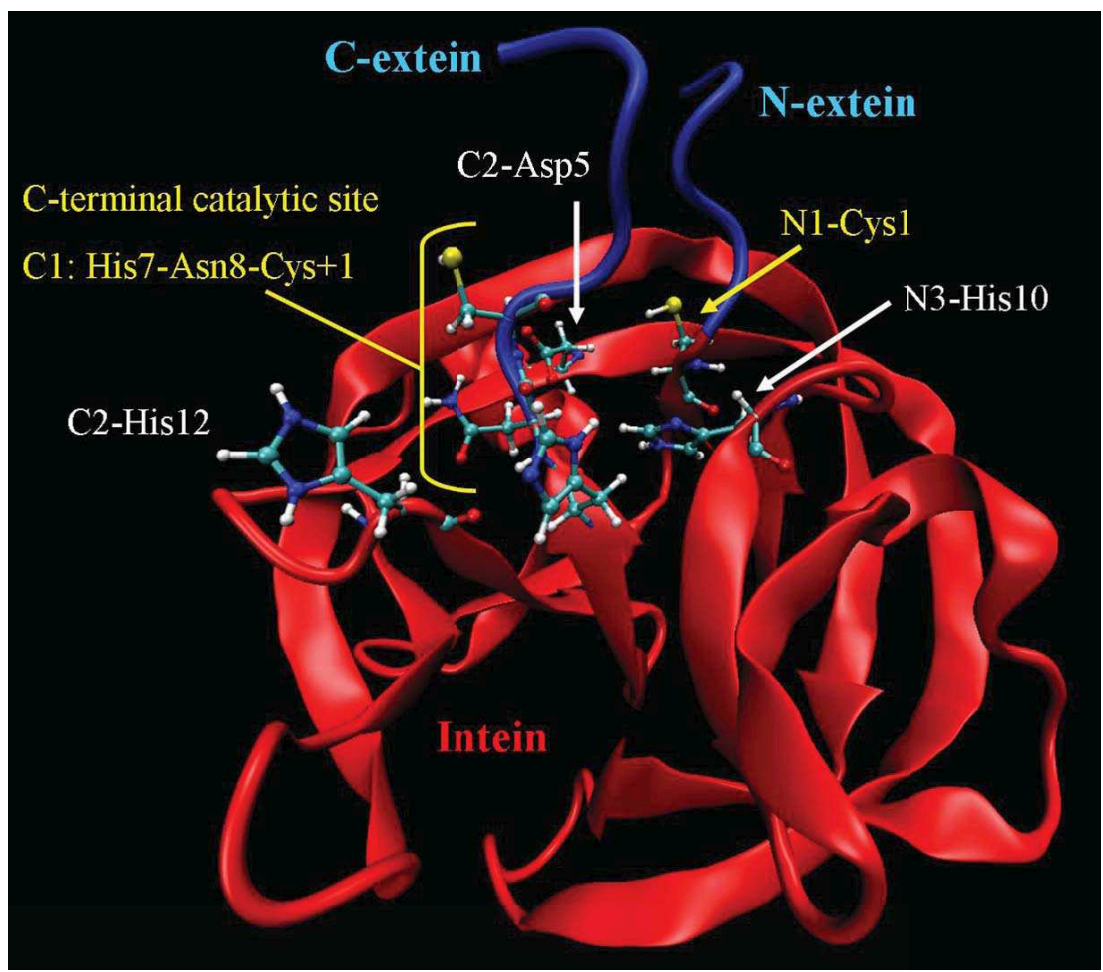}
\caption{The conserved amino acids treated as the quantum mechanical active site are highlighted in ribbon diagram of the the crystal structure.}
\end{figure}

During the process of protein splicing, an intein auto-catalytically cleaves both the N-terminus and C-terminus and simultaneously ligates the flanking peptides (exteins) \cite{paulus2000psa}.  Inteins are protein segments that catalyze splicing in a host protein; specifically, the N- and C-exteins that flank the intein\cite{belforts, perler1994pse}. From random and site-directed mutagenesis, the splicing reaction has been proposed to depend on several highly conserved residues \cite{wood1999gsy, southworth2000aps, chen2000psa, mills2004psp} located in non-gapped conserved regions (blocks) that are consistent for most canonical inteins. Conserved residues are highlighted in Figure \ref{fig:conserved}. Despite this information, an atomic-level description of the splicing mechanism is not available.

Intein crystal structures \cite{vanroey2007cam, mizutani2002psr, poland2000sii, klabunde1998csg, duan1997csp, ichiyanagi2000csa, sun2005csi} have shed light on the splicing and C-terminal cleavage mechanisms.  One limitation in using crystal structures for mechanistic studies is that they are mostly product proteins that do not include exteins, which were excised during the splicing reaction. Additionally, there is often increased spacing between N- and C-termini of the intein due to electrostatic interactions with solvent or metal ion binding.  In general, the intein termini of crystal structures are considered to be geometrically flexible. Despite these concerns with intein crystal structures, they are critical for understanding many aspects of intein reactions.  One intein (\textit{Ssp} dnaE) was crystallized both as a product protein as well as an engineered inactive precursor, with exteins \cite{sun2005csi}.  In this case, short exteins were crystallized with the intein, and the catalytic residues had to be substituted in order to inhibit the reaction.  Therefore, the exact state in which the inteins are found before splicing is not clear since to observe the precursor, mutations that inhibit splicing need to be engineered. Nevertheless, overlap between inhibited precursor and spliced product was reported for the protein backbone.

Experiments on the splicing reaction are exceedingly complex due to the various possible products.  For example, to measure the first step of splicing, namely the N-terminal thioester formation, a typical construct is used where the C-extein is not present (therefore no C-terminal cleavage is possible).  Dithiothreitol (DTT) is added in order to cleave the N-terminal thioester, although the exact process is not precise because without the C-extein, there is a proximal charged and flexible C-terminus which may affect the reaction steps and rates at the N-terminal. 
Because of this, experiments using DTT and measuring N-terminal cleavage most likely encounter different chemistry from N-terminal cleavage within the context of splicing.

\subsection{Splicing reaction}
The intein substrate is naturally tethered to its active site.  This makes the catalytic mechanism different than typical enzymes because there is binding between substrate and enzyme, and here the turnover number equals one. Because the substrate is tethered to the active site,  it has been extremely difficult to isolate molecular inhibitors for the splicing reaction \cite{lew2002vss}.  From \textit{in vitro} experiments, zinc and copper atoms are known to inhibit splicing \cite{mills2001rip, dassa2007tps}.  Due to the inability of zinc to enter the \textit{E. coli} cell, \textit{in vivo} experiments with precursor or intermediary intein constructs inhibited by zinc are not likely to occur.

Because of these factors, the splicing mechanism is not understood on the atomic level \cite{clarke1994pms}, and this has limited the biotechnological applications utilizing inteins \cite{amitai1999fte}.  Splicing is based on a four steps, and it is hoped that elucidation of the intein splicing mechanism will lead to new schemes to either trigger the reaction or to inhibit it.  Biotechnological applications include switches and delivery devices for bioseparations \cite{miao2005ssa,banki2005sbu}, drug development \cite{paulus2003itp}, and molecular sensors \cite{mootz2002pst, muralidharan2006ple}.

To date, the splicing reaction scheme, based on crystal structures \cite{vanroey2007cam, mizutani2002psr, poland2000sii, klabunde1998csg, duan1997csp, ichiyanagi2000csa, sun2005csi} and mutagenesis information, is as follows:  The thiol group (-SH) of the first intein residue (N1-Cys1) is involved in an N-S shift, where it attacks the downstream carbonyl carbon thus releasing the peptide nitrogen.   This intermediate state is called the N-terminal thioester. After N-terminal thioester formation, transesterification occurs.  In this step, the C-terminal Cys (first residue of C-extein and not present in the crystal structures) attacks the carbonyl of the thioester, thus combining the N- and C-exteins.  Once transesterification is complete, the C-terminal Asn residue undergoes cyclization into a succinimide and C-terminal cleavage occurs. Now, the N- and C-exteins are fused and the intein is released.  The ligated product (N- and C-extein) then undergoes the final reaction which involves an S-N shift, replacing the thioester with a more stable peptide bond.  Interestingly, the two Cys residues may be replaced by Ser\footnote{The C-extein Cys residue can also be replaced with Thr} with similar reactions steps, although the reaction usually occurs at a slower rate due to decreased acidity of the Ser –OH side chain compared to the –SH of Cys \cite{shingledecker2000rcr}.

The reaction scheme described above does not include the explicit role and pathway of protons. Prior to N-terminal thioester formation, the thiol group must lose its proton and become a thiolate (-S$^-$). Also, prior to transesterification, the C-extein thiol must similarly be ionized.  To be an adequate leaving group, the peptide nitrogen must have two protons (-NH$_2$). To be a stable N-terminus, at biological pH levels the nitrogen will have three protons (-NH$_3^+$).

\subsection{Effects of mutation at highly conserved locations}
The limited understanding of the role of protons is based from mutagenesis experiments \cite{wood1999gsy, southworth2000aps, chen2000psa, mills2004psp} and X-ray crystallography studies \cite{vanroey2007cam, mizutani2002psr, poland2000sii, klabunde1998csg, duan1997csp, ichiyanagi2000csa, sun2005csi}.  The splicing reaction has been proposed to proceed \textit{via} the following highly conserved residues for canonical inteins, which are shown in Figure \ref{fig:conserved}:  
\begin{itemize}
  \item N1-block Cys (N1-Cys1), which undergoes an N-terminal N-S acyl shift.  This results in the N-terminal thioester.  Ser or Thr may be used in certain constructs with a noticeable impact on reaction rate.  
  \item N3-block His (N3-His10) regulates N-terminal hydrolysis and is the most conserved of all intein residues.
  \item C2-block Asp (C2-Asp5) links the N- and C-termini at the active site. This coupling leads to successful splicing and not independent N- and C-terminal cleavage events. When mutated to Gly, the intein is converted to the C-terminal cleavage mutant (CM), a highly efficient and pH dependent mutant \cite{wood1999gsy,wood2000oss, shemella, shemella2}.  One major question exists: does CM have inhibited splicing or is cleavage just more efficient? A catalytic role for C2-Asp5 would explain why this residue is necessary chemically.
  \item C2-block His (C2-His12) may be a proton acceptor/donor for C-terminal cleavage reaction in the context of splicing.  
  \item C1-block C-terminal catalytic site and first C-extein residue (C1: His7 - Asn8 - Cys(+1)) are important for C-terminal cleavage.  The penultimate His side chain has a highly conserved hydrogen bond with the carbonyl oxygen, which is considered to be structurally important.  One intein lacks the penultimate His and uses an Arg residue at a different position for a similar role \cite{sun2005csi}.   The role of Asn and the C-extein Cys are described in the C-terminal cleavage chapters.
  \item An important yet not understood mutation is N3-Val4Leu (V67L), which converts the deletion variant mini-intein into a faster splicing mutant (SM).
\end{itemize}

Extein sequence, too, may play a role in intein reactions \cite{mizutani2002psr, klabunde1998csg, sun2005csi}. The importance of conserved residues in the exteins was shown for split inteins, specifically for residues near the scissile peptide bond at both the N- and C-termini \cite{iwai2006hep}.  Split inteins undergo trans-splicing, where the intein is split into an N-intein and C-intein that are connected to the N- and C-exteins, respectively.  In this case, the separate intein fragments must come into contact and enter a conformation that is conducive to splicing, and the exteins may play a role in this conformational search.

Here we present the intein splicing mechanism with an atomic-level description based on first principles density functional calculations.  

\section{Description of model system}
The splicing mutant (SM) crystal structure (PDB code 2IMZ) was computationally amended to include both an N- and C-extein.  The solvated system was equilibrated with classical molecular dynamics (MD) simulations with exteins present.  To test the N-S shift, QM/MM calculations were performed.  The protein consisted of 2351 atoms and there were 1321 water molecules present, for a system total of 6314 atoms.  The N-terminal active site is based on the N-extein residue Lys(-1) and the intein residues N1: Cys1-Leu2.  Also included are N3-Thr7, N3-His10, C2-Asp5.  From the C1 block and C-extein (C-terminal active site): Val6-His7-Asn8-Cys(+1)-Ser(+2), where the peptide bond between C1-Asn8 and the C-extein Cys is cut during C-terminal cleavage.  For non-mechanistic residues, those that affect the QM system \textit{via} polarization, the entire amino acid may not be chosen for inclusion in the QM system -- an example of such a residue is N3-Thr7, where only the side chain is included.  Nine water molecules were included explicitly in the quantum mechanical calculations.  Because of the extremely large QM active site (130+ atoms), it should be noted that for various calculations this QM system is either appended or truncated\footnote{The absolute energy is dependent on many factors including densinty functional, basis set, system size, and polarization environment. Because the splicing reaction occurs over a large portion of protein, it is necessary to modify the active site model system for different steps along the reaction path.  All energies presented are relative to analogue systems.}, but for all energy calculations and comparisons, the systems are directly identical in all atomic constituents (number of electrons, charge, spin, and nuclei).

\section{First principles splicing mechanism}
Protein crystal structures allow for mechanistic predictions based on structural information. Random and site-directed mutagenesis studies are helpful for mechanistic prediction because they show the reaction may be inhibited, slowed, or sped up through mutation. Inteins are a unique protein because they can be thought of as their own enzyme, where the exteins can be thought of as the substrate.  The mechanistic details for their auto-catalytic behavior are coded in their amino acid sequence.  Because reactive intein crystal structures do not include exteins (with the exception of inteins engineered to be inhibited) and because exteins are an important part to the mechanism, structural data are not able to provide complete mechanistic understanding of the splicing reaction at the atomic level.  Indeed, in terms of protons, which are the driving force of chemical reactions and acid/base chemistry, splicing is not understood.  Instead, certain amino acids are determined to be important or critical for splicing.  We present the splicing mechanism that is based on first principles calculations and that utilizes those amino acids that are highly conserved.

\begin{figure}
\centering
\includegraphics[scale=0.8]{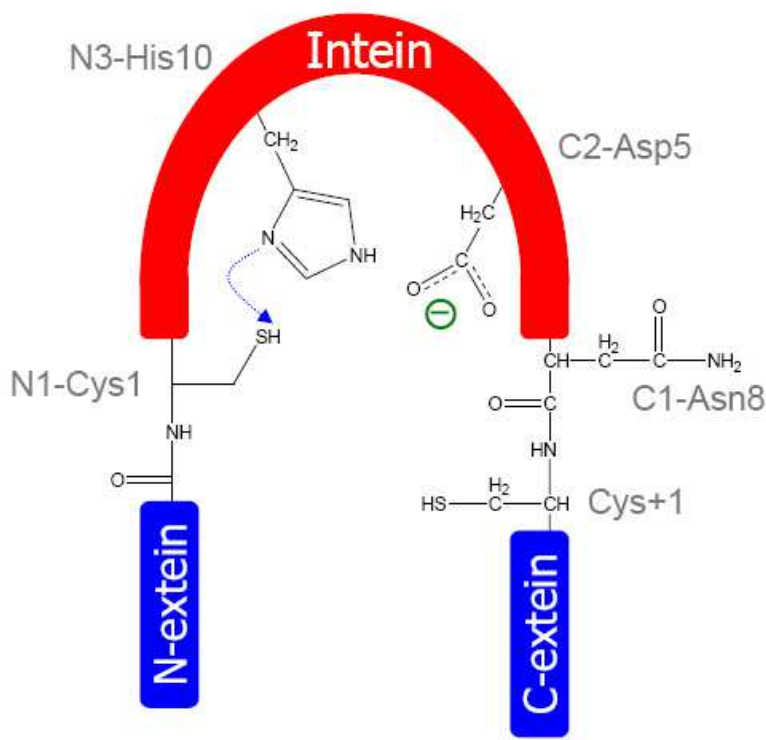}
\caption{\label{fig:splicing1} Ground state for the intein splicing mechanism.  Arrows indicate the direction of electrons.  Critical residues are shown, water molecules included in the calculation are not shown for clarity.}
\end{figure}

\subsection{N-terminal thioester}
The ground state of the intein splicing system is shown in Figure \ref{fig:splicing1}.  The proposed first step of splicing is the N-S shift, where the N1-Cys1 thiol group (-SH) is ionized and this thiolate group then attacks the carbonyl C of the upstream peptide bond. The newly formed thioester C(=O)-S bond causes the electrons shared between C and N (the peptide bond) to be unshared and the C-N bond breaks.  For this step, the peptide N atom requires protonation (-NH$_2$ terminal) in order to be a sufficient leaving group.
	
\begin{figure}
\centering
\includegraphics[scale=0.8]{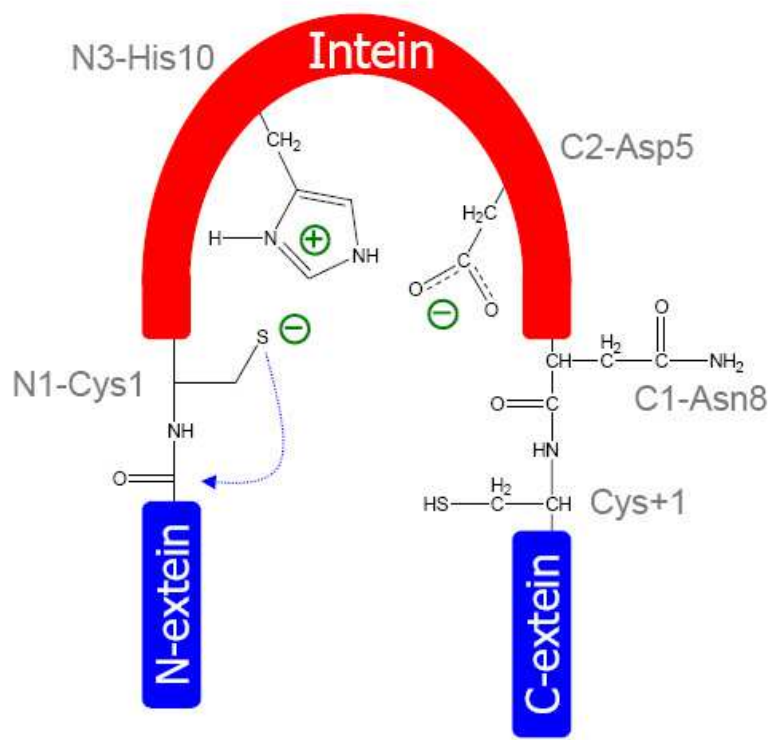}
\caption{\label{fig:splicing2} The N1-Cys1 side chain is ionized to a thiolate and gave its proton to N3-His10.}
\end{figure}

The role of N3/His10 is first as a base and then as an acid.  This residue, when in its neutral configuration, is in position to accept a proton from the N1-Cys1 thiol group, \textit{via} water. This step in the splicing reaction is shown in Figure \ref{fig:splicing2}. The energy for this proton transfer (N1-Cys1 to N3-His10) is an upper limit for the N1-Cys1 ionization energy and was computationally found to be 19.27 kcal/mol. The ionization energy is a computational maximum because the thiol group is not likely to exist as a charged group without being somewhere in the process of the N-S shift reaction, and the protein was not equilibrated to accommodate this proton transfer. Once the proton has moved from N1-Cys1 to N3-His10, and during the thioester formation, the N3-His10 residue, now positively charged, is in position to donate a proton to the scissile N-terminal peptide nitrogen (without the need for a transitory water molecule).

The energy difference between the system where N3-His10 is positively charged and the system where there is a terminal –NH$_2$ on the newly formed thioester is 25.05 kcal/mol.  These energy values are approximate due to the static nature of these calculations, which does not allow for full-protein equilibration of the structure with the intermediate N1-Cys1 thiolate and positively charged N3-His10. There exists an additional intermediate state that has an energy of 18.95 kcal/mol, where a ring structure is present and N is singly protonated and the carbonyl O is also protonated.  Although this state does not show bond breakage between C and N, it does likely undergo C(=O)-S cleavage with DTT despite not necessarily being on the splicing reaction path.  If this state were experimentally realized and more stable than the intermediate structure on the reaction pathway, then N-terminal cleavage \textit{via} DTT may occur independently and outside the context of splicing.

From these calculations, the N-terminal thioester formed due to the combined acid/base properties at near-neutral pH levels of the N3-His10 residue, which first accepts a proton from solvent and then donates a proton back to the N-terminus.  
	
NH$_2$ termination is necessary for an N-terminal leaving group, but is not optimally stable.  Based on typical pK$_a$ values, at neutral pH, solvent exposed –NH$_2$ terminal groups will be again protonated and positively charged (-NH$_3^+$).  The average pK$_a$ for a solvent exposed N-terminus depends on the downstream side chain and ranges from 8.8 to 10.8 \cite{voet1995b}.  For this reason, an additional proton was included in the system (charge and spin are constant input parameters), and was tested at several positions (N3-His10 being one of them).  With the thioester present, a stable state was found for -NH$_3^+$ termination, which was important for understanding the splicing reaction at the atomic level.

\begin{figure}
\centering
\includegraphics[scale=0.8]{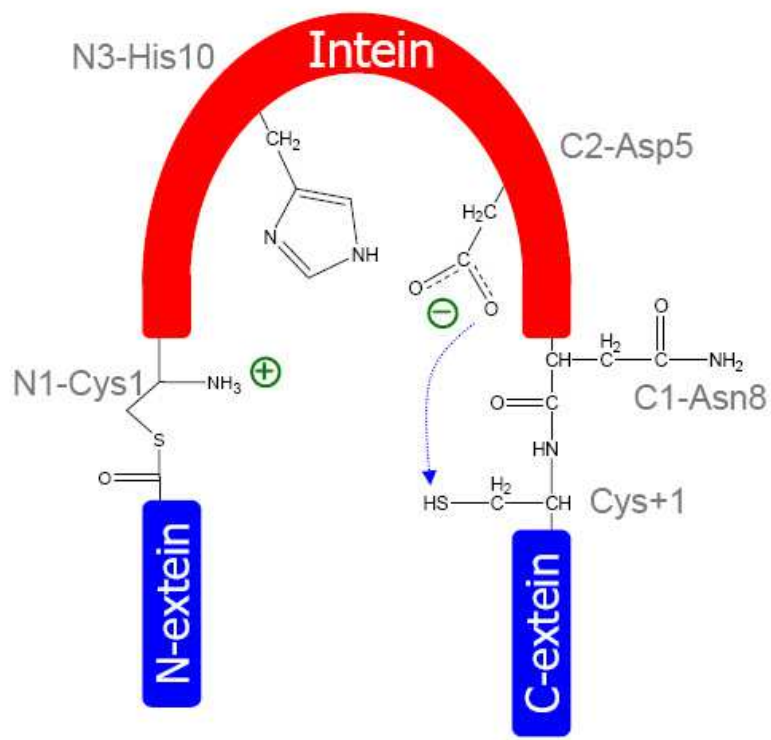}
\caption{\label{fig:splicing4} N-terminal thioester shown after protonation of -NH$_2$ (-NH$_3^+$) from solvent, which is in electrostatic contact with Cys+1 and C2-Asp5 and catalyzes the transesterification step.}
\end{figure}

The spacing between the termini was too distant for the C-extein Cys+1 residue to attack the N-terminus.  This could be because the crystal structure is the product protein segment, and the ends are flexible, charged, solvent exposed, and possibly metal-chelated. For this reason, the N- and C-terminal active sites were adiabatically combined after N-terminal thioester formation using both fully classical and semi-empirical/classical simulations (QM/MM).  After equilibration, a stable minimum was found with first principles methods; no constraint was necessary between the N- and C- termini with spacing of approximately 4.5 \AA{}, which is sufficiently close for the transesterification step.

The intermediate structures described here are accessible only with simulations.  With these, the atomic-level details of the splicing mechanism were observed.  In particular  with this intermediate state (N-terminal thioester), the N-terminal -NH$_3^+$ group was in direct contact with both the highly conserved C2-Asp5 and the Cys+1 thiol group, which turned out to be an essential step in the splicing reaction.  We will call these three groups the catalytic triad (N-terminal -NH$_3^+$, C2-Asp5 -COO$^-$, C-extein Cys+1 -SH), shown in Figure \ref{fig:splicing4}.

\begin{figure}
\centering
\includegraphics[scale=0.8]{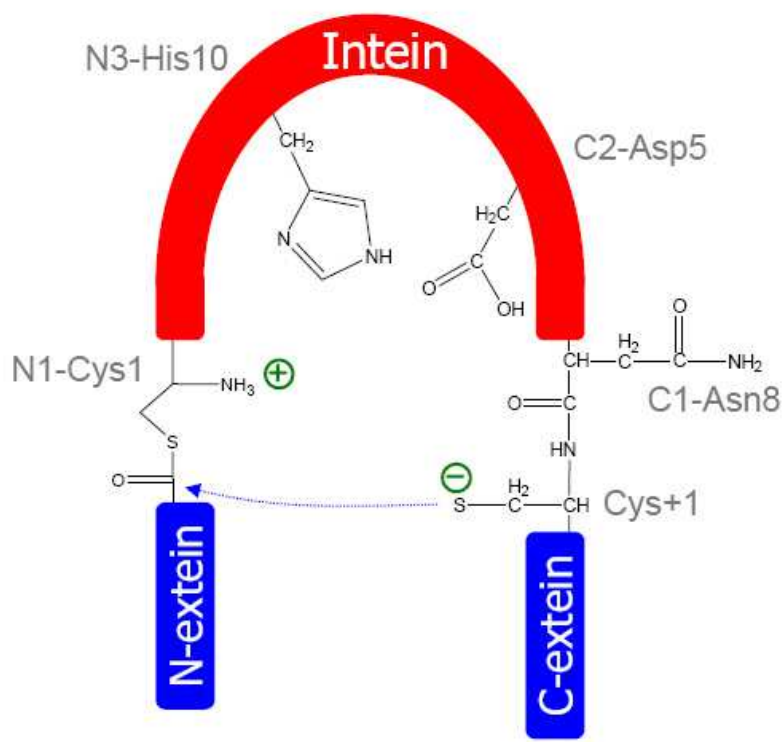}
\caption{\label{fig:splicing5} The C-terminal Cys+1 experiences electrostatic repulsion due to the N-terminal -NH$_3^+$, and therefore donates a proton to C2-Asp5.}
\end{figure}

\subsection{Mutation of C2-Asp5}
The C2-Asp5Gly mutation\footnote{The mutation formerly known as D422G.} was experimentally concluded to inhibit splicing activity and enhance C-terminal cleavage activity, the latter having an improved reaction rate with low pH \cite{wood1999gsy,wood2000oss}.  The presence of Gly was proposed to have one of two roles. First, the smaller volume of Gly may allow solvent to access the C-terminal active site which would enhance C-terminal cleavage and show strong rate dependence on the pH of solvent.  Second, the freedom of additional torsional rotation of Gly residues might allow the folded protein more flexibility for C-terminal cleavage.   Neither of these hypotheses explained that splicing was inhibited with the C2-Asp5Gly mutation.  Our experimental mutagenesis results indicate that when Ala or Asn is present instead of C2-Asp5, splicing is not observed and C-terminal cleavage is enhanced (although still less rapid than with Gly) \cite{vanroey2007cam, brian}.  From the fact that splicing is observed with Asp or Glu, we can conclude that the side chain at position 5 in the C2-block is chemically/mechanistically involved in the splicing reaction and actually essential as a proton acceptor and donor.

Using first principles methods, the role of the C2-Asp5 was determined. The Cys+1 thiol group (SH) was in close contact with both the C2-Asp5 and the N-terminal -NH$_3^+$ group.  It was observed that the Cys+1 –SH group was in direct contact with the N-terminal –NH$_3^+$ group, which is an unfavorable electrostatic interaction between (partially) positively charged hydrogen atoms (see Figure \ref{fig:splicing5}).  From this, the thiol (-SH) group donated its proton to the carboxylate group of Asp.  The triad was stabilized by having -S$^-$, -NH$_3^+$, and -COOH, in direct contact.

The C-terminal Cys+1 thiol group was located in a position in contact with the -NH$_3^+$ group that was newly formed after N-terminal thioester formation.  This electrostatic interaction was unfavorable, and Cys+1 donated its proton to C2-Asp5, which goes from negatively charged to neutral.  The thiolate (-S$^-$) was then activated for the transesterification step.

\subsection{Decreased stabilization of -NH$_3^+$ with C2/Asp5Gly}
\begin{figure}
\centering
\includegraphics[scale=0.8]{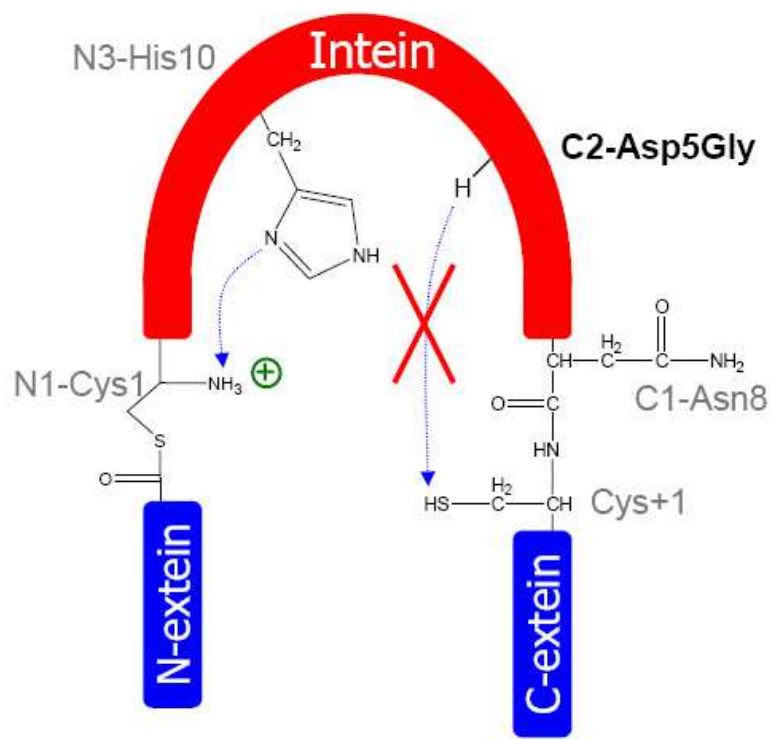}
\caption{\label{fig:Asp5Gly} With the C2-Asp5Gly mutation (D422G), the C-terminal Cys+1 cannot donate its proton to the C2-Asp5 side chain.  To stabilize the unfavorable electrostatic interaction between Cys+1 the the N-terminal -NH$_3^+$, the N-terminal group gives a proton back to the N3-His10 side chain.  This result indicates that the cleavage mutant (CM) is unable to complete the transesterification step of splicing and that the overall splicing reaction is inhibited.  N-terminal hydrolysis should still be possible, although it may occur much less frequently than pH enhanced C-terminal cleavage.}
\end{figure}

In addition to the catalytic role of C2-Asp5, we have studied the effect of Gly at that position.  It was observed (see Figure \ref{fig:Asp5Gly}) that with the C2-Asp5Gly mutation, the Cys+1 thiol group (-SH) is unable to donate a proton to the chemically inert Gly, as it did with Asp.  Because of the need for electrostatic stabilization, it was observed that the -NH$_3^+$ terminal group spontaneously gives its proton to N3-His10, and becomes the non-charged -NH$_2$ group.  This is a step backwards in the reaction, and confirms that splicing is impossible with the C2-Asp5Gly mutation.  In addition, the presence of Glu (which is chemically similar to Asp) in that position results in some splicing products (albeit reduced reaction rate), consistent with this mechanistic assertion.

At this reaction step, C2-Asp5 is protonated, despite the typical pK$_a$ of the aspartate side chain being 3.9 \cite{voet1995b}.  At biological pH levels, the side chain should be negatively charged (COO-).  We have shown that due to the electrostatic effect of the -NH$_3^+$ terminal group, the Cys+1 thiol donated a proton to the Asp group thus making it a carboxylic acid (COOH), despite the biological pH level.  This is important for the upcoming reaction step of C-terminal cleavage, which occurs after transesterification, and was shown independently to occur more rapidly at sub-biological pH levels \cite{wood1999gsy,wood2000oss}.
	
\subsection{Transesterification}
\begin{figure}
\centering
\includegraphics[scale=0.8]{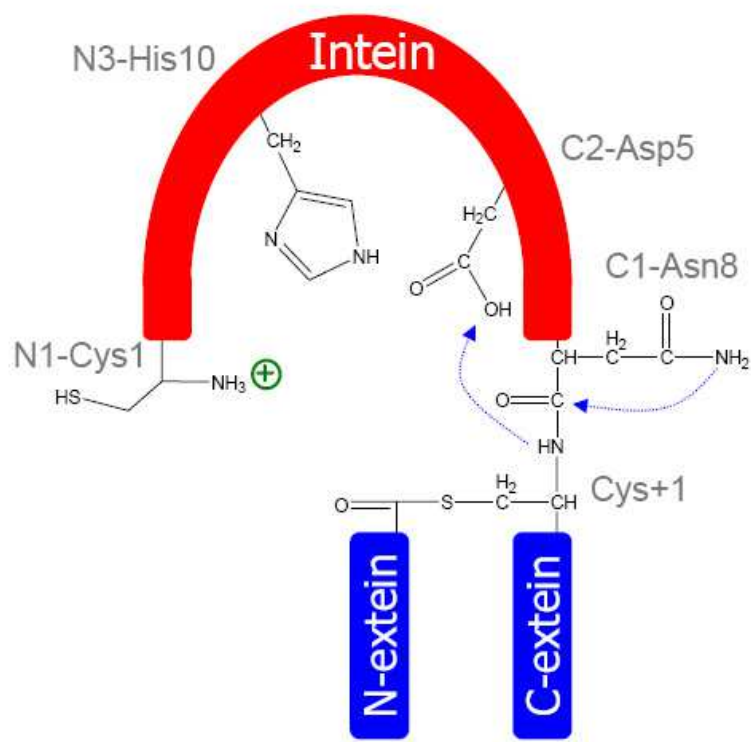}
\caption{\label{fig:splicing6} Joining of the N- and C-termini \textit{via} transesterification.  This reaction step is catalyzed once the C-terminal Cys+1 is a thiolate, which depends on the pK$_a$ of the side chain (discussed in text).}
\end{figure}

Transesterification occurs when the Cys+1 thiolate attacks the carbon N-terminal thioester (see Figure \ref{fig:splicing6}).  This reaction step bridges the N- and C-exteins and breaks the bond between the N-extein and the intein. Most likely because the crystal structure of the product protein does not include exteins, there is a spacing of approximately 8 \AA{} between N- and C-termini.  Because intein splicing does occur, it can be concluded that the spacing between N- and C-termini will increase in the crystal structure compared to the precursor due to charged end groups, solvent exposure, and possibly metal binding \cite{vanroey2007cam, mizutani2002psr, poland2000sii, klabunde1998csg, duan1997csp, ichiyanagi2000csa, sun2005csi}.  After N-terminal thioester formation, the activated C-extein Cys+1 thiolate was capable of attacking the N-terminal carbonyl, thus completing the transesterification step and fusing the N- and C-exteins.  The C-extein and intein remain chemically bonded until the C1-Asn8 cyclization reaction.

At this point, both the Cys residues (N-terminal N1-Cys1 and the C-extein Cys+1) have been independently ionized and have undergone a rearrangement reaction as a thiolate.   Shingledecker \textit{et al.} have shown that also for the \textit{Mtu} recA intein \cite{shingledecker2000rcr}, the pK$_a$ for the N1-Cys1 was 8.2, which is comparable to average values for the Cys side chain \cite{voet1995b}.  For Cys+1, the pK$_a$ was 5.8, which is considerably lower and would indicate the increased probability of finding Cys+1 as a negatively charged thiolate. The pK$_a$ of N1-Cys1 is not necessarily important for the splicing mechanism because the entire reaction will wait for this initial step to commence.  However, the low pK$_a$ for Cys+1 is of extreme interest, because the increased probability of finding a thiolate at Cys+1 indicates that there is some electrostatic explanation why the side chain is so likely to be ionized, which is consistent with the mechanism proposed in this work.  The Cys+1 attacking group must wait for the N-terminal thioester, which is much less stable than a peptide bond.  The low pK$_a$ indicates that the side chain ionization should be an energetically easy process, which should facilitate transesterification.

\subsection{C-terminal cleavage}
After transesterification, the subsequent reaction is C-terminal cleavage. In this reaction step, C1-Asn8 undergoes cyclization into a succinimide ring. Intein sequences and crystal structures indicate that either penultimate C1-His7 residue or  a C2-block Arg residue is necessary for stabilizing the oxyanion hole during the C-terminal cleavage step within the splicing reaction \cite{sun2005csi, gorbalenya1998nci, dassa2004psa, amitai2004psi}.  A hydrogen bond between the side chain -NH of either C1-His7 or N3-block Arg and the C=O of the scissile peptide bond is highly conserved in crystal structures, although it is not expected to be proton transfer from this residue that catalyzes C-terminal cleavage.

Based on previous experimental mutagenesis studies \cite{wood1999gsy}, splicing may be inhibited and C-terminal cleavage may be isolated by a C2-Asp5Gly mutation, for which the role of Asp or Gly is explained in the preceding section.  The C2-Asp5Gly mutant, termed the cleavage mutant (CM), also exhibited a higher activity in low pH, for which a mechanism was proposed in earlier chapters. Within the context of splicing, no striking pH dependence was observed, and although C-terminal cleavage still occurs, it occurs as a final reaction step and after the splicing reaction described above.  It is an extraordinary feature that C-terminal cleavage, which is extremely rapid in the CM, delays itself until after the completion of splicing's N-terminal reaction steps.  This can be explained by returning our focus to C2-Asp422, which was integral as a proton acceptor in triggering transesterification and the overall splicing reaction rate.

After transesterification, there is a proton on C2-Asp422 making the amino acid side chain neutral in charge (COOH).  Average and solvent exposed pK$_a$ values for aspartic acid are $\sim$3.9, indicating that at pH levels considered in splicing are much higher than typical average Asp pK$_a$ values.  The presence of the additional proton makes C2-Asp422 an acid (aspartic acid) and since it is protonated, an effect of a low pH environment is obtained, despite the overall biological pH level.  The ``effect'' of a low pH environment is important for C-terminal cleavage and consistent with experimental results. The COOH group on the side chain, which was only formed after the N-terminal N-S shift, acts as a catalytic acid and donates a proton to trigger the C-terminal cleavage reaction, the first truly irreversible step in the splicing reaction mechanism.

\begin{figure}
\centering
\includegraphics[scale=0.8]{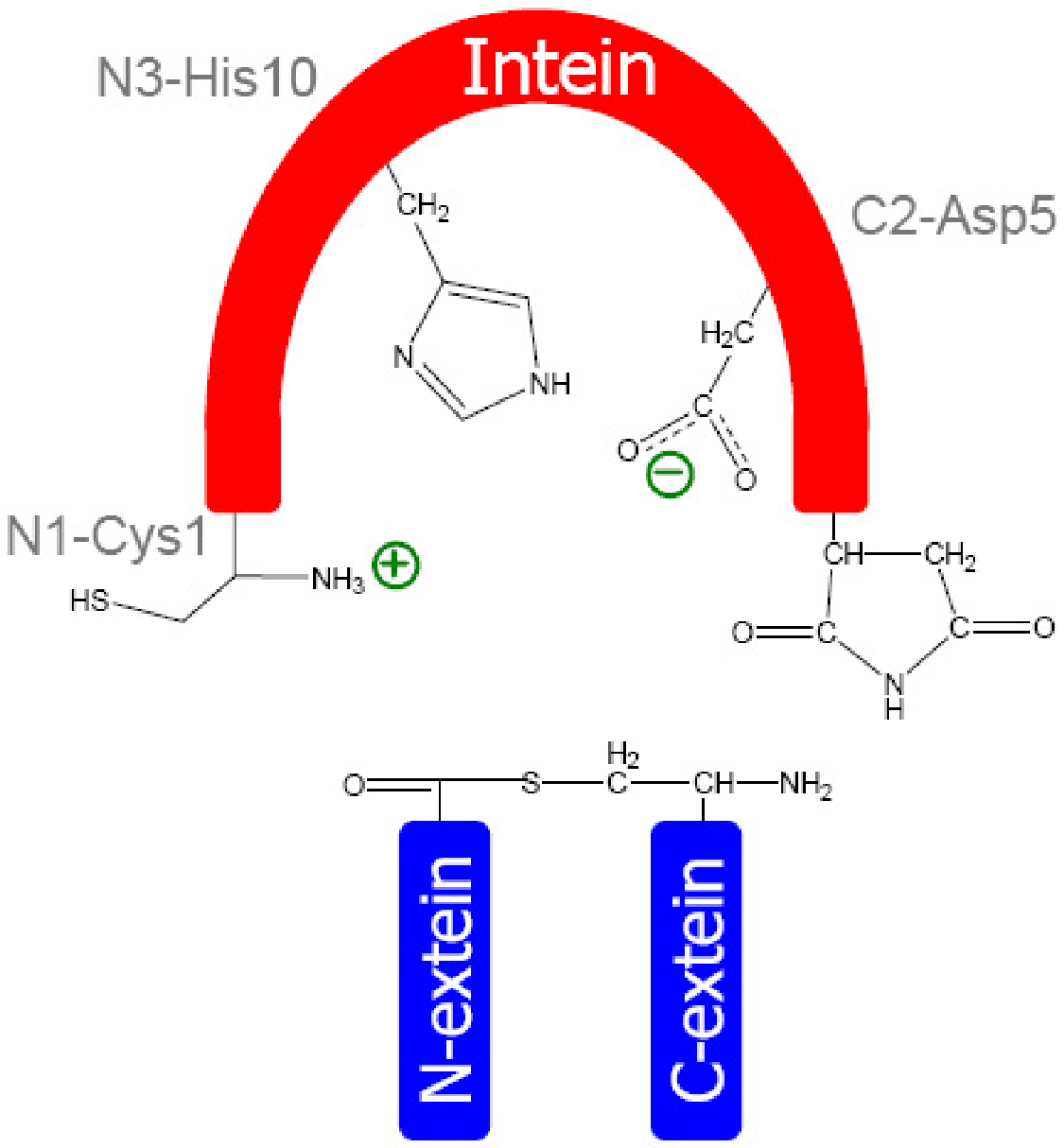}
\caption{\label{fig:splicing7} After C-terminal cleavage and before the finishing reactions, the N- and C-exteins are ligated and the intein is released from the host protein.  The finishing reaction will replace the thioester with a peptide bond.}
\end{figure}

\subsection{S-N shift (finishing reactions with succinimide hydrolysis)}
	After C1-Asn8 cyclization triggered by proton donation from C2-Asp5, a succinimide ring is formed and the ligated N- and C-exteins are released.  Two finishing reactions are expected (see Figure \ref{fig:splicing7}.  First, the branched thioester undergoes an S-N shift and rearranges to form a peptide bond, which is energetically more stable.  Second, succinimide may undergo hydrolysis and which would create a typical C-terminus (-COO$^-$) and a normal Asn or an iso-Asp residue, although in some crystal structures succinimide is observed.  For example, the final position of the intein in the \textit{Mtu} recA crystal structure \cite{vanroey2007cam} was observed in the crystal structure to be either a succinimide \cite{geiger1987dia, stephenson1989sfa} or a negatively charged carboxyl group.  The presence of succinimide gave insight to possible mechanisms for the isolated C-terminal cleavage reaction \cite{shemella, shemella2}.

\section{Conclusions: Splicing}
The mechanism for protein splicing of inteins was predicted based on first principles calculations.  Using electronic structure methods, particularly density functional theory, the energies and structures were studied along the reaction pathway.  Interesting observations include the catalyzing effect of the -NH$_3^+$ group of the N-terminal on the Cys+1 thiol group.  In addition, the highly conserved N1-Cys1, N3-His10, C2-Asp5, C1-Asn8 and Cys+1 residues are involved in the splicing mechanism.  An explanation for the lack of splicing activity with the C2-Asp5Gly mutation is discussed and the catalytic and essential role of C2-Asp5 as an acid and a base is explained.



\chapter{CONCLUSIONS}
\section{Discussion}
Intein-based technology is promising because a useful function is directly encoded into the intein sequence.  An intein cuts itself away from the host protein and joins its left and right neighbors in a predictable way.  Depending on sequence mutations and environmental stimuli, the function of inteins may be tuned, controlled, inhibited, or modified.  Inteins may be used as a nano-sensor device, where the ligation of the exteins brings two fluorescent proteins (as exteins) together that are easily detected.  Also, split inteins may be used for single molecule detection.  In this case, a small molecule can trigger a detection device, which would combine N- and C-intein fragments which would then allow/catalyze splicing.  

Because the enzyme-like intein is a component of the host protein (substrate) sequence, typical molecular inhibitors is not effective.  For standard enzyme catalysis, the enzyme is usually a small molecule which can be replaced with a molecule that is chemically similar but not active and therefore inhibits the reaction.  For inteins, the only evidence of chemical inhibition is from metal ions, which are unable to remain in the cell.  From this complication, using inteins for biotechnological purposes has been difficult.

For this reason, it is important to understand the reaction mechanism on the atomic level.  The role of protons is closely related to experimental crystal structures, random and site-directed mutagenesis, and NMR results.  For example, NMR can compliment the crystal structure by understanding the protonation states of side chains.  A complication with intein is that catalytic side chains such as Cys+1 are part of the exteins and are not included in the crystal structures. Mutagenesis experiments usually replace a normally active amino acid with something non-reactive, and the effect of this mutation on the reaction rate is measured.  The splicing and cleavage mechanisms are important to understand on the atomic-level because this is how the intein will be used as a nanoswitch, molecular sensor, or reporting device.  

\section{Results}
To investigate the reaction we have used first principles density functional theory methods to investigate the reaction mechanisms of inteins:  N-terminal thioester formation, and transesterification, and C-terminal cleavage.  Indeed, quantum mechanical methods allow for accurate prediction of energies related to biological systems, such as ionization potentials, electron affinities, binding energies, and energy barriers which could compliment experimental investigations.  We have used these methods to analyze the structure and energetics of intein systems.

\subsection{C-terminal cleavage}
For the C-terminal cleavage mutant, we have proposed a mechanism for C-terminal cleavage based on C1-Asn7 cyclization and succinimide formation.  Based on experimental results that showed the reaction to occur more rapidly in a low pH environment, our proposed mechanism takes this pH effect into account. We have also performed energy barrier calculations using QM, semi-empirical QM, implicit solvent QM, and QM/MM simulations to accurately recreate the reaction profile.

\subsection{Splicing}
In addition to C-terminal cleavage, we have studied the reaction steps of intein splicing, and found that the highly conserved N3-His10 plays an important role in accepting, then donating, a proton from the N1-Cys1.  After N-terminal thioester formation, the polarization effect of the N-terminal -NH$_3^+$ group forces C-terminal Cys+1 to donate a proton to C2-Asp5.  This step triggers transesterification, and the extra proton on C2-Asp5 catalyzes the C-terminal cleavage reaction in a way that mimics a low pH environment.  With the C2-Asp5Gly mutation (CM), the N-terminal -NH$_3^+$ group is not stabilized, and returns to the neutral -NH$_2$ terminal group which is unable to catalyze the transesterification step, thus inhibiting the splicing reaction.

\section{Future Research}
Future research will include understanding specific mutations affecting intein splicing, and how these mutations may make the reaction occur more rapidly or stimuli dependent.  For the \textit{Mtu} recA intein, the mutation of C2-Asp5Gly decouples the N- and C-termini leading to isolated and rapid C-terminal cleavage.  Also, the C-terminal Cys+1 residue may be mutated with the entire series of amino acids, and correlation with experiment tested. 

Splicing may still occur with the mutations of the catalytic Cys to Ser/Thr residues, although with lower reaction rate due to the decreased acidity of the Ser side chain (-OH) compared to Cys (-SH) \cite{shingledecker2000rcr}.  The effect of single and double Cys to Ser mutations can be studied with first principles calculations, and the intein splicing reaction may be tuned in order to control molecular docking related to inhibition or for a nanosensor device.

For the \textit{Ssp} dnaE split intein, the effect of mutation in the extein (Tyr to Asp) mutation will be studied, especially on its effect on the protonation state of N3-His10, a supposed proton acceptor/donor for N-terminal thioester formation.  One method to understand the effect of mutation is to calculate NMR chemical shifts for specific atoms based on a perturbed electrostatic environment due to mutation.


 
%
\specialhead{REFERENCES}
\bibliographystyle{unsrt} 
\begin{singlespace}
\bibliography{thesis} 

\begin{thebibliography}{10}

\bibitem{belforts}
M.~Belfort, B.L. Stoddard, D.W. Wood, and V.~Derbyshire.
\newblock {\em Homing Endonucleases and Inteins (Nucleic Acids and Molecular
  Biology)}.
\newblock Springer, 2005.

\bibitem{perler1994pse}
F.B. Perler, E.O. Davis, G.E. Dean, F.S. Gimble, W.E. Jack, N.~Neff, C.J.
  Noren, J.~Thorner, and M.~Belfort.
\newblock Protein splicing elements: inteins and exteins-—a definition of terms
  and recommended nomenclature.
\newblock {\em Nucleic Acids Research}, 22(7):1125--1127, 1994.

\bibitem{hiraga2005mas}
K.~Hiraga, V.~Derbyshire, J.T. Dansereau, P.~Van~Roey, and M.~Belfort.
\newblock {Minimization and stabilization of the Mycobacterium tuberculosis
  recA intein}.
\newblock {\em Journal of Molecular Biology}, 354(4):916--926, 2005.

\bibitem{shingledecker2000rcr}
K.~Shingledecker, S.~Q. Jiang, and H.~Paulus.
\newblock {Reactivity of the cysteine residues in the protein splicing active
  center of the Mycobacterium tuberculosis recA intein}.
\newblock {\em Archives of Biochemistry and Biophysics}, 375(1):138--144, 2000.

\bibitem{paulus2000psa}
H.~Paulus.
\newblock {Protein splicing and related forms of protein autoprocessing}.
\newblock {\em Annual Review of Biochemistry}, 69(1):447--496, 2000.

\bibitem{miao2005ssa}
J.~Miao, W.~Wu, T.~Spielmann, M.~Belfort, V.~Derbyshire, and G.~Belfort.
\newblock {Single-step affinity purification of toxic and non-toxic proteins on
  a fluidics platform}.
\newblock {\em Lab on a Chip}, 5(3):248--253, 2005.

\bibitem{banki2005sbu}
M.R. Banki, L.A. Feng, and D.W. Wood.
\newblock {Simple bioseparations using self-cleaving elastin-like polypeptide
  tags}.
\newblock {\em Nature Methods}, 2(9):659--661, 2005.

\bibitem{paulus2003itp}
H.~Paulus.
\newblock {Inteins as targets for potential antimycobacterial drugs}.
\newblock {\em Frontiers in Bioscience}, 8:s1157--s1165, 2003.

\bibitem{mootz2002pst}
H.D. Mootz and T.W. Muir.
\newblock {Protein splicing triggered by a small molecule}.
\newblock {\em Journal of the American Chemical Society}, 124(31):9044--9045,
  2002.

\bibitem{muralidharan2006ple}
V.~Muralidharan and T.W. Muir.
\newblock {Protein ligation: an enabling technology for the biophysical
  analysis of proteins}.
\newblock {\em Nature Methods}, 3:429--438, 2006.

\bibitem{vanroey2007cam}
P.~Van~Roey, B.~Pereira, Z.~Li, K.~Hiraga, M.~Belfort, and V.~Derbyshire.
\newblock {Crystallographic and mutational studies of Mycobacterium
  tuberculosis recA mini-inteins suggest a pivotal role for a highly conserved
  aspartate residue}.
\newblock {\em Journal of Molecular Biology}, 367(1):162--173, 2007.

\bibitem{mizutani2002psr}
R.~Mizutani, S.~Nogami, M.~Kawasaki, Y.~Ohya, Y.~Anraku, and Y.~Satow.
\newblock {Protein-splicing reaction via a thiazolidine intermediate: crystal
  structure of the VMA1-derived endonuclease bearing the N and C-terminal
  propeptides.}
\newblock {\em Journal of Molecular Biology}, 316(4):919--929, 2002.

\bibitem{poland2000sii}
B.W. Poland, M.Q. Xu, and F.A. Quiocho.
\newblock {Structural insights into the protein splicing mechanism of PI-SceI}.
\newblock {\em Journal of Biological Chemistry}, 275(22):16408--16413, 2000.

\bibitem{klabunde1998csg}
T.~Klabunde, S.~Sharma, A.~Telenti, W.R. Jacobs, and J.C. Sacchettini.
\newblock {Crystal structure of GyrA intein from Mycobacterium xenopi reveals
  structural basis of protein splicing}.
\newblock {\em Nature Structural Biology}, 5(1):31--36, 1998.

\bibitem{duan1997csp}
X.Q. Duan, F.S. Gimble, and F.A. Quiocho.
\newblock {Crystal structure of PI-SceI, a homing endonuclease with protein
  splicing activity}.
\newblock {\em Cell}, 89(4):555--564, 1997.

\bibitem{ichiyanagi2000csa}
K.~Ichiyanagi, Y.~Ishino, M.~Ariyoshi, K.~Komori, and K.~Morikawa.
\newblock {Crystal structure of an archaeal intein-encoded homing endonuclease
  PI-PfuI}.
\newblock {\em Journal of Molecular Biology}, 300(4):889--901, 2000.

\bibitem{sun2005csi}
P.~Sun, S.~Ye, S.~Ferrandon, T.C. Evans, M.Q. Xu, and Z.H. Rao.
\newblock {Crystal structures of an intein from the split dnaE gene of
  Synechocystis sp PCC6803 reveal the catalytic model without the penultimate
  histidine and the mechanism of zinc ion inhibition of protein splicing}.
\newblock {\em Journal of Molecular Biology}, 353(5):1093--1105, 2005.

\bibitem{johnson2007nsk}
M.A. Johnson, M.W. Southworth, T.~Herrmann, L.~Brace, F.B. Perler, and
  K.~Wuthrich.
\newblock {N}{M}{R} structure of a klba intein precursor from methanococcus
  jannaschii.
\newblock {\em Protein Science}, 2007.

\bibitem{wood1999gsy}
D.W. Wood, W.~Wu, G.~Belfort, V.~Derbyshire, and M.~Belfort.
\newblock {A genetic system yields self-cleaving inteins for bioseparations}.
\newblock {\em Nature Biotechnology}, 17(9):889--892, 1999.

\bibitem{southworth2000aps}
M.W. Southworth, J.~Benner, and F.B. Perler.
\newblock {An alternative protein splicing mechanism for inteins lacking an
  N-terminal nucleophile}.
\newblock {\em EMBO Journal}, 19(18):5019--5026, 2000.

\bibitem{chen2000psa}
L.X. Chen, J.~Benner, and F.B. Perler.
\newblock {Protein splicing in the absence of an intein penultimate histidine}.
\newblock {\em Journal of Biological Chemistry}, 275(27):20431--20435, 2000.

\bibitem{mills2004psp}
K.V. Mills, J.S. Manning, A.M. Garcia, and L.A. Wuerdeman.
\newblock {Protein splicing of a Pyrococcus abyssi intein with a C-terminal
  glutamine}.
\newblock {\em Journal of Biological Chemistry}, 279(20):20685--20691, 2004.

\bibitem{brian}
B.~Pereira.
\newblock Unpublished.

\bibitem{wood2000oss}
D.W. Wood, V.~Derbyshire, W.~Wu, M.~Chartrain, M.~Belfort, and G.~Belfort.
\newblock {Optimized single-step affinity purification with a self-cleaving
  intein applied to human acidic fibroblast growth factor}.
\newblock {\em Biotechnology Progess}, 16(6):1055--1063, 2000.

\bibitem{belfort2005gsa}
M.~Belfort, G.~Belfort, V.~Derbyshire, D.~Wood, and W.~Wu.
\newblock Genetic system and self-cleaving inteins derived therefrom,
  bioseparations and protein purification employing same, and methods for
  determining critical, generalizable amino acid residues for varying intein
  activity.
\newblock U.S. Patent \#6,933,362, 2005.

\bibitem{woodthesis}
D.W. Wood.
\newblock {\em {Generation and application of a self-cleaving protein linker
  for use in single-step affinity fusion based protein purification}}.
\newblock PhD thesis, Rensselaer Polytechnic Institute, 2000.

\bibitem{wu2002imp}
W.~Wu, D.W. Wood, G.~Belfort, V.~Derbyshire, and M.~Belfort.
\newblock {Intein-mediated purification of cytotoxic endonuclease I-TevI by
  insertional inactivation and pH-controllable splicing}.
\newblock {\em Nucleic Acids Research}, 30(22):4864--4871, 2002.

\bibitem{shemella}
P.~Shemella, B.~Pereira, Y.~Zhang, P.~Van~Roey, G.~Belfort, S.~Garde, and S.K.
  Nayak.
\newblock {Mechanism for intein C-Terminal cleavage: A proposal from quantum
  mechanical calculations}.
\newblock {\em Biophysical Journal}, 92(3):847--853, 2007.

\bibitem{shemella2}
P.~Shemella, B.~Pereira, P.~Van~Roey, G.~Belfort, S.~Garde, and S.K. Nayak.
\newblock {Controlling C-terminal cleavage rate of an intein through extein
  mutation: A quantum mechanical study}.
\newblock Submitted.

\bibitem{lew1998psv}
B.M. Lew, K.V. Mills, and H.~Paulus.
\newblock {Protein splicing in vitro with a semisynthetic two-component minimal
  intein}.
\newblock {\em Journal of Biological Chemistry}, 273(26):15887--15890, 1998.

\bibitem{hohenberg1964ieg}
P.~Hohenberg and W.~Kohn.
\newblock {Inhomogeneous electron gas}.
\newblock {\em Physical Review B}, 136(3B):864--871, 1964.

\bibitem{kohn1965sce}
W.~Kohn and L.J. Sham.
\newblock {Self-consistent equations including exchange and correlation
  effects}.
\newblock {\em Physical Review}, 140(4A):1133--1138, 1965.

\bibitem{feynman1998sm}
R.P. Feynman.
\newblock {\em {Statistical mechanics}}.
\newblock Addison-Wesley, 1998.

\bibitem{ceperley1980gse}
D.M. Ceperley and B.J. Alder.
\newblock {Ground state of the electron gas by a stochastic method}.
\newblock {\em Physical Review Letters}, 45(7):566--569, 1980.

\bibitem{martin2004esb}
R.M. Martin.
\newblock {\em {Electronic structure: Basic theory and practical methods}}.
\newblock Cambridge University Press, 2004.

\bibitem{wu2001tea}
X.~Wu, M.C. Vargas, S.K. Nayak, V.~Lotrich, and G.~Scoles.
\newblock {Towards extending the applicability of density functional theory to
  weakly bound systems}.
\newblock {\em Journal of Chemical Physics}, 115(19):8748--8757, 2001.

\bibitem{vanschilfgaarde2006qsc}
M.~van Schilfgaarde, T.~Kotani, and S.~Faleev.
\newblock {Quasiparticle self-consistent GW theory}.
\newblock {\em Physical Review Letters}, 96(22):226402, 2006.

\bibitem{becke1993}
A.D. Becke.
\newblock {Density-functional thermochemistry. III. The role of exact
  exchange}.
\newblock {\em Journal of Chemical Physics}, 98:5648--5652, 1993.

\bibitem{becke1988dfe}
A.D. Becke.
\newblock {Density-functional exchange-energy approximation with correct
  asymptotic behavior}.
\newblock {\em Physical Review A}, 38(6):3098--3100, 1988.

\bibitem{perdew1992aas}
J.P. Perdew and Y.~Wang.
\newblock {Accurate and simple analytic representation of the electron-gas
  correlation energy}.
\newblock {\em Physical Review B}, 45(23):13244--13249, 1992.

\bibitem{lee1988dcs}
C.T. Lee, W.T. Yang, and R.G. Parr.
\newblock {Development of the Colle-Salvetti correlation-energy formula into a
  functional of the electron density}.
\newblock {\em Physical Review B}, 37(2):785--789, 1988.

\bibitem{miehlich1989roc}
B.~Miehlich, A.~Savin, H.~Stoll, and H.~Preuss.
\newblock {Results obtained with the correlation energy density functionals of
  Becke and Lee, Yang and Parr}.
\newblock {\em Chemical Physics Letters}, 157(3):200--206, 1989.

\bibitem{gauss03}
M.~J. Frisch, G.~W. Trucks, H.~B. Schlegel, G.~E. Scuseria, M.~A. Robb, J.~R.
  Cheeseman, J.~A. Montgomery, Jr., T.~Vreven, K.~N. Kudin, J.~C. Burant, J.~M.
  Millam, S.~S. Iyengar, J.~Tomasi, V.~Barone, B.~Mennucci, M.~Cossi,
  G.~Scalmani, N.~Rega, G.~A. Petersson, H.~Nakatsuji, M.~Hada, M.~Ehara,
  K.~Toyota, R.~Fukuda, J.~Hasegawa, M.~Ishida, T.~Nakajima, Y.~Honda,
  O.~Kitao, H.~Nakai, M.~Klene, X.~Li, J.~E. Knox, H.~P. Hratchian, J.~B.
  Cross, V.~Bakken, C.~Adamo, J.~Jaramillo, R.~Gomperts, R.~E. Stratmann,
  O.~Yazyev, A.~J. Austin, R.~Cammi, C.~Pomelli, J.~W. Ochterski, P.~Y. Ayala,
  K.~Morokuma, G.~A. Voth, P.~Salvador, J.~J. Dannenberg, V.~G. Zakrzewski,
  S.~Dapprich, A.~D. Daniels, M.~C. Strain, O.~Farkas, D.~K. Malick, A.~D.
  Rabuck, K.~Raghavachari, J.~B. Foresman, J.~V. Ortiz, Q.~Cui, A.~G. Baboul,
  S.~Clifford, J.~Cioslowski, B.~B. Stefanov, G.~Liu, A.~Liashenko, P.~Piskorz,
  I.~Komaromi, R.~L. Martin, D.~J. Fox, T.~Keith, M.~A. Al-Laham, C.~Y. Peng,
  A.~Nanayakkara, M.~Challacombe, P.~M.~W. Gill, B.~Johnson, W.~Chen, M.~W.
  Wong, C.~Gonzalez, and J.~A. Pople.
\newblock Gaussian 03, \uppercase{R}evision \uppercase{C}.02.
\newblock \uppercase{G}aussian, Inc., Wallingford, CT, 2004.

\bibitem{zhang2005dmc}
X.D. Zhang, X.H. Zhang, and T.C. Bruice.
\newblock {A definitive mechanism for chorismate mutase}.
\newblock {\em Biochemistry}, 44(31):10443--10448, 2005.

\bibitem{zhang2006rmg}
X.D. Zhang and T.C. Bruice.
\newblock {Reaction mechanism of guanidinoacetate methyltransferase, concerted
  or step-wise}.
\newblock {\em Proceedings of the National Academy of Sciences},
  103(44):16141--16146, 2006.

\bibitem{moeller1934nat}
C.~Moeller and M.S. Plesset.
\newblock {Note on an approximation treatment for many-electron systems}.
\newblock {\em Physical Review}, 46(7):618--622, 1934.

\bibitem{headgordon1988mee}
M.~Head-Gordon, J.A. Pople, and M.J. Frisch.
\newblock {MP2 energy evaluation by direct methods}.
\newblock {\em Chemical Physics Letters}, 153(6):503--506, 1988.

\bibitem{frisch1990dmg}
M.J. Frisch, M.~Head-Gordon, and J.A. Pople.
\newblock {A direct MP2 gradient-method}.
\newblock {\em Chemical Physics Letters}, 166(3):275--280, 1990.

\bibitem{frisch1990sda}
M.J. Frisch, M.~Head-Gordon, and J.A. Pople.
\newblock {Semidirect algorithms for the MP2 energy and gradient}.
\newblock {\em Chemical Physics Letters}, 166(3):281--289, 1990.

\bibitem{hehre2003scm}
W.J. Hehre, R.~Ditchfield, and J.A. Pople.
\newblock {Self—-consistent molecular orbital methods. XII. Further extensions
  of Gaussian—type basis sets for use in molecular orbital studies of organic
  molecules}.
\newblock {\em Journal of Chemical Physics}, 56(5):2257--2261, 1972.

\bibitem{mclean2006cgb}
A.D. McLean and G.S. Chandler.
\newblock {Contracted Gaussian basis sets for molecular calculations. 1. Second
  row atoms, Z=11-18}.
\newblock {\em Journal of Chemical Physics}, 72(10):5639--5648, 2006.

\bibitem{stewart1989ops}
J.J.P. Stewart.
\newblock {Optimization of parameters for semiempirical methods I. Method}.
\newblock {\em Journal of Computational Chemistry}, 10(2):209--220, 1989.

\bibitem{ferguson1992vzap}
D.M. Ferguson, I.R. Gould, W.A. Glauser, S.~Schroeder, and P.A. Kollman.
\newblock {Comparison of ab initio, semiempirical, and molecular mechanics
  calculations for the conformational--analysis of ring--systems}.
\newblock {\em Journal of Computational Chemistry}, 13(4):525--532, 1992.

\bibitem{csonka1993acr}
G.I. Csonka.
\newblock {Analysis of the core-repulsion functions used in AM1 and PM3
  semiempirical calculations: conformational analysis of ring systems}.
\newblock {\em Journal of Computational Chemistry}, 14(8):895--898, 1993.

\bibitem{miertus1981eis}
S.~Miertus, E.~Scrocco, and J.~Tomasi.
\newblock {Electrostatic interaction of a solute with a continuum. A direct
  utilization of ab initio molecular potentials for the prevision of solvent
  effects}.
\newblock {\em Chemical Physics}, 55(11):117--129, 1981.

\bibitem{cances1997ese}
M.T. Cances, B.~Mennucci, and J.~Tomasi.
\newblock {A new integral equation formalism for the polarizable continuum
  model: Theoretical background and applications to isotropic and anisotropic
  dielectrics}.
\newblock {\em Journal of Chemical Physics}, 107(8):3032--3041, 1997.

\bibitem{maseras1995ini}
F.~Maseras and K.~Morokuma.
\newblock {IMOMM: A new integrated ab initio plus molecular mechanics geometry
  optimization scheme of equilibrium structures and transition states}.
\newblock {\em Journal of Computational Chemistry}, 16(9):1170--1179, 1995.

\bibitem{vreven2003goq}
T.~Vreven, K.~Morokuma, O.~Farkas, H.B. Schlegel, and M.J. Frisch.
\newblock {Geometry optimization with QM/MM, ONIOM, and other combined methods.
  I. Microiterations and constraints}.
\newblock {\em Journal of Computational Chemistry}, 24(6):760--769, 2003.

\bibitem{thompson1995esb}
M.A. Thompson and G.K. Schenter.
\newblock {Excited states of the Bacteriochlorophyll b dimer of
  Rhodopseudomonas viridis: A QM/MM study of the photosynthetic reaction center
  that includes MM polarization}.
\newblock {\em Journal of Physical Chemistry}, 99(17):6374--6386, 1995.

\bibitem{schoneboom2002eos}
J.C. Schoneboom, H.~Lin, N.~Reuter, W.~Thiel, S.~Cohen, F.~Ogliaro, and
  S.~Shaik.
\newblock {The elusive oxidant species of cytochrome P450 enzymes:
  Characterization by combined quantum mechanical/molecular mechanical (QM/MM)
  calculations}.
\newblock {\em Journal of the American Chemical Society}, 124(27):8142--8151,
  2002.

\bibitem{gao2002qmm}
J.L. Gao and D.G. Truhlar.
\newblock {Quantum mechanical methods for enzyme kinetics}.
\newblock {\em Annual Review of Physical Chemistry}, 53:467--505, 2002.

\bibitem{cui2002tap}
Q.~Cui, M.~Elstner, and M.~Karplus.
\newblock {A theoretical analysis of the proton and hydride transfer in liver
  alcohol dehydrogenase (LADH)}.
\newblock {\em Journal of Physical Chemistry B}, 106(10):2721--2740, 2002.

\bibitem{torrent2002epe}
M.~Torrent, T.~Vreven, D.G. Musaev, K.~Morokuma, O.~Farkas, and H.B. Schlegel.
\newblock {Effects of the protein environment on the structure and energetics
  of active sites of metalloenzymes. ONIOM study of methane monooxygenase and
  ribonucleotide reductase}.
\newblock {\em Journal of the American Chemical Society}, 124(2):192--193,
  2002.

\bibitem{dellago2006tps}
C.~Dellago, P.G. Bolhuis, F.S. Csajka, and D.~Chandler.
\newblock {Transition path sampling and the calculation of rate constants}.
\newblock {\em Journal of Chemical Physics}, 108(5):1964--1977, 1998.

\bibitem{irikura1998ctp}
K.K. Irikura and D.J. Frurip.
\newblock {\em {Computational thermochemistry: Prediction and estimation of
  molecular thermodynamics}}.
\newblock American Chemical Society, 1998.

\bibitem{ochterski2000tg}
J.W. Ochterski.
\newblock {\em {Thermochemistry in Gaussian}}.
\newblock Gaussian Inc., 2000.

\bibitem{geiger1987dia}
T.~Geiger and S.~Clarke.
\newblock {Deamidation, isomerization, and racemization at asparaginyl and
  aspartyl residues in peptides. Succinimide-linked reactions that contribute
  to protein degradation}.
\newblock {\em Journal of Biological Chemistry}, 262(2):785--794, 1987.

\bibitem{capasso1996kam}
S.~Capasso, L.~Mazzarella, G.~Sorrentino, G.~Balboni, and A.J. Kirby.
\newblock {Kinetics and mechanism of the cleavage of the peptide bond next to
  asparagine}.
\newblock {\em Peptides}, 17(6):1075--1077, 1996.

\bibitem{catanzano1997tae}
F.~Catanzano, G.~Graziano, S.~Capasso, and G.~Barone.
\newblock {Thermodynamic analysis of the effect of selective monodeamidation at
  asparagine 67 in ribonuclease A}.
\newblock {\em Protein Science}, 6(8):1682--1693, 1997.

\bibitem{peters2006adp}
B.~Peters and B.L. Trout.
\newblock {Asparagine deamidation: pH-dependent mechanism from density
  functional theory}.
\newblock {\em Biochemistry}, 45(16):5384--5392, 2006.

\bibitem{ding2003csm}
Y.~Ding, M.Q. Xu, I.~Ghosh, X.H. Chen, S.~Ferrandon, G.~Lesage, and Z.H. Rao.
\newblock {Crystal structure of a mini-intein reveals a conserved catalytic
  module involved in side chain cyclization of asparagine during protein
  splicing}.
\newblock {\em Journal of Biological Chemistry}, 278(40):39133--39142, 2003.

\bibitem{milnerwhite1997pcn}
E.J. Milner-White.
\newblock {The partial charge of the nitrogen atom in peptide bonds}.
\newblock {\em Protein Science}, 6(11):2477--2482, 1997.

\bibitem{voet1995b}
D.~Voet and J.G. Voet.
\newblock {\em {Biochemistry}}.
\newblock Wiley, 1995.

\bibitem{trylska2004rhb}
J.~Trylska, P.~Grochowski, and J.A. McCammon.
\newblock {The role of hydrogen bonding in the enzymatic reaction catalyzed by
  HIV-1 protease}.
\newblock {\em Protein Science}, 13(2):513--528, 2004.

\bibitem{krug1992tsn}
J.P. Krug, P.L.A. Popelier, and R.F.W. Bader.
\newblock {Theoretical study of neutral and of acid and base-promoted
  hydrolysis of formamide}.
\newblock {\em Journal of Physical Chemistry}, 96(19):7604--7616, 1992.

\bibitem{eppler2006wda}
R.K. Eppler, R.S. Komor, J.~Huynh, J.S. Dordick, J.A. Reimer, and D.S. Clark.
\newblock {Water dynamics and salt-activation of enzymes in organic media:
  Mechanistic implications revealed by NMR spectroscopy}.
\newblock {\em Proceedings of the National Academy of Sciences},
  103(15):5706--5710, 2006.

\bibitem{davis1991nsr}
E.O. Davis, S.G. Sedgwick, and M.J. Colston.
\newblock {Novel structure of the recA locus of Mycobacterium tuberculosis
  implies processing of the gene product}.
\newblock {\em Journal of Bacteriology}, 173(18):5653--5662, 1991.

\bibitem{cornell1995sgf}
W.D. Cornell, P.~Cieplak, C.I. Bayly, I.R. Gould, K.M. Merz, D.M. Ferguson,
  D.C. Spellmeyer, T.~Fox, J.W. Caldwell, and P.A. Kollman.
\newblock {A second generation force field for the simulation of proteins,
  nucleic acids, and organic molecules}.
\newblock {\em Journal of the American Chemical Society}, 117(19):5179--5197,
  1995.

\bibitem{vanderspoel2005gff}
D.~Van~der Spoel, E.~Lindahl, B.~Hess, G.~Groenhof, A.E. Mark, and H.J.C.
  Berendsen.
\newblock {GROMACS: Fast, flexible, and free}.
\newblock {\em Journal of Compututational Chemistry}, 26(16):1701--1718, 2005.

\bibitem{shemella1}
P.~Shemella and S.K. Nayak.
\newblock Unpublished.

\bibitem{reed1988iin}
A.E. Reed, L.A. Curtiss, and F.~Weinhold.
\newblock {Intermolecular interactions from a natural bond orbital,
  donor-acceptor viewpoint}.
\newblock {\em Chemical Reviews}, 88(6):899--926, 1988.

\bibitem{bai1993pse}
Y.W. Bai, J.S. Milne, L.~Mayne, and S.W. Englander.
\newblock {Primary structure effects hydrogen exchange}.
\newblock {\em PROTEINS: Structure, Function, and Genetics}, 17(1):75--86,
  1993.

\bibitem{vmd}
W.~Humphrey, A.~Dalke, and K.~Schulten.
\newblock {VMD}: {V}isual {M}olecular {D}ynamics.
\newblock {\em Journal of Molecular Graphics}, 14(1):33--38, 1996.

\bibitem{pereira2006qpc}
B.~Pereira, S.~Jain, and S.~Garde.
\newblock {Quantifying the protein core flexibility through analysis of cavity
  formation}.
\newblock {\em Journal of Chemical Physics}, 124:074704, 2006.

\bibitem{mulliken1955epa}
R.S. Mulliken.
\newblock {Electronic population analysis on LCAO [single bond] MO molecular
  wave functions. I}.
\newblock {\em Journal of Chemical Physics}, 23(10):1833--1840, 1955.

\bibitem{lew2002vss}
B.M. Lew and H.~Paulus.
\newblock {An in vivo screening system against protein splicing useful for the
  isolation of non-splicing mutants or inhibitors of the RecA intein of
  Mycobacterium tuberculosis}.
\newblock {\em Gene}, 282(1-2):169--177, 2002.

\bibitem{mills2001rip}
K.V. Mills and H.~Paulus.
\newblock {Reversible inhibition of protein splicing by zinc ion}.
\newblock {\em Journal of Biological Chemistry}, 276(14):10832--10838, 2001.

\bibitem{dassa2007tps}
B.~Dassa, G.~Amitai, J.~Caspi, O.~Schueler-Furman, and S.~Pietrokovski.
\newblock {Trans protein splicing of cyanobacterial split inteins in endogenous
  and exogenous combinations.}
\newblock {\em Biochemistry}, 46(1):322--330, 2007.

\bibitem{clarke1994pms}
N.D. Clarke.
\newblock {A proposed mechanism for the self-splicing of proteins}.
\newblock {\em Proceedings of the National Academy of Sciences of the United
  States of America}, 91(23):11084--11088, 1994.

\bibitem{amitai1999fte}
G.~Amitai and S.~Pietrokovski.
\newblock {Fine-tuning an engineered intein}.
\newblock {\em Nature Biotechnology}, 17(9):854--855, 1999.

\bibitem{iwai2006hep}
H.~Iwai, S.~Zuger, J.~Jin, and P.H. Tam.
\newblock {Highly efficient protein trans-splicing by a naturally split DnaE
  intein from Nostoc punctiforme}.
\newblock {\em FEBS Letters}, 580(7):1853--1858, 2006.

\bibitem{gorbalenya1998nci}
A.E. Gorbalenya.
\newblock {Non-canonical inteins}.
\newblock {\em Nucleic Acids Research}, 26(7):1742, 1998.

\bibitem{dassa2004psa}
B.~Dassa, H.~Haviv, G.~Amitai, and S.~Pietrokovski.
\newblock {Protein splicing and auto-cleavage of bacterial intein-like domains
  lacking a C'-flanking nucleophilic residue}.
\newblock {\em Journal of Biological Chemistry}, 279(31):32001--32007, 2004.

\bibitem{amitai2004psi}
G.~Amitai, B.~Dassa, and S.~Pietrokovski.
\newblock {Protein splicing of inteins with atypical glutamine and aspartate
  C-terminal residues}.
\newblock {\em Journal of Biological Chemistry}, 279(5):3121--3131, 2004.

\bibitem{stephenson1989sfa}
R.C. Stephenson and S.~Clarke.
\newblock {Succinimide formation from aspartyl and asparaginyl peptides as a
  model for the spontaneous degradation of proteins}.
\newblock {\em Journal of Biological Chemistry}, 264(11):6164--6170, 1989.

\end{thebibliography}
\end{singlespace}

\appendix    
\addcontentsline{toc}{chapter}{APPENDICES}             

\chapter{Amino Acid Chart}
Twenty natural amino acids (source http://www.neb.com).
\begin{figure*}[h]
 \centering    
 \includegraphics[scale=0.5]{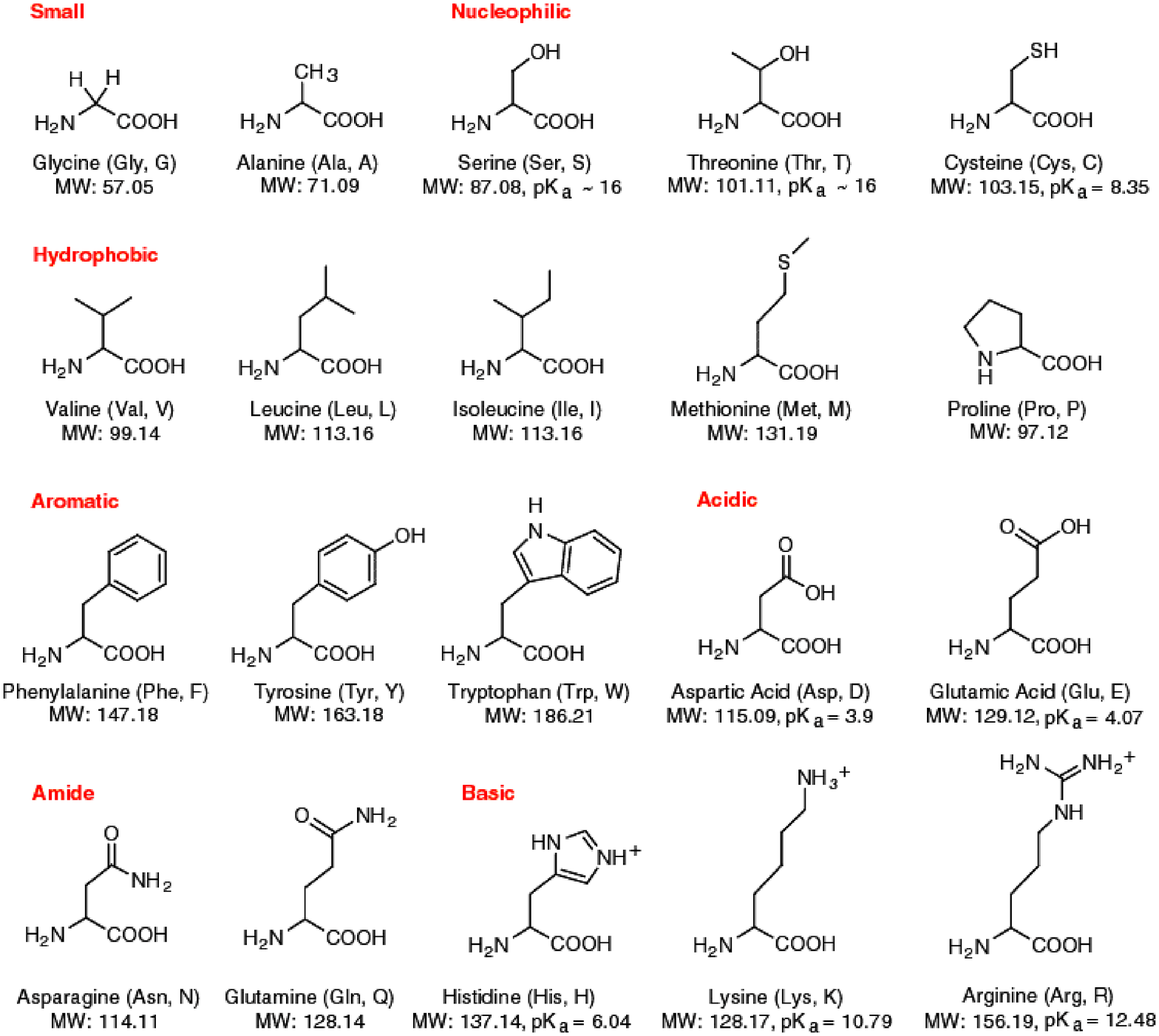}
\end{figure*}

\chapter{Intein sequences, motifs, and numbering scheme}
\begin{table}[h]
\vspace*{-12pt} 
\label{tab:sequence} 
\begin{center}
\tiny
\renewcommand\arraystretch{.91} 
\begin{tabular}{ccccccccc}
\hline
\hline
Old block & New block & Position	&	\textit{Mtu} recA	&	WT+EN	&	\textit{Mtu} recA 	&	\textit{Mtu} recA	&	\textit{Ssp}6803 	&	\textit{Ssp}6803 	\\
& &  on block	& &	&	110$\bigtriangleup$383 & 94$\bigtriangleup$403 &	dnaE &	dnaE	\\
\hline
A	&	N1	&	1	&	Cys	&	1	&	1	&	1	&	Cys	&	1	\\
A	&	N1	&	2	&	Leu	&	2	&	2	&	2	&	Leu	&	2	\\
A	&	N1	&	3	&	Ala	&	3	&	3	&	3	&	Ser	&	3	\\
A	&	N1	&	4	&	Glu	&	4	&	4	&	4	&	Phe	&	4	\\
A	&	N1	&	5	&	Gly	&	5	&	5	&	5	&	Gly	&	5	\\
A	&	N1	&	6	&	Thr	&	6	&	6	&	6	&	Thr	&	6	\\
A	&	N1	&	7	&	Arg	&	7	&	7	&	7	&	Glu	&	7	\\
A	&	N1	&	8	&	Ile	&	8	&	8	&	8	&	Ile	&	8	\\
A	&	N1	&	9	&	Phe	&	9	&	9	&	9	&	Leu	&	9	\\
A	&	N1	&	10	&	Asp	&	10	&	10	&	10	&	Thr	&	10	\\
A	&	N1	&	11	&	Pro	&	11	&	11	&	11	&	Val	&	11	\\
A	&	N1	&	12	&	Val	&	12	&	12	&	12	&	Glu	&	12	\\
A	&	N1	&	13	&	Thr	&	13	&	13	&	13	&	Tyr	&	13	\\
-	&	N2	&	1	&	His	&	17	&	17	&	17	&	Leu	&	16	\\
-	&	N2	&	2	&	Arg	&	18	&	18	&	18	&	Pro	&	17	\\
-	&	N2	&	3	&	Ile	&	19	&	19	&	19	&	Ile	&	18	\\
-	&	N2	&	4	&	Glu	&	20	&	20	&	20	&	Gly	&	19	\\
-	&	N2	&	5	&	Asp	&	21	&	21	&	21	&	Lys	&	20	\\
-	&	N2	&	6	&	Val	&	22	&	22	&	22	&	Ile	&	21	\\
-	&	N2	&	7	&	Val	&	23	&	23	&	23	&	Val	&	22	\\
-	&	N2	&	8	&	Asp	&	24	&	24	&	24	&	Ser	&	23	\\
B	&	N3	&	1	&	Gly	&	64	&	64	&	64	&	Gly	&	63	\\
B	&	N3	&	2	&	Ala	&	65	&	65	&	65	&	Ser	&	64	\\
B	&	N3	&	3	&	Ile	&	66	&	66	&	66	&	Val	&	65	\\
B	&	N3	&	4	&	Val	&	67	&	67	&	67	&	Ile	&	66	\\
B	&	N3	&	5	&	Trp	&	68	&	68	&	68	&	Arg	&	67	\\
B	&	N3	&	6	&	Ala	&	69	&	69	&	69	&	Ala	&	68	\\
B	&	N3	&	7	&	Thr	&	70	&	70	&	70	&	Thr	&	69	\\
B	&	N3	&	8	&	Pro	&	71	&	71	&	71	&	Ser	&	70	\\
B	&	N3	&	9	&	Asp	&	72	&	72	&	72	&	Asp	&	71	\\
B	&	N3	&	10	&	His	&	73	&	73	&	73	&	His	&	72	\\
B	&	N3	&	11	&	Lys	&	74	&	74	&	74	&	Arg	&	73	\\
B	&	N3	&	12	&	Val	&	75	&	75	&	75	&	Phe	&	74	\\
B	&	N3	&	13	&	Leu	&	76	&	76	&	76	&	Leu	&	75	\\
B	&	N3	&	14	&	Thr	&	77	&	77	&	77	&	Thr	&	76	\\
-	&	N4	&	1	&	Trp	&	81	&	81	&	81	&	Leu	&	81	\\
-	&	N4	&	2	&	Arg	&	82	&	82	&	82	&	Leu	&	82	\\
-	&	N4	&	3	&	Ala	&	83	&	83	&	83	&	Ala	&	83	\\
-	&	N4	&	4	&	Ala	&	84	&	84	&	84	&	Ile	&	84	\\
-	&	N4	&	5	&	Gly	&	85	&	85	&	85	&	Glu	&	85	\\
-	&	N4	&	6	&	Glu	&	86	&	86	&	86	&	Glu	&	86	\\
-	&	N4	&	7	&	Leu	&	87	&	87	&	87	&	Ile	&	87	\\
-	&	N4	&	8	&	Arg	&	88	&	88	&	88	&	Phe	&	88	\\
-	&	N4	&	9	&	Lys	&	89	&	89	&	89	&	Ala	&	89	\\
-	&	N4	&	10	&	Gly	&	90	&	90	&	90	&	Arg	&	90	\\
-	&	N4	&	11	&	Asp	&	91	&	91	&	91	&	Gln	&	91	\\
-	&	N4	&	12	&	Arg	&	92	&	92	&	92	&	Leu	&	92	\\
-	&	N4	&	13	&	Val	&	93	&	93	&	93	&	Asp	&	93	\\
-	&	N4	&	14	&	Ala	&	94	&	94	&	94	&	Leu	&	94	\\
-	&	N4	&	15	&	Gln	&	95	&	95	&	-	&	Leu	&	95	\\
-	&	N4	&	16	&	Pro	&	96	&	96	&	-	&	Thr	&	96	\\
F	&	C2	&	1	&	Ala	&	418	&	146	&	119	&	Gln	&	136	\\
F	&	C2	&	2	&	Arg	&	419	&	147	&	120	&	Arg	&	137	\\
F	&	C2	&	3	&	Thr	&	420	&	148	&	121	&	Ile	&	138	\\
F	&	C2	&	4	&	Phe	&	421	&	149	&	122	&	Phe	&	139	\\
F	&	C2	&	5	&	Asp	&	422	&	150	&	123	&	Asp	&	140	\\
F	&	C2	&	6	&	Leu	&	423	&	151	&	124	&	Ile	&	141	\\
F	&	C2	&	7	&	Glu	&	424	&	152	&	125	&	Gly	&	142	\\
F	&	C2	&	8	&	Val	&	425	&	153	&	126	&	Leu	&	143	\\
F	&	C2	&	9	&	Glu	&	426	&	154	&	127	&	Pro	&	144	\\
F	&	C2	&	10	&	Glu	&	427	&	155	&	128	&	Gln	&	145	\\
F	&	C2	&	11	&	Leu	&	428	&	156	&	129	&	Asp	&	146	\\
F	&	C2	&	12	&	His	&	429	&	157	&	130	&	His	&	147	\\
F	&	C2	&	13	&	Thr	&	430	&	158	&	131	&	Asn	&	148	\\
F	&	C2	&	14	&	Leu	&	431	&	159	&	132	&	Phe	&	149	\\
F	&	C2	&	15	&	Val	&	432	&	160	&	133	&	Leu	&	150	\\
G	&	C1	&	1	&	Ala	&	433	&	161	&	134	&	Ala	&	152	\\
G	&	C1	&	2	&	Glu	&	434	&	162	&	135	&	Asn	&	153	\\
G	&	C1	&	3	&	Gly	&	435	&	163	&	136	&	Gly	&	154	\\
G	&	C1	&	4	&	Val	&	436	&	164	&	137	&	Ala	&	155	\\
G	&	C1	&	5	&	Val	&	437	&	165	&	138	&	Ile	&	156	\\
G	&	C1	&	6	&	Val	&	438	&	166	&	139	&	Ala	&	157	\\
G	&	C1	&	7	&	His	&	439	&	167	&	140	&	Ala	&	158	\\
G	&	C1	&	8	&	Asn	&	440	&	168	&	141	&	Asn	&	159	\\
G	&	C1	&	1	&	Cys	&	441	&	169	&	142	&	Cys	&	160	\\
\hline
\end{tabular}
\end{center}
\end{table}


\end{document}